\documentclass[twocolumn,trackchanges,twocolappendix]{aastex631} 
\shorttitle{Ultra-Diffuse Galaxies – a Distinct Population?}
\shortauthors{Zöller et al.}
\graphicspath{{./}{figures/}}
\usepackage{amsmath}
\usepackage{todonotes}

\begin{document}

\title{Ultra-Diffuse Galaxies -- A Distinct Population?\\
Dwarf Galaxies in the Coma Cluster and A262 from Deep $u'$--$g'$--$r'$ Wendelstein Imaging Data}

\author[0000-0002-0938-5686]{Raphael Zöller}
\thanks{rzoeller@mpe.mpg.de}
\affiliation{University Observatory, Faculty of Physics, Ludwig-Maximilians-Universität München, Scheinerstr. 1, 81679 Munich, Germany}
\affiliation{Max Planck Institute for Extraterrestrial Physics, Giessenbachstrasse, D-85748 Garching, Germany}
\author[0000-0002-9618-2552]{Matthias Kluge}
\affiliation{University Observatory, Faculty of Physics, Ludwig-Maximilians-Universität München, Scheinerstr. 1, 81679 Munich, Germany}
\affiliation{Max Planck Institute for Extraterrestrial Physics, Giessenbachstrasse, D-85748 Garching, Germany}
\author[0009-0003-1182-147X]{Benjamin Staiger}
\affiliation{University Observatory, Faculty of Physics, Ludwig-Maximilians-Universität München, Scheinerstr. 1, 81679 Munich, Germany}
\author[0000-0001-7179-0626]{Ralf Bender}
\affiliation{University Observatory, Faculty of Physics, Ludwig-Maximilians-Universität München, Scheinerstr. 1, 81679 Munich, Germany}
\affiliation{Max Planck Institute for Extraterrestrial Physics, Giessenbachstrasse, D-85748 Garching, Germany}



\begin{abstract}
In this study, we compare the structural parameters of ultra-diffuse galaxies (UDGs) to those of other dwarf galaxies and investigate whether UDGs form a distinct population. We observed deep $u'$-, $g'$-, and $r'$-band images (maximum limiting surface brightness [3$\sigma$, $10\arcsec\times10\arcsec$]  $u'$ and $g'$: $\mathrm{\approx 30\,mag\,arcsec^{-2}}$;  $r'$: $\mathrm{\approx 29\,mag\,arcsec^{-2}}$) of A1656 (Coma cluster) and A262 with the Wendelstein Wide Field Imager at the 2.1\,m-Fraunhofer Wendelstein Telescope at the Wendelstein Observatory. We measure $u'-g'$ and $g'-r'$ colors and structural parameters using parametric fitting of tens of thousands of potential UDGs and other dwarf galaxies. Cluster members are identified and separated from diffuse background galaxies based on red sequence membership and location in the $u'-g'$ vs. $g'-r'$  color--color diagram. We find 11 UDGs in A262 and 48 UDGs in A1656. The latter is six times more than van Dokkum et al. found in the overlapping region. By comparing the structural parameters of UDGs to non-UDGs in our sample and to spheroidals from the literature, we do not find any separation in all tested parameter spaces. Instead, UDGs form the diffuse end of the already well-known spheroidal population and slightly extend it. Furthermore, we find that the UDG definition used by Koda et al. and Yagi et al. mainly extends the definition by van Dokkum et al. toward ordinary spheroidals. 
\end{abstract}

\keywords{Galaxy photometry(611) -- Low surface brightness galaxies(940) -- Dwarf galaxies(416) -- Dwarf spheroidal galaxies(420) -- Galaxies(573) -- Galaxy structure(622) -- Galaxy evolution(594) --  Galaxy clusters(584) -- Coma Cluster(270) -- Abell clusters(9) -- Extragalactic astronomy(506) }


\section{Introduction} \label{sec:intro}
Ultra-diffuse galaxies (UDGs) are faint but unusually large galaxies. Some of them have effective radii ($R_e$) comparable to the Milky Way but only $\approx 1/1000$ of its stellar mass \citep{vanDokkum2015}. They were first studied and defined by \citet{vanDokkum2015} who found 47 UDGs in the Coma cluster (A1656) using the Dragonfly (DF) Telephoto Array \citep{Abraham2014}. UDGs are defined via their extremely faint central surface brightness ($\mu_0 >24\,g\,\mathrm{mag\,arcsec^{-2}}$) and large effective radii ($R_e>1.5\,\mathrm{kpc}$). \citet{Koda} and \citet{Yagi2016} also studied UDGs in the Coma cluster but using their own UDG definition: total absolute magnitudes of $- 17 < M_R < - 9$, $\mathrm{FWHM>1.9\,kpc}$, $R_e>0.7\,\mathrm{kpc}$, a faint mean surface brightness within $R_e$ of $\langle\mu_{\rm e}\rangle>24\,R\,\mathrm{mag\,arcsec^{-2}}$, and shallow central light profiles specified as the difference between the surface brightness at $R_e$ and the mean surface brightness within $R_e$ ($\mu_e-\langle\mu_{\rm e}\rangle<0.8\,\mathrm{mag\,arcsec^{-2}}$). According to their definition, they found 854 UDGs in the Coma cluster. In this paper, we discuss the impact of using this different UDG definition on the number and type of galaxies classified as UDGs.

First estimates of the dark matter fraction of UDGs were based on the argument that such diffuse galaxies could only survive the cluster central tidal forces when a large fraction of their total mass is in the form of dark matter ($>98\%$, \citealt{vanDokkum2015}; $>99\%$, \citealt{Koda}). This raised the question of whether those galaxies with this presumably high dark matter fraction could solve or at least significantly reduce the missing satellite problem \citep[e.g.,][]{Mateo1998}. However, the number of newly discovered galaxies is too low to solve this problem \citep{Yagi2016}.

Since UDGs were first defined, their dark matter content has been debated. The results range from  undermassive dark matter halos or even dark-matter-free UDGs \citep{vanDokkum2019DF4,Danieli2019,vanDokkum2022} to overmassive dark matter halos \citep{Beasley,vanDokkum2016,vanDokkum2019DF44,Forbes2021,Gannon2023}. Furthermore, the number of globular clusters (GC) correlates with the dark matter halo mass \citep{Harris2013}, and UDGs have significantly varying GC counts ranging from GC-poor to GC-rich systems which indicates a strongly varying dark matter content of UDGs \citep[see, e.g.][]{BeasleyTrujillo2016,Beasley,vanDokkum2017,Amorisco2018,Forbes2020,Gannon2022}.

Such UDGs with overmassive dark matter halos are dark-matter-dominated even in the center and, hence, provide an extreme probe to study the dark matter profiles in the center of galaxies with (nearly) no direct influence of baryonic matter \citep{vanDokkum2019DF44}. However, due to their shallow surface brightness, studying their spatially resolved stellar kinematics and inferring the underlying gravitational potential requires an enormous amount of telescope time and, hence, is rarely done. \citet{vanDokkum2019DF44} measured the velocity dispersion profile of the Coma cluster UDG DF44 and found the profile to be consistent with either a \citet{DiCintio2014} core profile or with a Narvarro–Frenk–White (NFW) profile \citep{NFW}, but the latter requires a high tangential orbit anisotropy. Furthermore, \citet{Forbes2021} showed that the halo mass within $R_e$ of NGC 5846\_UDG1 favors a cored \citet{DiCintio2014} or \citet{Burkert1995} mass profile over a cuspy NFW profile.
However, UDGs are not the sole providers of such a probe of dark-matter-dominated centers. Also, multiple spheroidals from the Local Group are known to be dark-matter-dominated within $R_e$, some of them even with significantly larger $M_{\mathrm{dyn}}/L$ ratios \citep{BattagliaNipoti2022}. Note, that spheroidals are also frequently referred to as dwarf spheroidals (dSph) or dwarf ellipticals (dE). Similar to UDGs, the $M_{\mathrm{dyn}}/L$ ratios of Local Group spheroidals vary strongly \citep{BattagliaNipoti2022}. 

In addition to the dark matter content, the formation and survival of such diffuse galaxies remain two of the main puzzles of UDGs. The first potential formation scenario proposed by \citet{vanDokkum2015} suggests that UDGs might be failed $L_*$ galaxies that were quenched (e.g., by ram pressure stripping) at high redshift before forming a second generation of stars. Another possibility could be that they were formed in the tail of such stripped gas \citep{Poggianti2019} which could explain the existence of UDGs with undermassive halos. Moreover, UDGs could be created by gas outflows due to star formation feedback and subsequent expansion of the galaxy \citep{diCintio}. \citet{Wright2021} showed that field UDGs can be formed by early mergers that severely increase the spin temporarily, causing a migration of star formation to the galaxy outskirts leaving shallow centers behind. A further formation scenario predicted by \citet{Shin2020} is high-velocity collisions of galaxies. In a supersonic collision of two (or more) gas-rich dwarf galaxies, the gas is separated from the dark matter halos. The latter continue on their trajectories, whereas the gas is compressed by the shock and tidal interaction leading to the formation of stars and, subsequently, a UDG with an undermassive dark matter halo. Such a system was identified by \citet{vanDokkum2022} including the UDGs DF2 and DF4. \citet{amorisco} explained UDGs as normal dwarf galaxies with higher-than-average spin. The higher centrifugal force expands the galaxy, leading to the UDG-typical extended size and low surface brightness compared to normal dwarfs. Furthermore, UDGs might not exist despite the tidal force in the centers of galaxy clusters but actually because of it \citep{Sales,Tremmel2020}. Such puffed-up tidal dwarf galaxies were already found by \cite{Duc} in galaxy groups and explicitly described as galaxies with a ``\textit{low central surface brightness and large effective radius, compared to other dwarf galaxies of similar luminosity/mass and even gas content}." 

All of those different formation scenarios raise the question of how so many different formation paths could lead to a distinct population. On the contrary, neither \citet{amorisco} nor \citet{Tremmel2020} predict a separation of UDGs from the rest of the dwarf cluster population. \citet{vanDokkum2015}  stressed that the term UDG, ``\textit{does not imply that these objects are distinct from the general galaxy population.}" However, some treat UDGs like a new galaxy type. Also, \citet{Conselice} mentioned that similar galaxies were already found in previous studies \citep[e.g.,][]{SandageBinggeli1984,CaldwellBothun1987,Impey1988,Binggeli1994,Conselice2003} and that UDGs overlap with low-mass cluster galaxies analyzed by \citet{Conselice2003} in the $M_{\mathrm{tot}}-R_e$ parameter space. Furthermore, \citet{Chamba2020} questioned that UDGs are actually Milky Way-sized, using $R_1$, which is tracing the in situ star formation \citep{Trujillo2020} instead of $R_e$ as size indicator.

Galaxy families can be distinguished in structural parameter spaces \citep[e.g.,][]{Kormendy1985,Bender1992,Binggeli1994,Kormendy2009}.
\citet{Kormendy2009} showed a dichotomy between ellipticals and classical bulges on the one hand and spheroidals on the other hand in the $R_e-\mu_e$ (the \citealt{KormendyRelation1977} relation), $M_{\mathrm{tot}}-\mu_e$, and $M_{\mathrm{tot}}-R_e$ parameter spaces. Furthermore, \citet{kluge} found that brightest cluster galaxies (BCGs) show a distinct scaling relation from ellipticals and classical bulges. 

The goal of this work is to identify whether UDGs populate another distinct region in these parameter spaces or whether they are indistinguishable from one of the already known populations. In addition to those three parameter relations, we investigate which region in the $M_{\mathrm{tot}}-\mu_0$ parameter space UDGs populate, and we compare our results to the findings of \citet{Binggeli1994}. Unlike previous studies, we do not solely probe UDGs, but measure and study the structural parameters for a large number of cluster members ranging from UDGs to the normal spheroidal galaxy regime, giving us a direct comparison sample from the same data, without a selection bias, and without potential systematic differences in the analysis.

For this, we measure and study the structural parameters for a large number of cluster members ranging from UDGs to the normal dwarf spheroidal galaxy regime. We have chosen A1656 (Coma cluster) and A262 for this study. A1656 is a rich cluster and also allows us to directly compare our measurements to \citet{vanDokkum2015} and \citet{Yagi2016}. A262 is a poorer cluster but even closer with a redshift of  $z=0.0162$ \citep{redshiftA262} compared to A1656 with a redshift of  $z=0.0231$ \citep{redshiftA1656} and thus, the UDGs appear brighter and larger in A262. Throughout this paper, we use the cosmology calculator by \citet{wrightcosmocalc2006} assuming a flat universe, $H_0=69.6\,\mathrm{km\,s^{-1}\,Mpc^{-1}}$, and $\Omega_m=0.286$ \citep{Bennetcosmology} to calculate physical scales and distance moduli. For A262, this gives a physical scale of $\mathrm{0.33\,kpc\,arcsec^{-1}}$ and a distance modulus of 34.25\,mag, and for A1656 a physical scale of $\mathrm{0.47\,kpc\,arcsec^{-1}}$ and a distance modulus of 35.03\,mag.

\section{Data} \label{sec:data}
Our observations have been carried out with the 2.1\,m-Fraunhofer Wendelstein Telescope \citep{Hopp2014} using the Wendelstein Wide Field Imager  \citep[WWFI;][]{WWFI}. The WWFI covers a field of view of $27.6\arcmin\times28.9\arcmin$ and consists of four CCDs aligned in a $2\times2$ mosaic. Each of these CCDs has $4096\times4109$ pixels with a pixel scale of $\mathrm{0.2\,arcsec\,px^{-1}}$.

\subsection{Sample and Observing Strategy}
Both clusters are part of the sample from \cite{kluge,Kluge2021}. For our study, we use their imaging data for A1656 ($\approx4-5$\,hr in $g'$) and A262 ($\approx8$\,hrs in $u'$, $\approx4-5$\,hr in $g'$, $\approx1$\,hr in $r'$) plus new observations in the $u'$ and $r'$ bands. The color information is used to select the cluster members (see Sections \ref{sec:bi colorsequenceselection} and \ref{sec:redsequence}).

Furthermore, we observed a reference field to investigate the sample contamination by interloping galaxies. For this, we chose a pointing around the lensed quasar SDSSJ1433+6007 as we already had deep $g'$-band data available for the time-delay cosmography studies of this quasar \citep{Queirolo}. The pointing is centered at R.A. = 14:32:29.41 decl. = 60:12:26.82.

All of our observations were carried out in photometric conditions and dark time with a zenith sky brightness fainter than $21.3\,V\mathrm{\,mag\,arcsec^{-2}}$. For the individual exposures, we chose the exposure time such that the photon noise of the sky is dominant over the readout noise. For the $g'$ and the $r'$ bands, we used an exposure time of 60\,s in the fast readout mode, whereas for the $u'$ band we used 600\,s in the slow readout mode, which results in only $\approx 1/4$ of the readout noise compared to the fast readout mode but at the cost of a four-times-higher readout time. For the $g'$-band observations of our reference field, the individual exposure time was 240\,s.

We stick to the dithering strategy from \citet{kluge}, so that our $u'$- and $r'$-band data are consistent with the archival $g'$-band data. This strategy was optimized to measure the faint intracluster light (ICL) around local BCGs but also provides a large spatial coverage, allowing us to study other galaxy populations in these clusters.
\begin{figure}[t]
    \begin{center}
        \includegraphics[width=0.4\textwidth]{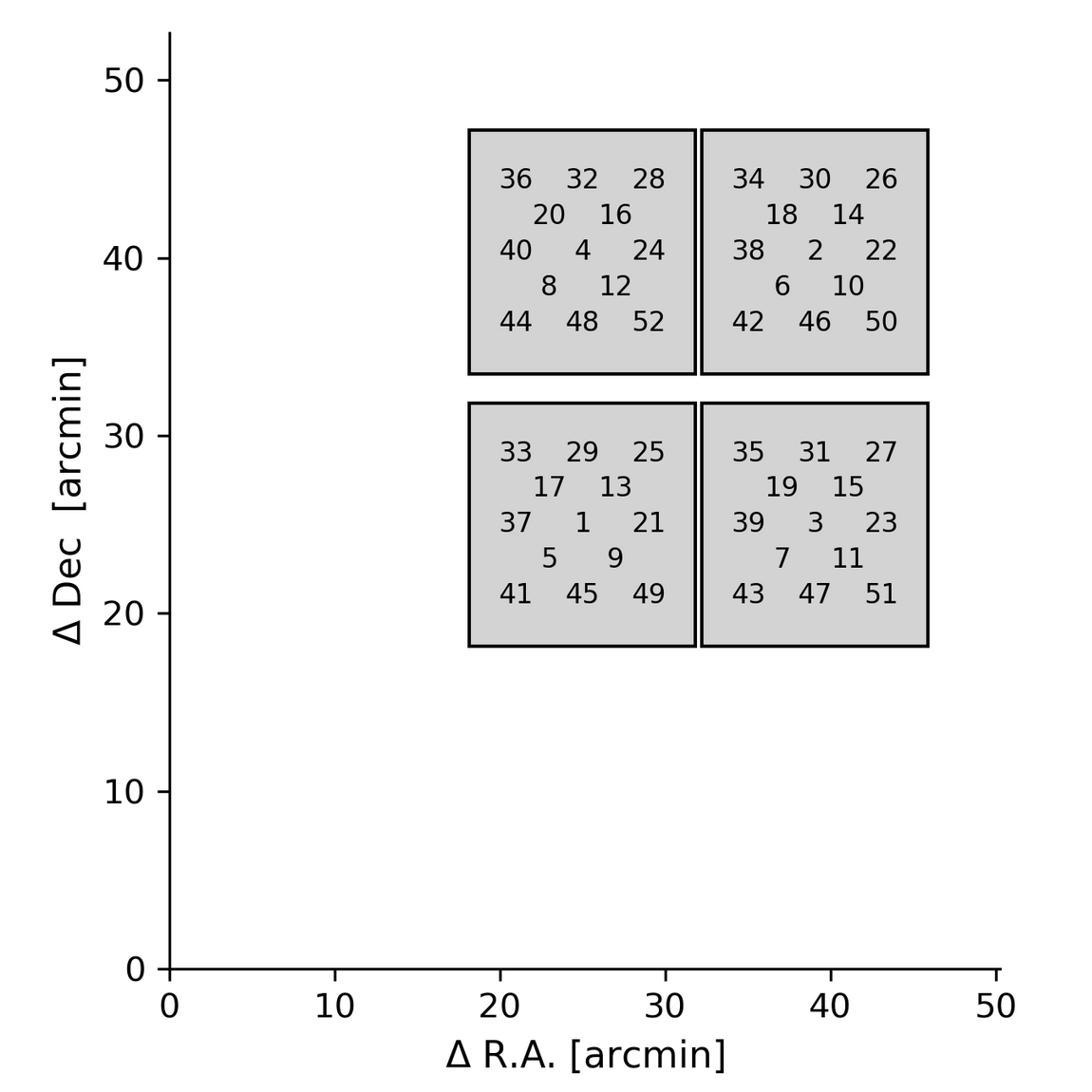} 
    \caption{Illustration of the dither pattern. The four CCDs are represented by gray squares. The illustrated pointing corresponds to the first element of the dither pattern. The position of the center on the detectors is indicated by the number $i$ for each dither element $i$. Figure adapted from \citet{kluge}.}
    \label{fig:ditherstrategy}
    \end{center}
\end{figure}
The full 52-step dither pattern is illustrated in Figure \ref{fig:ditherstrategy}. Our observations are centered on the BCG or, in the case of A1656, between the two BCGs. For the first four exposures, the middle of the cluster is centered on each of the CCDs. The following positions are shifted by 2\arcmin\, in R.A. or decl. direction, where the four large dither steps are repeated off-center. This procedure is repeated 13 times. For A1656, we observed 1.5 full dither patterns in the $u'$ band, four in the $g'$ band, and two in the $r'$ band. Additionally, we obtained sky pointings (centered at R.A. = 2:56:38 and decl. = 28:08:27) in between each of these dither steps to create night-sky flats from these sky pointings. The ICL of A1656 covers nearly the full field of view with the large dither pattern such that an accurate night-sky flat cannot be determined using these target exposures.

For the archival A262 $g'$-band data, only half of the dither pattern with the cluster center on the upper-right and lower-left CCD chip was performed. The archival $u'$- and $r'$-band data were taken with the full dither pattern. We took further $u'$- and $r'$-band data with the same dither-strategy as the $g'$-band to increase the depth.

For our reference pointing, we stick to the dither pattern applied by \citet{Queirolo}. Here, we dither only 8\arcsec~ per dither step without centering the pointing on the different CCDs. This gives us a relatively uniform depth over the field of view, whereas the larger dither patterns of the two clusters result in a nonuniform depth. Due to this varying depth over the field of view in the A262 and A1656 images, we can only qualitatively compare them to the reference pointing (see section \ref{sec:referencefield}).

Of all of the data taken, we reject some due to low sky transparency, bad seeing, or significantly varying night-sky patterns. 
\begin{deluxetable}{c|cccc}[t]
    \tabletypesize{\small}
    \tablecaption{Exposure Times and Depths}
    \label{tab:exposuretimes}
    \tablehead{
        \colhead{}& \colhead{} & \colhead{A262} & \colhead{A1656} & \colhead{Ref.}   
    }
    \startdata
        exp. time  & $u'$ & 780 & 790 & 570 \\
        (min) & $g'$ &321 & 210 & 216 \\
        & $r'$ &133 & 79 & 113\\
        \hline
        maximum   & $u'$ & 30.2 & 30.7 & 30.1\\
        $3\sigma$ depth & $g'$ & 30.0 & 29.9 & 30.0\\
        $(\mathrm{mag\,arcsec^{-2}})$& $r'$ & 29.2 & 28.8 &29.0\\
        \hline median  & $u'$ &  29.7 & 30.1 & 30.1\\
        $3\sigma$ depth& $g'$ & 29.5 & 29.4 & 30.0 \\
        $(\mathrm{mag\,arcsec^{-2}})$& $r'$ & 28.7 &  28.2 & 29.0\\
    \enddata
    \tablecomments{Total exposure time and maximum, as well as median $3\sigma$ depth on a $10\arcsec\times10\arcsec$ scale of our A262, A1656, and reference field observations for the individual filter bands.}
\end{deluxetable}    
\begin{figure*}
    \centering
    \includegraphics[width=0.94\textwidth, trim= 0cm -0.3cm 0cm 0cm]{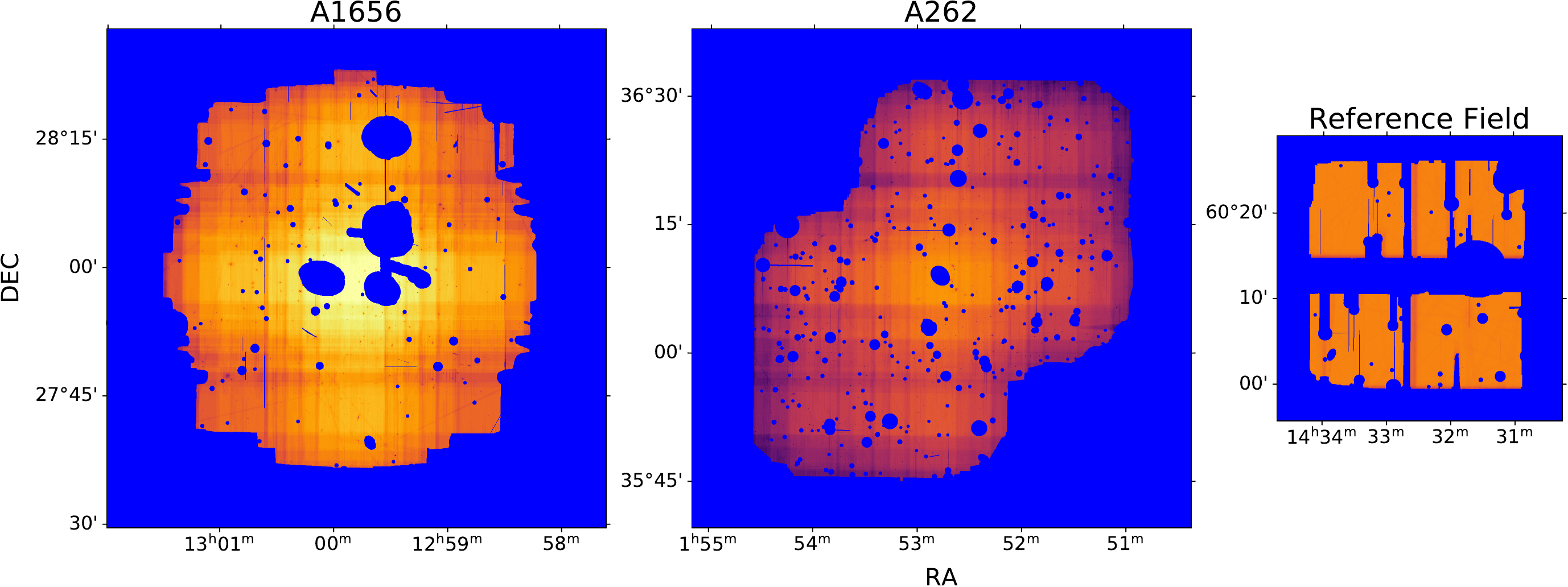}
    \includegraphics[width=0.8\textwidth, trim= 0cm -0.5cm 0cm 0cm]{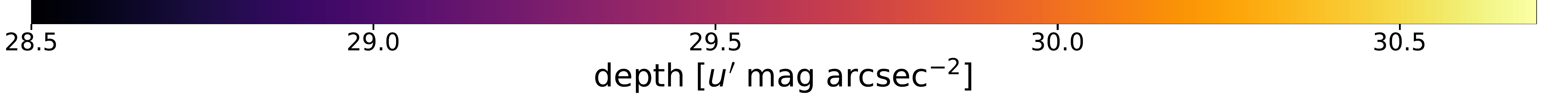}
    \includegraphics[width=0.94\textwidth, trim= 0cm -0.3cm 0cm 0cm]{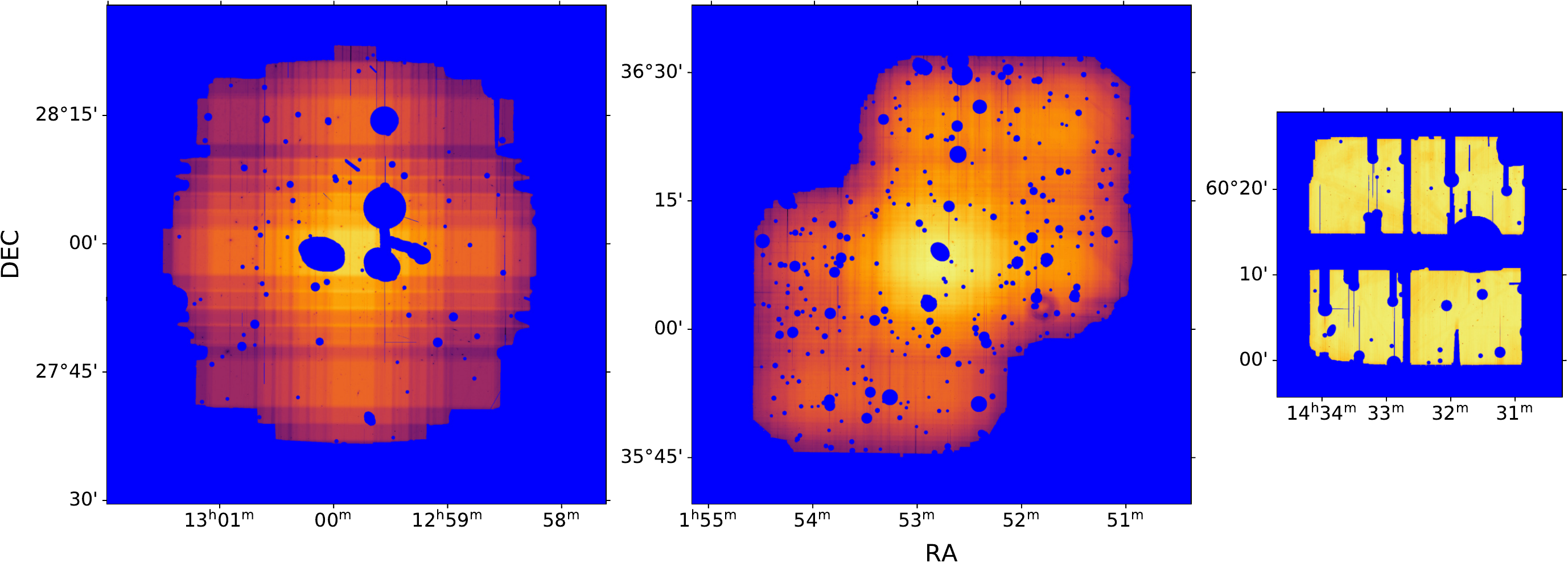}
    \includegraphics[width=0.8\textwidth, trim= 0cm -0.5cm 0cm 0cm]{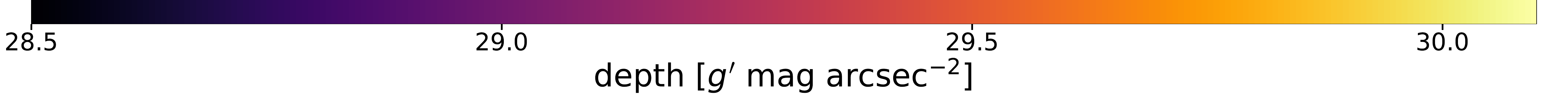}
    \includegraphics[width=0.94\textwidth, trim= 0cm -0.3cm 0cm 0cm]{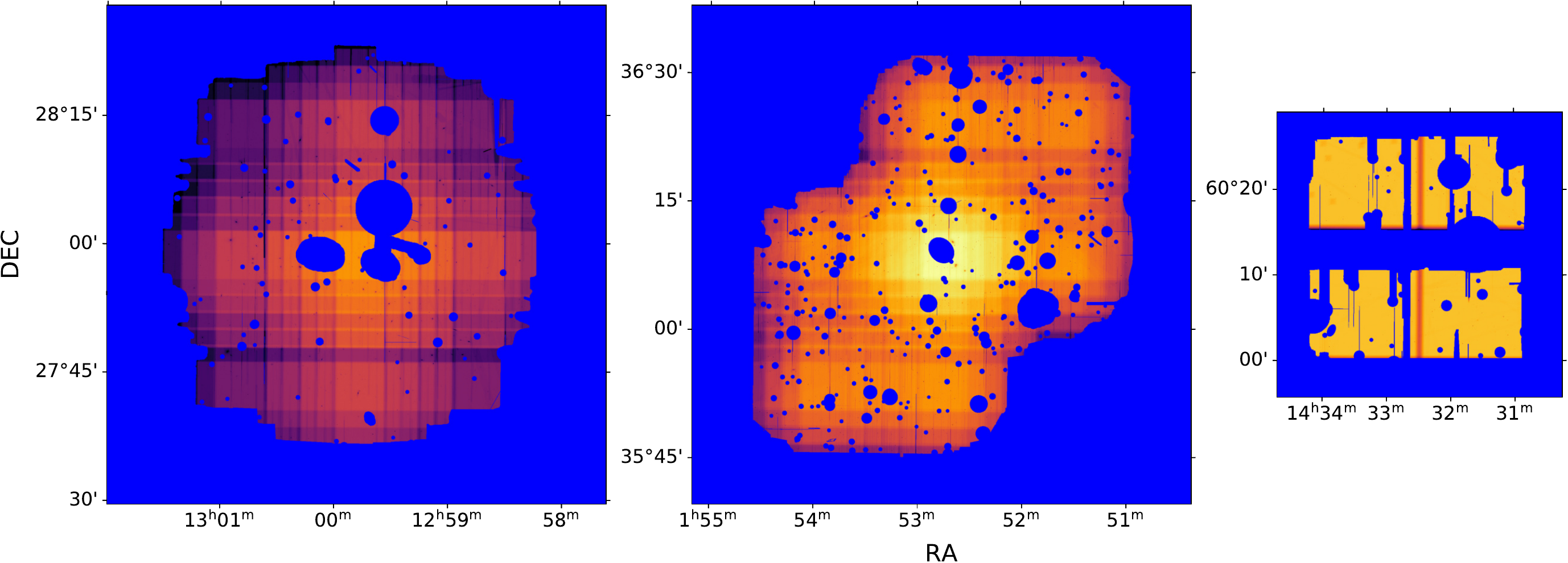}
    \includegraphics[width=0.8\textwidth, trim= 0cm -0.5cm 0cm 0cm]{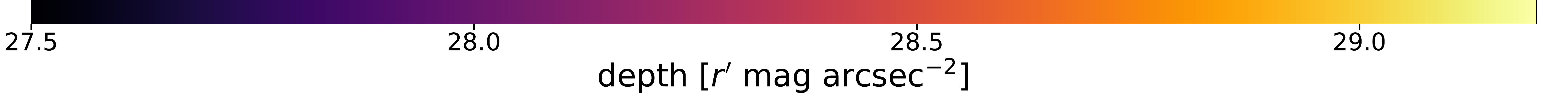}
    \caption{$3 \sigma$ depth on a 10"x10" scale of our A1656 (left), A262 (middle), and reference field (right) data in the $u'$ (top), $g'$ (middle), and $r'$-band (bottom).}
    \label{fig:depths}
\end{figure*}

The total exposure times, as well as the maximum and median $3\sigma$ depths on a $10\arcsec \times 10 \arcsec$ scale are given in Table \ref{tab:exposuretimes} for all pointings and filters. This depth gives the detection threshold in surface brightness at a $3\sigma$ level for a source with a size of $10\arcsec \times 10 \arcsec$. We calculate the depths following \citet{Roman2020}:
\begin{equation}
    \mu_{\mathrm{lim}}(3\sigma;10\arcsec\times10\arcsec)=-2.5\mathrm{log\left(\frac{3\sigma}{pxs\times10}\right)+ZP}
\end{equation}
where "pxs" is the pixel size in arcseconds. The distribution of the depths for the different pointings and filters is visualized in Figure \ref{fig:depths}. We clipped all regions of the images with < 40 minutes exposure time in the $g'$ band (corresponding to a surface brightness limit of about $29\,g'\,\mathrm{mag \,arcsec^{-2}}$) to reduce the number of false detections.

\subsection{Data Reduction} \label{sec:datareduction}
The data was reduced with the WWFI data reduction pipeline \citep{kluge,klugeDiss}. The WWFI pipeline frequently makes use of our in-house fitstools \citep{fitstools}, SExtractor \citep{Sextractor}, SCAMP \citep{scamp}, and SWarp \citep{swarp}.
It includes bias subtraction, flat-fielding, automatic masking of charge persistence, bad pixels, and cosmic rays, as well as manual satellite masking. Dark current is negligible for the WWFI at the operating temperature of $\mathrm{-115^{\circ}C}$ \citep{WWFI}. The photometric zero-points for the $g'$ and $r'$ bands are determined by matching the flux of point sources in apertures of 5\arcsec\, diameter ($\mathrm{ZP_{5}}$) to the Pan-STARRS DVO PV3 catalog \citep{panstars}. Our $u'$-band data is calibrated to the Sloan Digital Sky Survey (SDSS) photometric system because Pan-STARRS has no coverage in the $u'$ band. However, A262 is not covered by SDSS; hence we determine the $u'$-band zero-point from the zero-point for A1656 assuming that it only changes with airmass and correcting for galactic extinction as the data were taken under photometric conditions. 

Afterward, we subtract extended point-spread function (PSF) models and ghosts from bright stars to improve the flatness of the background. For A262 and A1656, we use all stars contained in the TYCHO-2 catalog \citep{Tycho2}. For the reference pointing, we subtract all stars in the GAIA EDR3 catalog \citep{GAIAeDR3} brighter than $14\,G\,\mathrm{mag}$.

Then, we create night-sky flats for each night, scale them to the individual exposures and subtract them. After that, we inspect the night-sky-corrected images for stray light contamination and, if necessary, mask it in the images before the night-sky subtraction. Then, new night-sky flats are created and subtracted.

Finally, all images are stacked, and accurate zero-points are calculated for the stacks with a larger aperture of 10\arcsec. Furthermore, new zero-points are calculated for the $g'$ band accounting for lost flux outside of the 10\arcsec\,  aperture following \citet{klugeDiss}:
\begin{equation}
    \mathrm{ZP_{inf}=ZP_{10}+0.1155}\,g'\,\mathrm{mag}
\end{equation}
Here, we can only correct the $g'$ band for this effect, as we only have an extended PSF model for this filter available. As the $g'$ band serves as our reference measurement band and we are using the $u'$- and $r'$-band data only for obtaining aperture colors, this correction is also not needed for the latter. Unless explicitly written, all $g'$-band total magnitudes and surface brightnesses are corrected for $\mathrm{ZP_{inf}}$. All colors and aperture magnitudes are corrected with $\mathrm{ZP_{10}}$. Magnitudes corrected with $\mathrm{ZP_{10}}$ are consistent with Pan-STARRS magnitudes \citep{kluge}.

Furthermore, a new astrometric solution is calculated for the final stacks using the GAIA EDR3 catalog \citep{GAIAeDR3}.

\section{Data Processing Pipeline} \label{sec:dataprocessingpipeline}
For the measurement of the structural parameters and  colors of the UDGs and for the necessary preparatory steps, we have developed a nearly automatic pipeline. The pipeline is highly parallelized using up to 512 cores simultaneously. We frequently make use of our in-house \verb+fitstools+ \citep{fitstools}. 

We intend to use this pipeline not only to study the UDG population in A262 and A1656 but also to investigate the whole galaxy population of many other galaxy clusters in the future. Its first part, the basic pipeline (Section \ref{sec:basicpipeline}), prepares the measurements for all types of galaxies in a galaxy cluster, except spiral galaxies. The second part (Section \ref{sec:EllS0}) is for measuring the total magnitudes and $g'-r'$ colors of bright galaxies such as ellipticals and S0's that are required for the red sequence selection (Section \ref{sec:redsequence}). The third part (Section \ref{sec:udgmeasurements}) is to measure the properties of UDG- and spheroidal-like galaxies and to select cluster members.
\subsection{Basic Pipeline} \label{sec:basicpipeline}
The basic pipeline includes accurate measurements of the inner 10\arcsec  of the PSF over the whole field of view using \verb+PSFEx+ \citep[see Section \ref{sec:psfmeasurements}]{PSFEx}. Furthermore, we improve the flattening of the background by subtracting BCG and ICL models and bright stars ($m_{tot} \lesssim 16\,g'$\,mag) using an extended PSF model (Section \ref{sec:starBCGsubtraction}). Source catalogs are created using \verb+SExtractor+ \citep{Sextractor}. The catalogs contain first estimates of the structural parameters and positions of the objects. They are in the following used to preselect dwarf galaxy candidates and as initial parameters for \verb+GALFIT+ \citep{galfit}. We model the UDG candidates simultaneously with overlapping objects using \verb+GALFIT+. Therefore, we need reasonable initial parameters for all types and sizes of galaxies. But as there are no perfect parameters for \verb+SExtractor+ to detect and measure all types and sizes of galaxies accurately simultaneously, we create two object catalogs. One for large and bright sources and one for relatively small and faint sources, and combine them afterward. The \verb+SExtractor+ parameters for faint objects were optimized to reliably detect UDGs in A1656 while avoiding obvious false detections in the low signal-to-noise ratio (S/N) regions of our images. For this run, we use smoothed images, as this significantly reduces false detections caused by noise peaks and, at the same time, increases the number of detected UDGs. The parameters for large objects were tuned to detect elliptical galaxies. For a detailed discussion about how the object catalogs are created, see Section \ref{sec:sextractor}.

\subsubsection{Preparations} \label{sec:preparations}
Before the pipeline can be started, bad regions, such as not perfectly masked charge persistence stripes, diffraction spikes of bright stars, over- or undersubtracted PSF wings, and ghosts are masked manually. Discrete star formation regions within spiral galaxies can erroneously be detected as individual objects. That problem also affects larger scales due to overshooting effects in the background subtraction. We overcome this issue by manually masking all spiral galaxies because we are only interested in UDGs and their transition to spheroidal galaxies, as well as to S0's and elliptical galaxies in this work.

After these masks are created, the first part of the pipeline can be started. Firstly, the object stacks and weight images are smoothed using a 2D Gaussian with a standard deviation of 2 pixels (equivalent to $0.\arcsec4$) and subsequently, the smoothed, as well as the original images and weight images are multiplied with the masks. Smoothing the images reduces false detections and simultaneously increases the number of correct detections of low surface brightness objects using \verb+SExtractor+. That is because noise peaks are smoothed out that would otherwise be detected as a source. 
Noise peaks inside of an object could also lead to one object being detected as multiple. This can be prevented by smoothing the images, too. Here we ensure that the smoothing is not too strong so that two real objects would not erroneously be detected as one. Smoothing the images also allows the minimum detection area to increase, as after smoothing, more pixels of the source are connected. Larger-sized noise peaks are then rejected using the increased minimum detection area.

The last preparatory step is to clip all low-S/N regions, as false detections would occur in these regions due to the constant detection threshold. We noticed that false detections due to noise peaks occur more frequently in regions with an exposure time of less than about 40 minutes in the $g'$ band. Hence, we mask those regions. Due to our dithering strategy and the gaps between the CCDs, multiple thin stripes are below this threshold. However, these thin low-S/N stripes do not cause a significant amount of false detections, whereas the same stripes can produce false detections due to edge effects if masked. Therefore, these thin regions are de-masked again manually.

\subsubsection{PSF Measurements} \label{sec:psfmeasurements}
An accurate PSF model and FWHM estimate are crucial for the reliability of \verb+SExtractor+`s star-galaxy classifier (S/G), which is later on used to decide which objects are point sources that are to be subtracted from the image stack (see Section \ref{sec:starBCGsubtraction}) and whether an object is modeled by \verb+GALFIT+ using a PSF model or a \citet{sersic} model (see Section \ref{sec:galfit}). Furthermore, a precise PSF model is inalienable to obtain the intrinsic structural parameters of the UDGs using \verb+GALFIT+  which fits a PSF-convolved Sérsic model to the data. Additionally, if a point source overlaps with a UDG-candidate, it will be modeled simultaneously by \verb+GALFIT+ using this PSF model. Lastly, this PSF model is used to convolve cutout images in the different filter bands to a target PSF so that their PSF shapes are identical. This is crucial to prevent systematic errors in the aperture  color measurements (see Section \ref{sec: colormeasurements}).

We measure the PSF's FWHM and determine the exact PSF over the whole field of view using PSFEx \citep{PSFEx}, which is crucial, as the PSF is varying over the field of view. This variation is caused by two effects. Firstly, the PSF of the WWFI varies in the single images over the field of view. Secondly and more importantly, the  object stack consists of observations taken over many years and with different observing strategies covering different parts of the field. As the seeing conditions can strongly vary in the different nights, this, in combination with the different spatial coverage, leads to a significantly varying PSF over the field of view in the object stack.

In order to create these PSF models, we first create star catalogs for each filter while detecting the sources in the $g'$ band with \verb+SExtractor+. Here, we use the parameters \verb+DETECT_THRESH=15+, \verb+DETECT_MINAREA=36+, and \verb+BACKGROUND_SIZE=128+ to detect mainly bright point sources. The \verb+VIGNET+ is measured in a box with 101 pixels ($=\mathrm{20.\arcsec2}$) side length. 

The next step is the actual selection of point sources. Here we orient ourselves on the automatic point-source selection by PSFEx \citep{PSFEx} but do the selection manually in order to take care of the varying PSF in our images. Thereby, we plot the effective radius ($R_e$) against the central surface brightness ($\mathrm{\mu_{max}}$). As effective radius, we use the model-independent \verb+FLUX_RADIUS+ with \verb+FLUX_FRAC=0.5+.

To discriminate between point sources and extended objects, we use the property that all point sources have nearly the same effective radius, independent of their brightness and therefore, form a narrow vertical line in a $\mathrm{\mu_{max}}-R_e$ plot. The upper and lower limits for $\rm{\mu_{max}}$ and $R_e$ for the point-source selection are chosen manually. They are chosen such that only nonsaturated point sources are included, by fulfilling the following criteria:
\begin{enumerate}
    \item a sufficient number of sources to cover the whole field of view;    
    \item bright to guarantee a high S/N;
    \item not too bright to discard saturated stars;
    \item a relatively narrow range in $R_e$  to discard extended objects ; and
    \item a broad enough range in $R_e$ to represent the variation in $R_e$ of the point sources over the field of view.
\end{enumerate}

The discrimination between point sources and extended sources improves with better seeing. The absolute value of the FWHM, its variation, and other PSF-shape parameters can influence the reliability.

Hence, the filter band where this classification is most reliable based on the $\mathrm{\mu_{max}}-R_e$ selection is chosen. Those objects identified as point sources in the chosen filter band are used later on to determine the PSF in the other filter bands. Here, saturated stars are rejected for each filter individually. Furthermore, objects deviating by more than $3\,\sigma$ from the median FWHM, as well as objects with \verb+FLAGS>0+ and \verb+IMAFLAGS>0+ are discarded.

Finally, the PSF models for the original and smoothed images are created with PSFex for all filter bands. The PSF is derived directly using the ``pixel vector basis." It does not rely on an analytic model or any assumption about the PSF shape and, hence, also supports the modeling of deformed PSF shapes. We use  2D fifth-degree polynomials to describe the spatial variation of each pixel of the PSF. After that, \verb+SExtractor+ is run again with the new FWHM estimate and PSF models to obtain an accurate S/G.

The last step of the PSF measurements is to fit a Moffat profile to the point sources. This is done, as the central part of the WWFI PSF is represented well by a Moffat profile and not by a Gaussian profile, as used by \verb+SExtractor+. Hence, this gives us a more accurate PSF estimate. Nevertheless, the FWHM determined by the previous Gaussian fit is still used for \verb+SExtractor+ as its S/G relies on the FWHM of a Gaussian fit. The FWHM of the Moffat fit is used to describe the seeing quality of our images (see Table \ref{tab:FWHM}).

Furthermore, for the aperture color measurement (Section \ref{sec: colormeasurements}), we convolve the cutout images with an optimized kernel to make the PSFs identical in all filter bands. This is crucial for the aperture color measurement, because otherwise the color of the objects would be biased due to the finite aperture. As the target PSF, we use a Moffat profile as given in Equation (\ref{eq:moffat}) with a $\beta$ value representing the shapes of the PSFs in all filter bands (see below).
\begin{equation}
    I(r)=I_0\,\frac{\beta-1}{\pi\alpha^2}\left[1+\left(\frac{r}{\alpha}\right)^2\right]^{-\beta}
    \label{eq:moffat}
\end{equation}
\begin{equation}
    \mathrm{FWHM=2\alpha\sqrt{2^{1/\beta}-1}}\\
    \label{eq:FWHMalpha}
\end{equation}
The target PSF FWHM, which is related to the $\alpha$ and $\beta$ parameters (Equation (\ref{eq:FWHMalpha})), must be chosen larger than the maximum FWHM in all filter bands. This is necessary, as reshaping the individual PSFs to the target PSF should not involve deconvolution but only convolution because deconvolution amplifies noise and introduces ringing artifacts.

In order to constrain $\beta$, we use the tool \verb+starphot+ \citep{fitstools} to fit Moffat profiles to the point sources that were previously used to create the PSF models. Here, we let both $\beta$ and the FWHM vary. 

Then, we run \verb+starphot+ once more keeping $\beta$ fixed at the previously determined median $\beta$ to overcome the degeneration between $\alpha$ and $\beta$. 
As the Moffat FWHM estimate, we use the median value. For the target PSF, we use the largest maximum reliable FWHM appearing in any of the filter bands. The maximum reliable PSF is estimated by the median FWHM plus three times the standard deviation. This ensures that the chosen FWHM is large enough to ensure an accurate convolution while excluding strong outliers.

The FWHM of our A262 and A1656 observations using a Gaussian as well as a Moffat fit is shown in Table \ref{tab:FWHM}. Using the Gaussian fit, we overestimate the true FWHM of our data by about 0.\arcsec12.

\begin{deluxetable*}{ccccccc}[ht]
    \tabletypesize{\small}
    \tablecaption{FWHM of the PSFs of Our Observations Determined Using a Gaussian Fit and a Moffat Fit}
    \label{tab:FWHM}
    \tablehead{
        \colhead{Filter} & \colhead{A262} & \colhead{A262} & \colhead{A1656} & \colhead{A1656} & \colhead{Reference Field} & \colhead{Reference Field} \\
        \colhead{} & \colhead{Gauss. FWHM} & \colhead{Moffat FWHM} & \colhead{Gauss. FWHM} & \colhead{Moffat FWHM} & \colhead{Gauss. FWHM} & \colhead{Moffat FWHM} \\
        \colhead{} & \colhead{(arcsec)} & \colhead{(arcsec)} & \colhead{(arcsec)}  & \colhead{(arcsec)} & \colhead{(arcsec)} & \colhead{(arcsec)}
    }
    \startdata
      $u'$ & 1.35 & 1.26 & 1.29 & 1.22 & 1.06&0.93 \\  
      $g'$ & 1.00 & 0.90 & 1.03 & 0.94 & 1.27&1.17 \\
      $r'$ & 1.06 & 0.89 & 0.89 & 0.74 & 1.17&1.03 
    \enddata
\end{deluxetable*}    

\subsubsection{Subtraction of Stars, BCGs, and ICL} \label{sec:starBCGsubtraction}
Similar to \cite{kluge}, we subtracted extended PSF models from bright stars in the data reduction to improve the background flatness. In this work, we require a higher local background flatness. Therefore, we select even fainter stars in order to further flatten the background to improve the object detection and object fits. This procedure is performed for all nonsmoothed images, and the resulting images are smoothed afterward.

First, we subtract all bright ($m\leq 16\,g'\,\mathrm{mag}$) stars selected using \verb+SExtractor+'s $\mathrm{S/G}\geq 0.97$ from the object stacks. Here, we use an extended PSF model from \citet{kluge} scaled with the total brightness of each star to create an image of all selected stars. This star stack also contains the far outer wings of the stars that were already subtracted in the data reduction as they were still present in the night-sky flats. Here, we manually set a flux threshold up to which the background is modeled. We choose the flux between two stars in the central region of the star stack as the threshold. This background is subtracted from the star stack, and the result is then subtracted from the object stack. As the center of the PSF is seeing dependent, the fixed extended WWFI PSF model usually does not fit well in the center. Hence, the centers of the stars are masked.

\begin{figure}[ht]
    \begin{center}
        \includegraphics[width=0.45\textwidth]{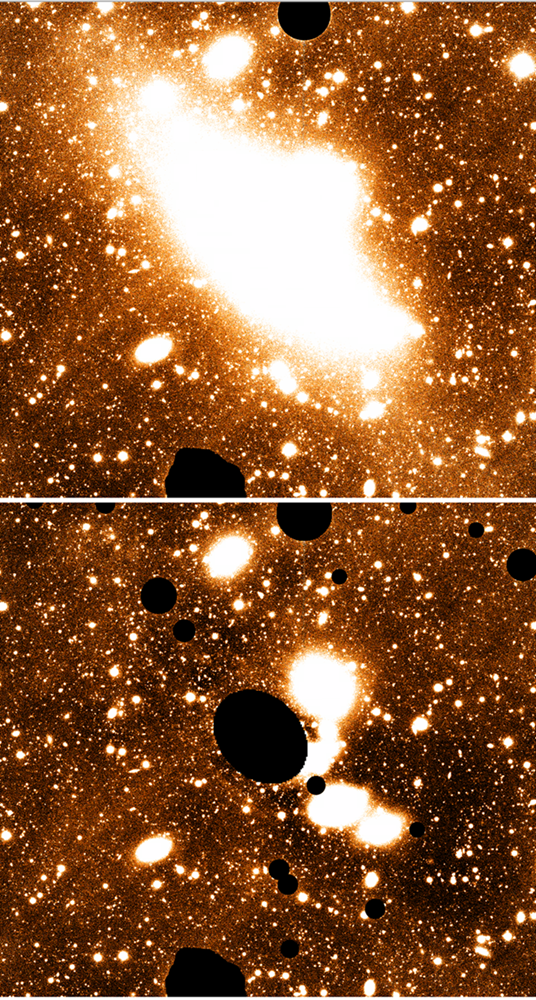}
        \includegraphics[width=0.47\textwidth]{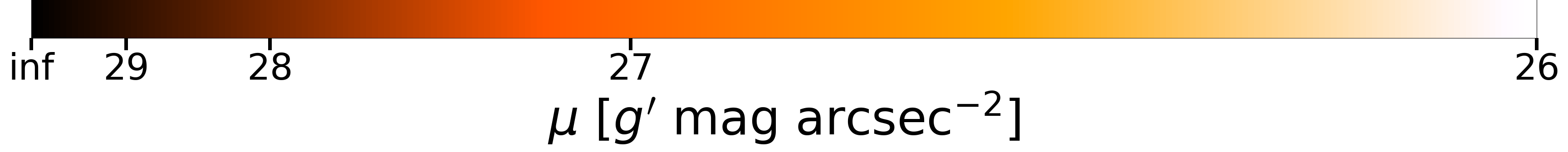}
        \caption{Two by two binned cutout ($\mathrm{12.\arcmin8\times11.\arcmin7}$) of the central region of the $g'$-band A262 object stack both before the star and BCG+ICL subtraction (top) and after the subtraction (bottom).}
        \label{fig:BCGsub}
    \end{center}
\end{figure}

In the next step, we subtract models of the BCGs and the ICL from the object stacks. The A262 $g'$-band BCG+ICL model is obtained from \citet{kluge}. For the $u'$ and $r'$ bands, we follow the procedure presented in \cite{klugeisophotespy} and \cite{Kluge2023rhea}. In brief, ellipses are fitted to the isophotes using the python tool \verb+photutils+ \citep{photutils}. Here, the ellipticity, position angle, and center of the isophotes can vary. Beyond the largest fitted radius, we fix all ellipse parameters apart from the radius. Model images are then generated by setting the flux along these ellipses to the median measured value. Masks are adopted from \cite{kluge} and manually improved for the different filter bands. As an example of the BCG+ICL subtraction, the central region of A262 in the $g'$ band is shown in Figure \ref{fig:BCGsub} before and after the BCG+ICL subtraction.

For A1656, we iteratively create the models for the two BCGs. For that, we first apply the masks from \citet{kluge}, manually mask NGC 4874, and then create a first model of NGC 4889. It is then subtracted from the star-subtracted-object stack, and the residuals are masked. Using the resulting image, the model of NGC 4874 is created and subtracted from the star-subtracted-object stack, and the residuals are masked. Then, we fit the final model of NGC 4889. The models of NGC 4889 and NGC 4874 are combined and subtracted from the star-subtracted-object stack. After that, residuals of the star and BCG+ICL subtraction are masked manually.

Finally, all images are smoothed again.

\subsubsection{SExtractor Object Catalogs and Segmentation Maps} \label{sec:sextractor}
To create our final object catalogs, we run \verb+SExtractor+ twice per filter. One run is tuned to detect faint and relatively small sources, and one is tuned for bright and relatively large sources. These two catalogs are matched afterward based on the central position of the objects. If an object is contained in both catalogs, we prioritize the bright source catalog. We always use the $g'$-band images to detect the sources.

\begin{figure}[ht]
    \centering
    \includegraphics[width=0.45\textwidth]{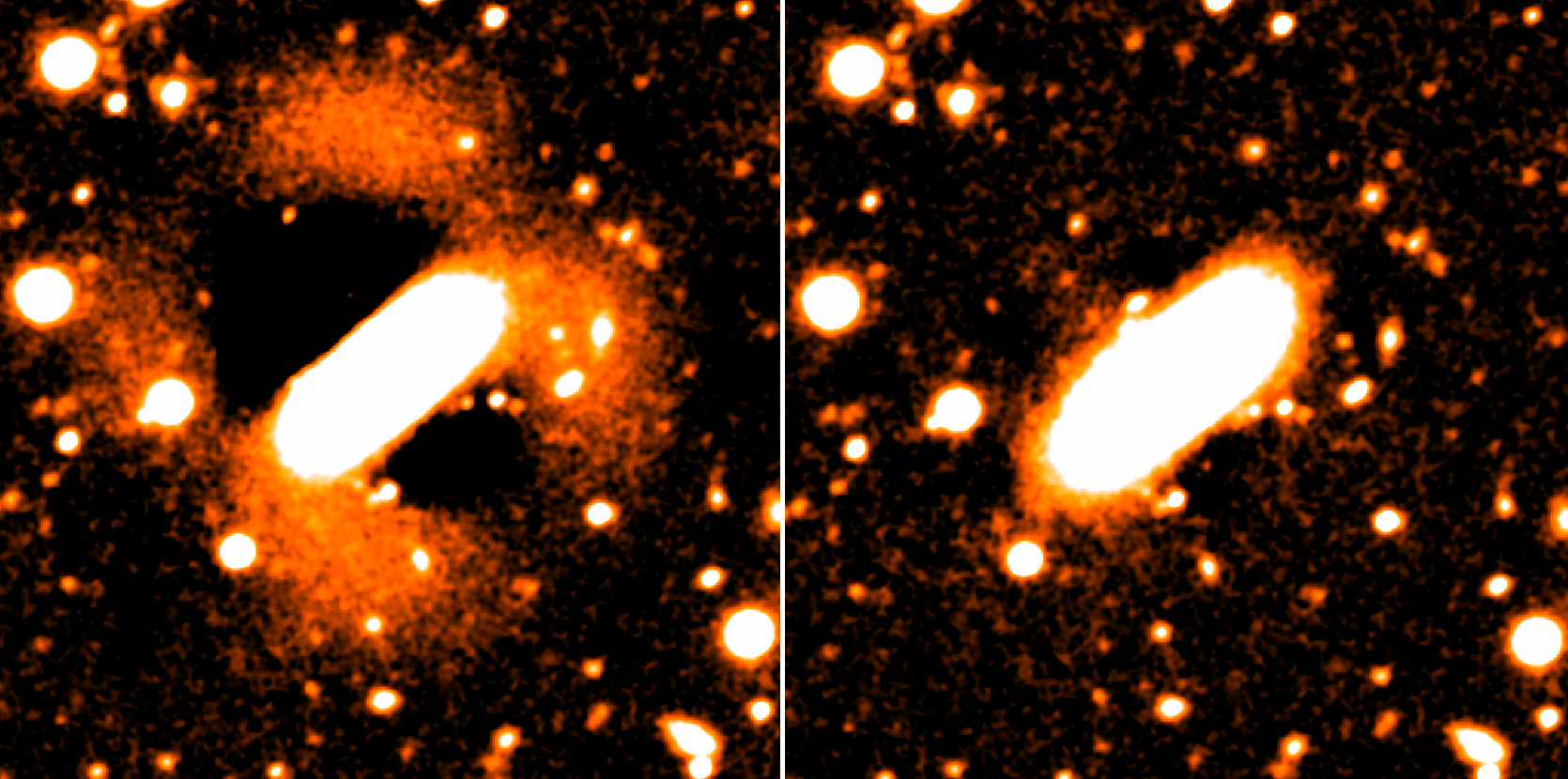}
    \includegraphics[width=0.47\textwidth]{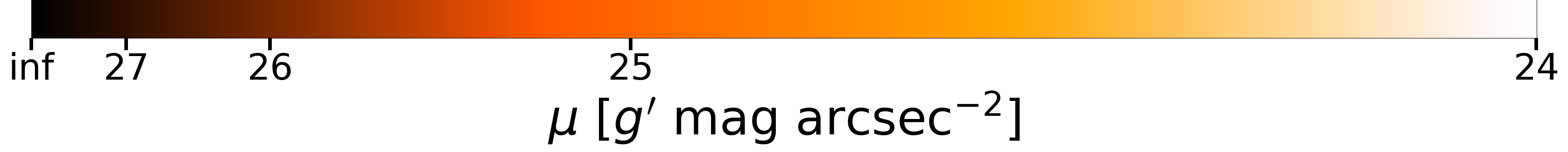}
    \caption{Cutout ($\mathrm{2.\arcmin3\times2.\arcmin3}$) of a region around an elliptical galaxy in A262 (smoothed $g'$-band data) after SExtractor's background subtraction with a background size of 128\,px in the left panel and 45\,px (corresponding to the optimal background size for UDG detection in A1656 scaled to the distance of A262) in the right panel.}
    \label{fig:UDGsBackground}
\end{figure}

To create the catalogs of the small and faint sources, we use the smoothed images. As detection parameters, we use a limiting surface brightness of $27.4\,g'\,\mathrm{mag \,arcsec^{-2}}$ and a minimum detection area of 7 $\times$ 7 px = 49 px (although we note that the detection area can take any shape) at the distance of A1656 scaled with the physical scale for A262. This relatively large minimum detection area reduces the number of false detections in low-S/N regions at the cost of missing faint compact objects. As we are interested in detecting UDGs and similar objects that are relatively large, missing faint small objects (presumably point-source-like background objects) is not a big issue. But as we model all detected objects overlapping with the main object of interest, this would also include modeling false detections alongside real sources which can lead to erroneous or even failing fits using \verb+GALFIT+. Our chosen background subtraction parameters are \verb+BACK_SIZE=32+ (32 px = 6.\arcsec4) at the distance of A1656, scaled with the physical scale for A262, as well as \verb+BACK_FILTER_SIZE=3+. Using the latter, the background is determined from the medians inside 3 $\times$ 3 background patches. Both the small background size and the background filtering do eliminate significant overshooting effects in the background subtraction. Such overshooting effects could even mimic real UDGs, as shown in Figure \ref{fig:UDGsBackground}. Here, the background was subtracted using two different sets of parameters: on the left-hand side with a too-large background size and without background filtering, and on the right-hand side with our best background subtraction parameters. Furthermore, we use the \verb+CLEAN+ option with \verb+CLEAN_PARAM=1.0+ to avoid many spurious detections.

The small background size also has a disadvantage. Due to the small background size, the outskirts of large galaxies get subtracted, which erroneously truncates their surface brightness profiles. Hence, we perform additional \verb+SExtractor+ runs with adjusted background subtraction parameters for relatively large and bright objects. For those runs, we use \verb+BACK_SIZE=225+ (at the distance of A1656, scaled for A262), as well as a detection threshold of $3\,\sigma$ above the background and a minimum detection area of 450 pixels (again at the distance of A1656, scaled for A262).

Finally, a third \verb+SExtractor+ run is performed. Its only purpose is to obtain a better mask (segmentation map) for the largest objects. As the \verb+SExtractor+ segmentation maps only provide masks down to the detection threshold, the segmentation maps are too shallow. For this run, we smooth the $g'$-band image strongly, using a 2D Gaussian with a standard deviation of $\sigma$ = 5 px. The background is subtracted just like for the initial \verb+SExtractor+ run for large and bright galaxies, but as detection threshold, we use $27.4\,g'\,\mathrm{mag \,arcsec^{-2}}$ and a minimum detection area of 8000 pixels at the distance of A262 (again scaled for each cluster). Note that all segmentation maps obtained from \verb+SExtractor+ runs on smoothed images actually do provide masks that cover even fainter surface brightness regions than the detection threshold when applied to the nonsmoothed images. 

\subsubsection{Source Masks} \label{sec:sourcemasks}
Still, the problem remains that we cannot mask significantly deeper than $27.4\,g'\,\mathrm{mag\,arcsec^{-2}}$ using \verb+SExtractor+'s segmentation maps. Another issue is that in the outer region of the objects, noise peaks are above the threshold, and noise valleys are below. Hence, the noise peaks are masked whereas the valleys are not.

To obtain more complete masks, we use the masking tool described in \citet{kluge}. It first smooths the image with a 2D Gaussian filter with a standard deviation $\sigma$ = 11 px. Then, all connected pixels above a certain local threshold $T(x,y)$ are masked if their area exceeds the detection area. As detection threshold, we use a median signal-to-noise threshold $T_0$ and the option to scale this threshold with the square root of local rms scatter rms$(x,y)$:  

\begin{equation}
    T\left(x,y\right)\geq T_0\times\left(\frac{\sqrt{\mathrm{rms}\left(x,y\right)}}{\mathrm{median}\left\{\sqrt{\mathrm{rms}\left(x,y\right)}\right\}}\right)
\end{equation}

Additionally, we also expand most masks by convolving them with circular tophat kernel with different expand diameters. This also reduces the effect of noise peaks in the outskirts of an object being masked, while the noise valleys are not. This way, we create seven masks for each filter band. The input parameters are listed in Table \ref{tab:maskparameters1}.

\begin{deluxetable}{ccccc}[ht]
    \tabletypesize{\small}
    \tablecaption{Mask Parameters}
    \label{tab:maskparameters1}
    \tablehead{
        \colhead{mask}& \colhead{$T_0$} &\colhead{Expand} & \colhead{Detect} & \colhead{Background} \\  
        \colhead{}& \colhead{} & \colhead{Diameter} & \colhead{Area} & \colhead{Box Size}\\
        \colhead{}& \colhead{(S/N)} & \colhead{(px)} & \colhead{(px)}& \colhead{(px)} 
    }
    \startdata
        1  & 0.15 & 9 & 5 & 301  \\
        2 & 0.15 & 7 & 5 & 301\\
        3 &  0.15 & 4 & 5 & 301\\
        4  & 0.15 & 11 & 21 & 301 \\
        5 & 1 & 1 &5 & 201 \\
        6 & 0.5 & 21 & 70 & 301 \\
        7 & 0.5 & 50 & 110 & 301 
    \enddata
    \tablecomments{List of mask parameters for A262. The background box size is scaled with the kiloparsec/arcsecond scale for each cluster. The other parameters are the same for all clusters.}
\end{deluxetable}

\begin{figure*}[hp]
    \centering
    \includegraphics[width=0.35\textwidth]{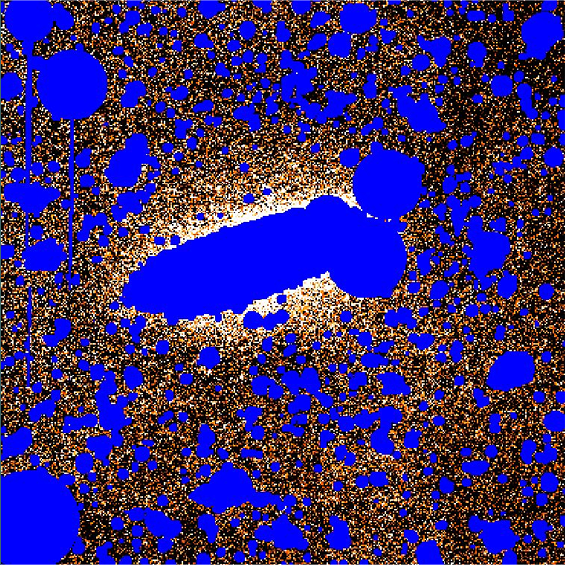}
    \includegraphics[width=0.35\textwidth]{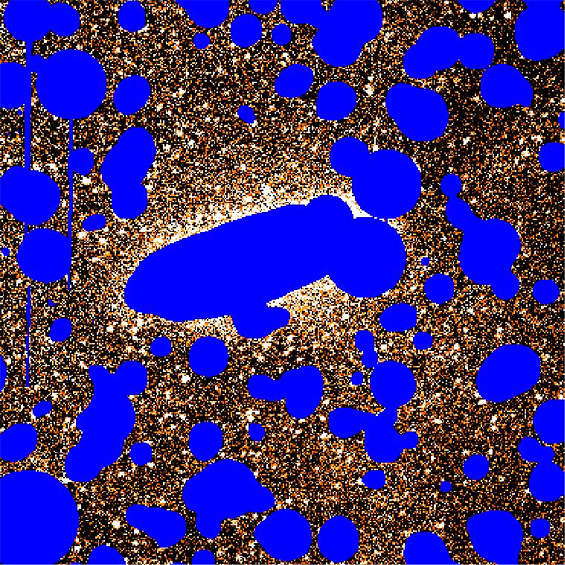}
    \includegraphics[trim = 0cm 0cm 0cm -0.2cm,width=0.35\textwidth]{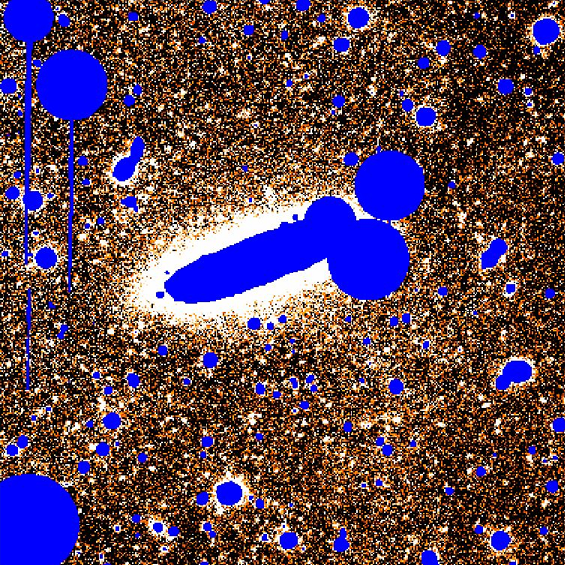}
    \includegraphics[trim = 0cm 0cm 0cm -0.2cm,width=0.35\textwidth]{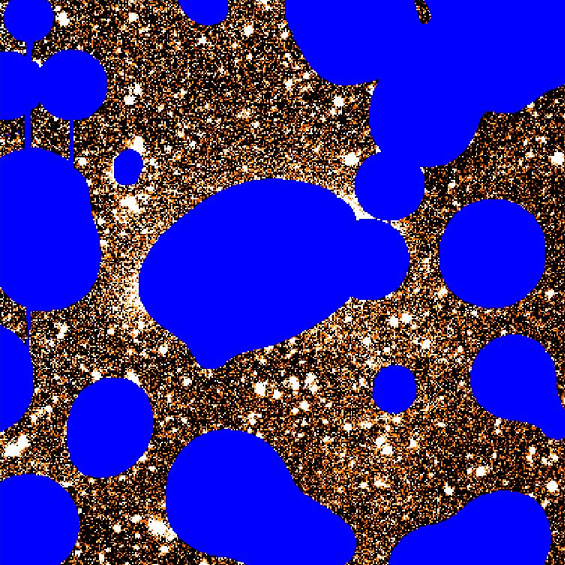}
    \includegraphics[trim = 0cm 0cm 0cm -0.2cm,width=0.35\textwidth]{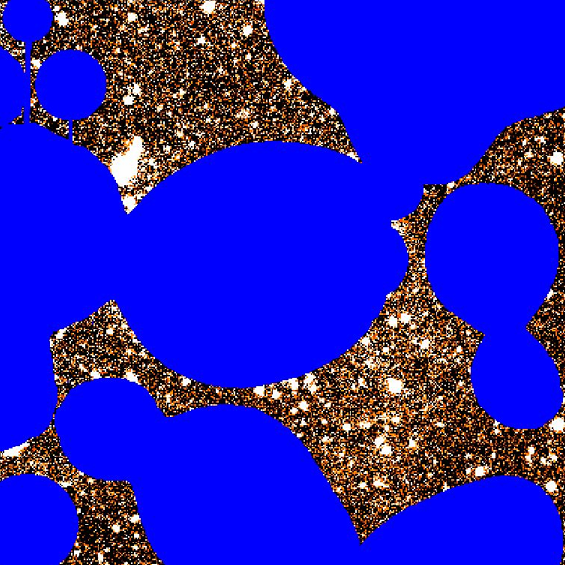}
    \includegraphics[trim = 0cm 0cm 0cm -0.2cm,width=0.35\textwidth]{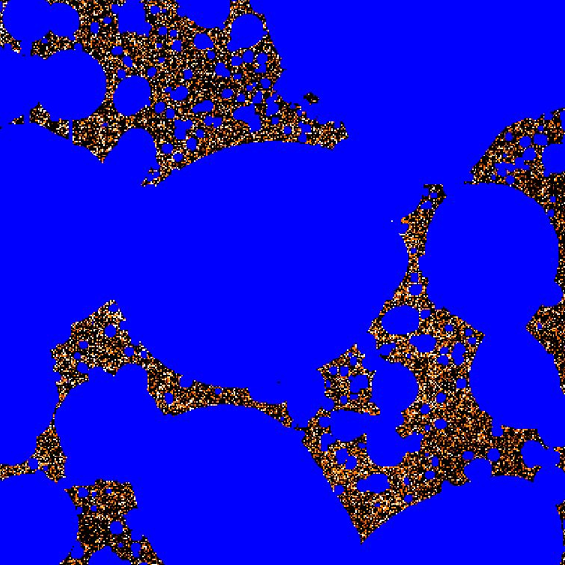}
    \includegraphics[width=0.8\textwidth]{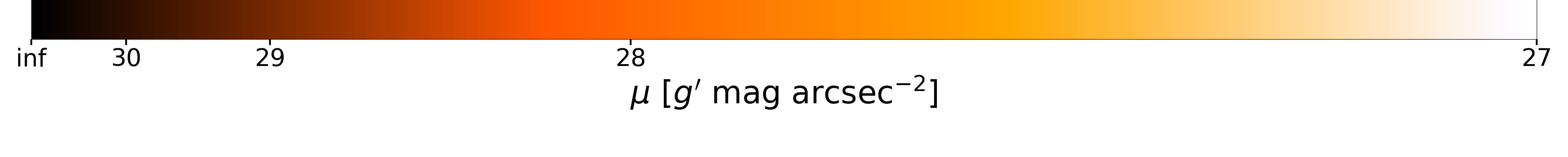}
    \caption{Mask 1, 4, 5, 6, and 7 (from left to right and top to bottom) and all masks together (bottom right) applied to a four by four binned cutout ($\mathrm{4.\arcmin9\times4.\arcmin9}$) of the $g'$-band data of A262. }
    \label{fig:masks}
\end{figure*}
\vspace{-4.5mm}

These seven masks are each optimized for differently sized objects. Mask 1, as well as the nearly identical masks 2 and 3, mask small sources, mask 4 and 5 mask medium-sized objects, and mask 6 and 7 mask large galaxies or the extended PSF wings of relatively bright stars. Figure \ref{fig:masks} shows these masks applied to a 4 $\times$ 4 pixel binned cutout image  of the $g'$-band stack of A262. Masks 2 and 3 are not shown here because the difference between them and mask 1 is not noticeable on this scale. 

Mask 1 is especially important for an accurate measurement of the UDG's structural parameters, as it also masks objects fainter than those detected by \verb+SExtractor+ and, hence, would not have been masked using only the \verb+SExtractor+ segmentation maps. Additionally, masks 1, 2, 3, and 4 include more of the faint wings than the \verb+SExtractor+ segmentation maps. Masking these outer wings of objects close to a UDG is also crucial, as they could not be modeled as a linear background gradient and, hence, would contaminate the Sérsic fits to the UDG`s outer profile. The masks for the larger objects do not cover the outermost wings of those objects, but this is not crucial for measuring UDG candidates, as these outer wings can be modeled with a linear background by \verb+GALFIT+. In the bottom-right panel of Figure \ref{fig:masks}, the cutout is shown with all masks applied. All sources are reliably masked.

Nevertheless, the largest elliptical galaxies in the cluster are not sufficiently masked. To improve the masks, the stack with all masks applied is masked manually. Note here that we de-mask the target galaxy in each mask before combining them (see Section \ref{sec:individualobjectmasks}).

The de-masking procedure is the reason for the slight variations between masks 1, 2, and 3. Thereby, all connected pixels of the mask, in which the central coordinate of the object is contained, are de-masked. In mask 1, it frequently occurs that masks of individual objects are only connected by very few pixels, which leads to these objects getting unintentionally de-masked, too. As long as those de-masked objects are included in the \verb+SExtractor+ catalogs, this is not a big issue, as they are modeled simultaneously with the target. On the other hand, de-masked objects that are not detected by \verb+SExtractor+ would significantly bias the measurements. To mitigate this issue, we create masks 2 and 3 with a slightly smaller expand diameter so that those objects that are connected only by a few pixels in mask 1 remain in the mask. We do not only use mask 3 with the smallest expand diameter, as the other two masks cover sources more conservatively. 



\begin{figure}[t]
    \centering
    \includegraphics[width=0.23\textwidth]{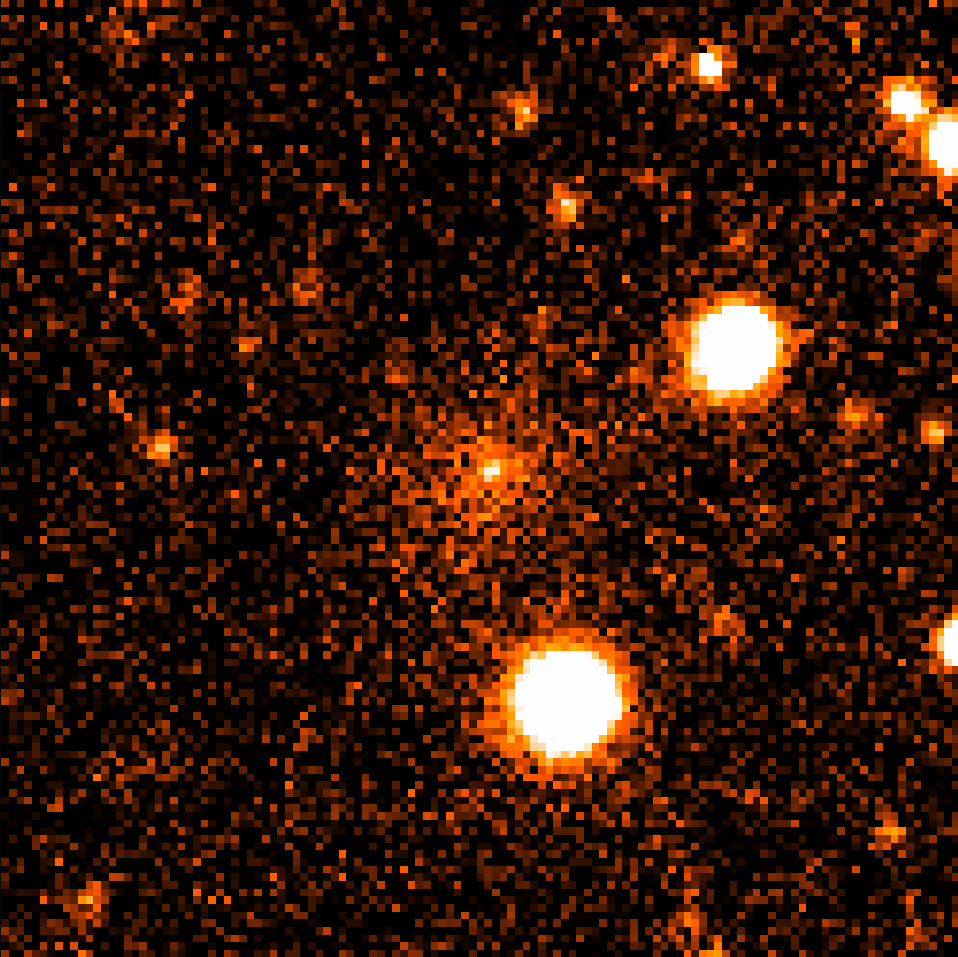}\\
    \includegraphics[width=0.23\textwidth]{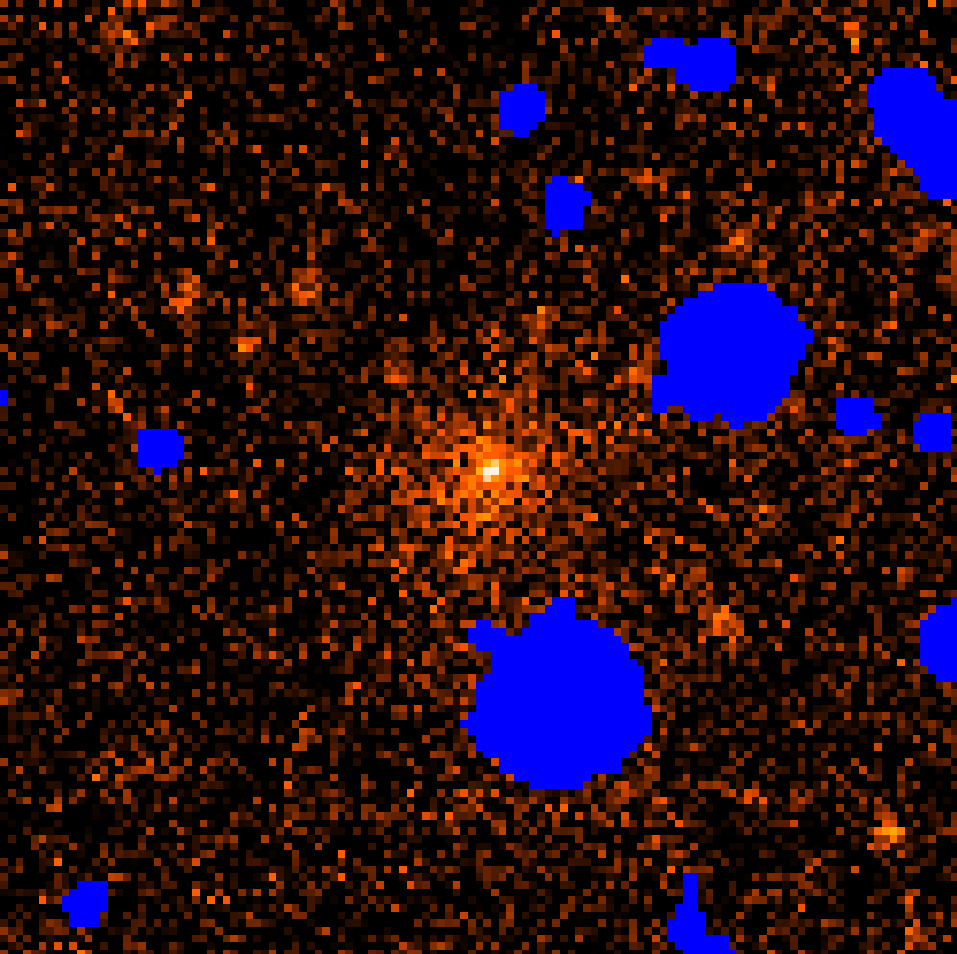}
    \includegraphics[width=0.23\textwidth]{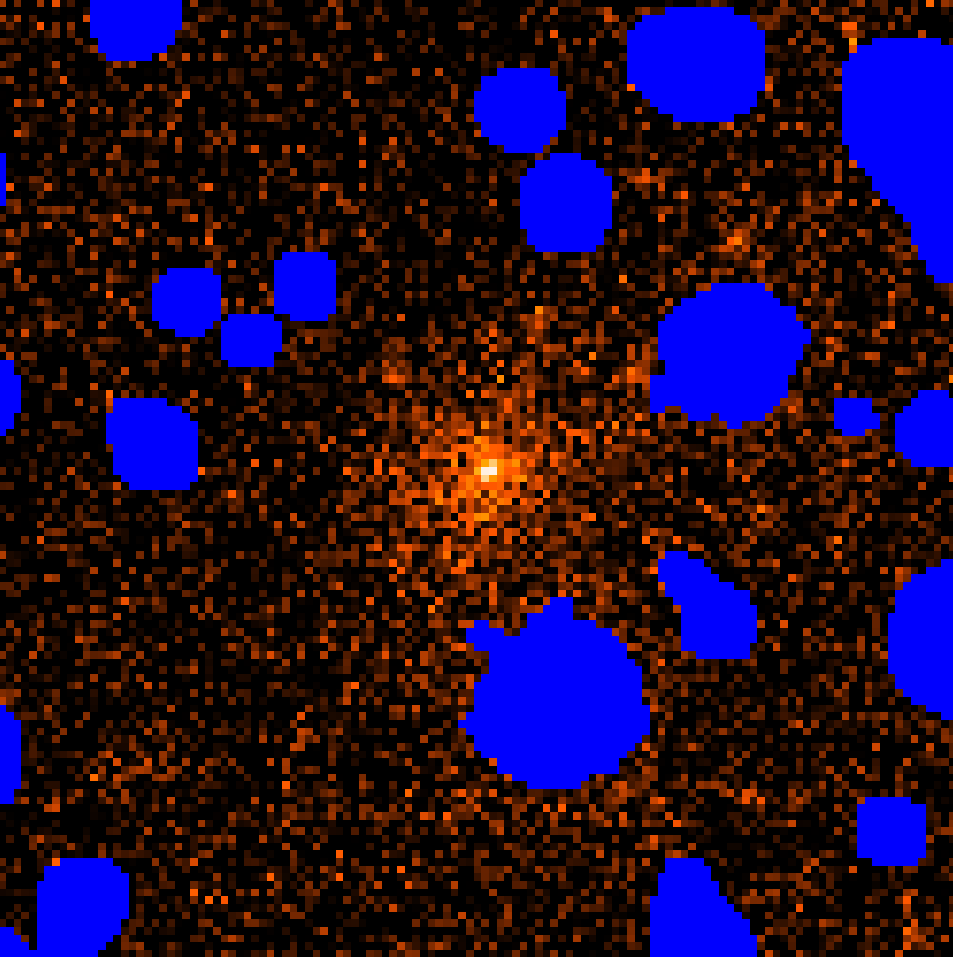}
    \includegraphics[width=0.47\textwidth]{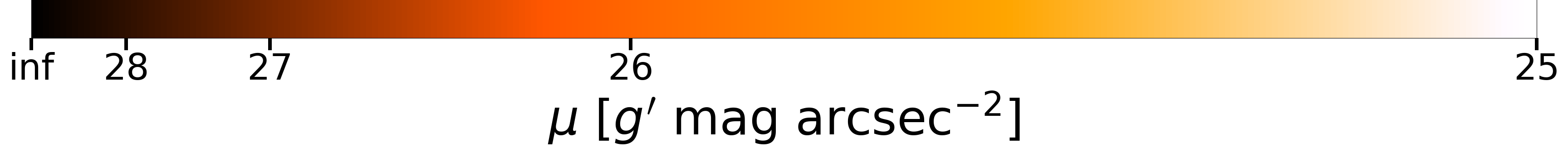}
    \caption{Two by two binned cutout ($\mathrm{50\arcsec\times50\arcsec}$) around a UDG in A262. The top panel shows those without masks, the bottom-left panel shows those masked with the SExtractor segmentation map and with the central object de-masked, and the bottom-right panel shows those using the masks from our masking routine combined with the SExtractor segmentation map. The masks are shown in blue.}
    \label{fig:SegmentationVSall}
\end{figure}

In Figure \ref{fig:SegmentationVSall},we show that our masking procedure in fact delivers more complete masks than the \verb+SExtractor+ segmentation maps. The top panel shows a cutout around a UDG in A262 ($g'$ band). The image size is as it is fed to \verb+GALFIT+ . The bottom-left panel is the same image but masked with only the de-masked \verb+SExtractor+ segmentation map. We see that many objects remain unmasked. The bottom-right panel shows the image with the demasked \verb+SExtractor+ segmentation map combined with our masks. By including our masks, significantly more small and faint objects are masked. Moreover, those of our masks that were not de-masked cover the sources more conservatively than \verb+SExtractor+`s segmentation maps do. 

\subsubsection{Error Images} \label{sec:errorimages}
In order to determine the errors of our \verb+GALFIT+ fits and our aperture color measurements accurately, we first have to calculate  error images. These images contain the uncertainty for each pixel error$(x,y)$. This calculation is done using the data in the object stacks $d(x,y)$, their weight images $w(x,y)$, and the global mean gain $g$. We approximate that the readout noise and the thermal noise are negligible, and hence, the error is purely the photon noise of the sources and of the sky:

\begin{equation}
     \mathrm{error}(x,y)=\sqrt{\frac{\mathrm{source}(x,y)+\mathrm{sky}(x,y)}{g(x,y)}}\\
\end{equation}

Furthermore, we approximate that the data equals the source flux, and the local background standard deviation std($x,y$) resembles the noise introduced by the sky. This gives: 
\begin{align}
    \mathrm{error}(x,y)\approx\sqrt{\frac{|d(x,y)|}{g(x,y)}+\mathrm{std}(x,y)^2}
\end{align}
The local gain $g(x,y)$ can be calculated using the global mean gain (provided by SWarp \citealt{swarp} in the data reduction) scaled with the ratio of the local weight $w(x,y)$ (also provided by SWarp) and the median weight (median\{$w$\}). The local background standard deviation is determined via the global minimum  of the spatially varying standard deviation ($\mathrm{std_{min}}$) scaled inversely with the square root of the local fraction of the total exposure time, which in turn is given by the ratio of the local weight and the maximum weight (max\{$w$\}). This gives:
\begin{align}
    \mathrm{error}(x,y)=\sqrt{\frac{|d(x,y)|}{g\times\frac{w(x,y)}{\mathrm{median}\{w\}}}+\left(\frac{\mathrm{std}_{\mathrm{min}}}{\sqrt{\frac{w(x,y)}{\mathrm{max}\{w\}}}}\right)^2}
\end{align}
To calculate $\mathrm{std_{min}}$, we apply all of the previously created masks to the science image (result: $d_{m}$) and to the weight image (result: $w_{m}$) first. Then, we rescale the masked science image  $d_{m}$ with the square root of the fraction of the local weight and the maximum weight. Here, the fraction of the weight approximates fraction of the exposure time. This resembles a background image with a constant global minimum standard deviation over the whole field of view. The global minimum standard deviation is given by the standard deviation of this background image and the final error image, is given by:

\begin{align}
    \begin{aligned}
         \mathrm{error}(x,y)&=\mathrm{sqrt}\Biggl\{\frac{|d(x,y)|}{g\times w(x,y)/\mathrm{median}\{w\}}\\
        &+\left(\frac{\mathrm{std}\bigl\{d_{m}\times\sqrt{w_{m}/\mathrm{max}(w)}\bigr\}}{\sqrt{w(x,y)/\mathrm{max}\{w\}}}\right)^2\Biggr\}
    \end{aligned}
\end{align}

\subsection{Brightness and Color Measurements of Bright Galaxies} \label{sec:EllS0}

\begin{figure*}[ht]
    \centering
    \includegraphics[width=\textwidth]{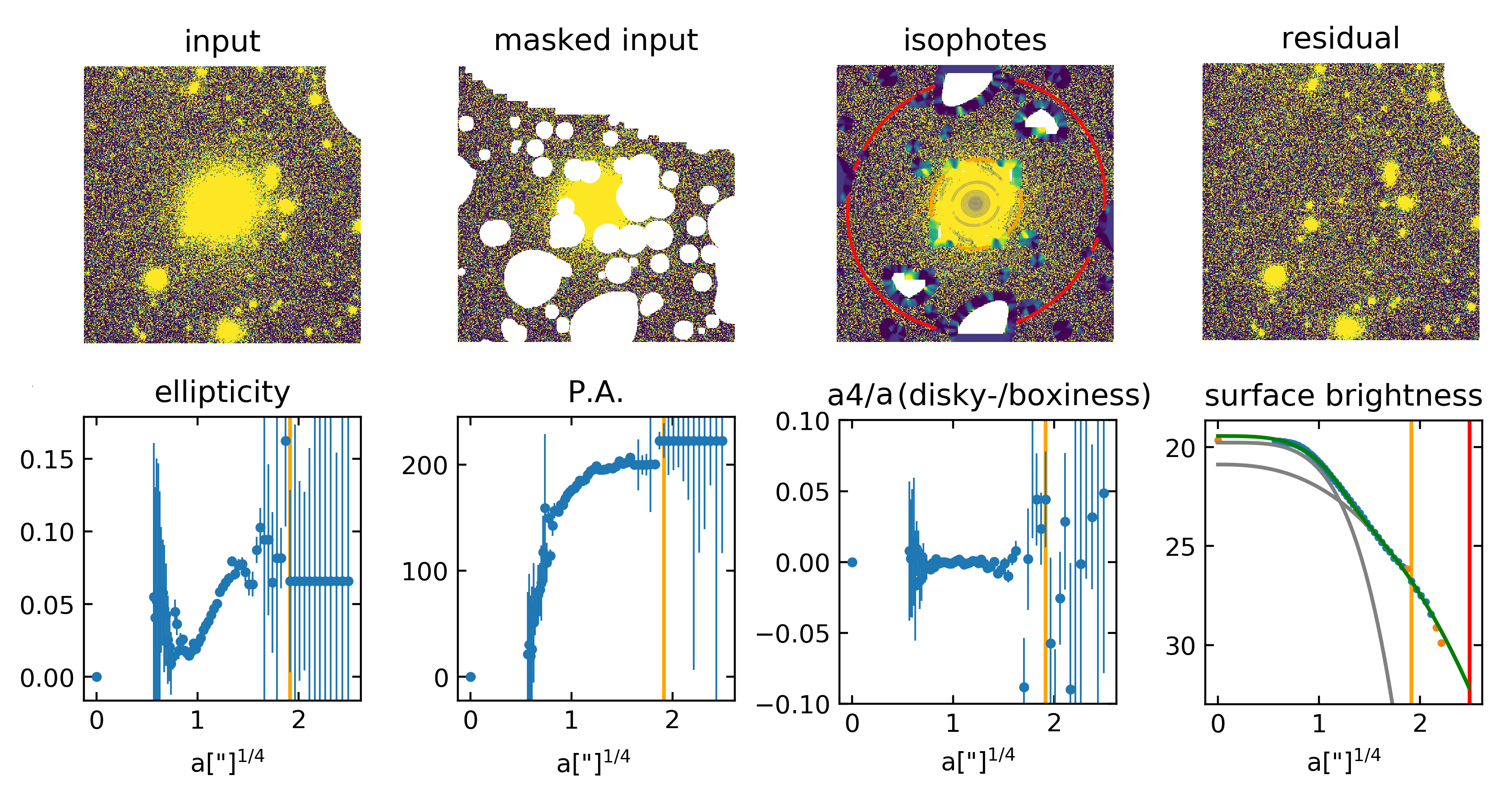}
    \caption{Selected outputs of the fitting routine for the galaxy 2MASX J01515160+3615027 in the $g'$ band. The ellipticity, position angle, and central coordinates are fixed after the orange marked isophote. The background is determined at the red marked position. In the surface brightness profile plot, the green dots are the data points used for the fit, the green line corresponds to the best-fit double Sérsic profile, and the two gray lines correspond to the two individual Sérsic profiles.}
    \label{fig:ELLexample}
\end{figure*}

Firstly, we preselect bright ($m_{\mathrm{tot}}<17\,g'$\,mag) possible cluster members. Therefore, we select all galaxies whose $g'-r'$ color deviates by less than five times the median absolute deviation from the median color of all bright galaxies in our sample. These galaxies are later-on used to fit a red sequence model (see sec. \ref{sec:redsequence}).

We create masks for all selected galaxies using the previously created masks 1--5 and the \verb+SExtractor+ segmentation maps. As we cannot distinguish between the mask of the target and an overlapping object in the masks created with our masking tool, we de-mask the target and connected objects. In the \verb+SExtractor+ segmentation maps, we only unmask the target. Then, all masks are combined, applied to the cutout images, and then manually improved.

For these bright galaxies, we are only interested in the total magnitudes and colors in order to determine the red sequence. To measure them, we directly integrate the flux down to $\mathrm{30\,mag\,arcsec^{-2}}$ and add the flux below that threshold by integrating an analytic best-fit Sérsic or double Sérsic function from $\mathrm{30\,mag\,arcsec^{-2}}$ to infinity. To do so, we follow the procedure presented in \cite{klugeisophotespy} and \cite{Kluge2023rhea}, which is based on the python package \verb+photutils+ \citep{photutils}. We create an isophote model of the galaxies in the $g'$ band with radially varying ellipticity, position angle, and center of the isophotes. We fit single or double Sérsic functions to the surface brightness profiles. As we only use these analytic functions to account for the flux below $\mathrm{30\,mag\,arcsec^{-2}}$, we do not require them to be accurate in the center, but only to trace the outer profile well. Hence, we fit single Sérsic profiles only between $\mathrm{22\,mag\,arcsec^{-2}}$ and $\mathrm{29\,mag\,arcsec^{-2}}$ \citep[see also][]{klugeisophotespy}. We fit double Sérsic functions to the full surface brightness profiles down to $\mathrm{29\,mag\,arcsec^{-2}}$ in order to better constrain the profile that has more degrees of freedom. We only use the double Sérsic profiles if the galaxy shows a clear double component profile; otherwise, we use the simpler single Sérsic profile. If both of these attempts fail to fit the galaxy profile, we adjust the surface brightness fit limits manually. As uncertainties for the total magnitudes, we use the deviation from the directly integrated total magnitudes. One example of those measurements is shown in Figure \ref{fig:ELLexample} for the galaxy 2MASX J01515160+3615027 in the $g'$ band.

\subsection{UDG and Dwarf Measurements} \label{sec:udgmeasurements}
\subsubsection{Preselection of UDG and Dwarf Candidates} \label{sec:preselectionudgcandidates}
We preselect our UDG sample just very roughly to also include similar galaxies such as spheroidals and classify the UDGs afterward based on the parameters measured in the \verb+GALFIT+ fits. 

We select galaxies with an apparent magnitude between 17\,$g'$\,mag and 27.2\,$g'$\,mag and a mean surface brightness within the effective radius between $15\,g'\,\mathrm{mag\,arcsec^{-2}}$ and $29.4\,g'\,\mathrm{mag\,arcsec^{-2}}$ for A1656 and adjust these criteria for other pointings, correcting for galactic extinction, cosmic dimming, K-correction, and distance modulus under the assumption that the object is at the distance of the respective galaxy cluster. The faint limits are set to mitigate false detections.  Furthermore, we remove objects from our sample with \verb+S/G>0.97+ or $R_e<2\,\mathrm{px}$ to reject point sources, as well as objects with \verb+FLAGS>4+ or \verb+PETRO_RADIUS=0+ to mitigate false detections.

\subsubsection{GALFIT Fits and Individual Object Masks} \label{sec:individualobjectmasks} \label{sec:galfit}
\begin{figure*}[ht]
    \centering   
    \includegraphics[width=\textwidth]{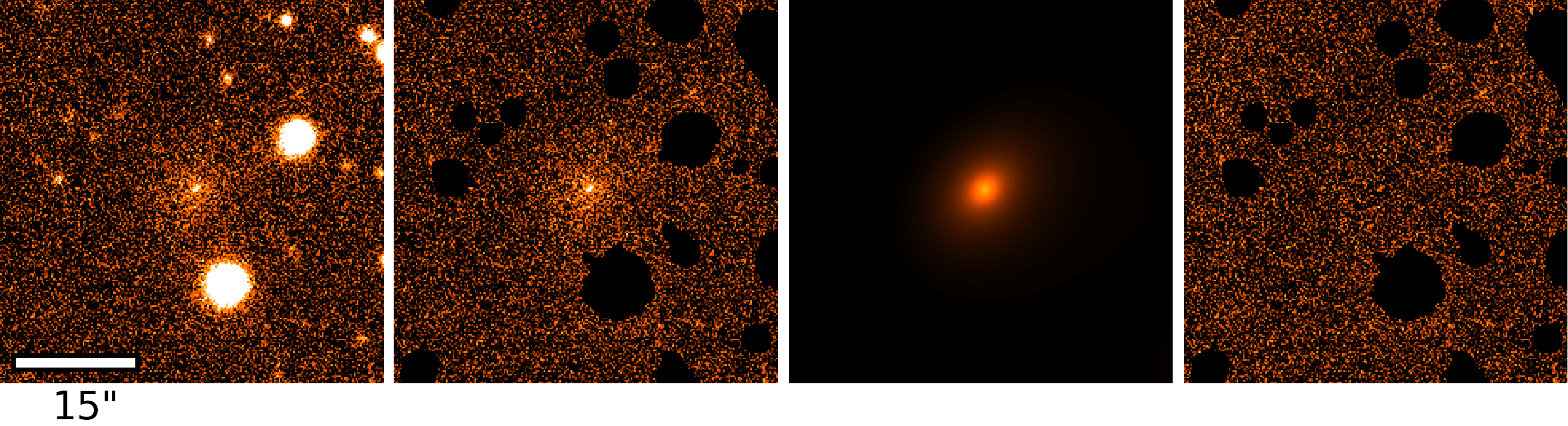}
    \includegraphics[width=0.8\textwidth]{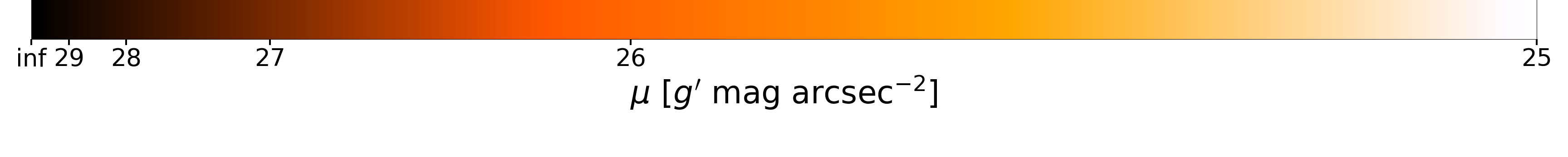}
    \caption{Original cutout image, automatically masked cutout image, best-fit GALFIT model, and residuum for an UDG in A262 from left to right.}
    \label{fig:UDGfirstfit}
\end{figure*}
For the creation of the masks for these galaxies, we use the previously created masks 1--5 and \verb+SExtractor+ segmentation maps. For the $g'$-band masks, we again remove the masks of the target and connected objects. Unlike for the bright and large galaxies, we also remove connected masks in the \verb+SExtractor+ segmentation maps, as these masks are not conservative enough. Instead of masking these nearby objects, we model them using \verb+GALFIT+. 

For the $u'$ and the $r'$ band, we combine all masks without demasking, as we use \verb+GALFIT+ only to fit the background as a gradient. The actual color measurement is done using more reliable aperture photometry (see section \ref{sec: colormeasurements}), as this is more stable.

The \verb+GALFIT+ fits are performed on cutouts around the target with a side length of $12R_e$, where $R_e$ refers to the directly integrated half-light radius obtained with \verb+SExtractor+. We also set a minimum side length of 101 px and a maximum of 251 px.
Then, we create a 101 $\times$ 101 px PSF model at the central position of the target from our \verb+PSFEx+ model.

As initial parameters for our \verb+GALFIT+ fits, we use the parameters measured with \verb+SExtractor+. All de-masked objects in the cutout are modeled either by a single Sérsic function or by a PSF. Objects with $\mathrm{S/G>0.97}$ or an $\mathrm{FWHM < (FWHM_{PSF} - 0.\arcsec1)}$ and a/b $<$ 1.3 are considered as point sources and hence modeled with a PSF model. The background is fitted by a linear gradient. Furthermore, we set the size of the convolution box to 99 $\times$ 99 px.

We found that using the total magnitude $M_{\mathrm{tot}}$ for the \verb+GALFIT+ fits leads to more converging fits than using $\mu_0$ or $\mu_e$ probably because the $\mathrm{mag_{auto}}$ that we use as the initial parameter is more reliable than $\mu_0$ or $\mu_e$. Hence, we use $M_{\mathrm{tot}}$ for our initial \verb+GALFIT+ fit. To determine $\mu_0$ and $\mu_e$ and the corresponding uncertainties, we use the parameters from the initial \verb+GALFIT+ run and fix all parameters, except either $\mu_0$ or $\mu_e$, respectively. As this does not provide reasonable errors, we rerun these fits again using the parameters determined in this way as initial parameters without fixing them.

An example of these \verb+GALFIT+ fits and automatic masking is shown in Figure \ref{fig:UDGfirstfit}. Here, we show from left to right the original cutout image, the masked cutout image, the model, and the residuum for a UDG in A262.

\subsubsection{Color Measurements} \label{sec: colormeasurements}
In order to measure the colors of our galaxy sample, we use aperture photometry. This is more reliable than a parametric fit, especially for the faint $u'$-band data. If the aperture is smaller than a few times the PSF, then the measurements can be affected by different PSFs for the different filter bands. Therefore, we convolve all cutouts to the same target PSF determined in section \ref{sec:psfmeasurements}.

In detail, we first subtract the background determined in the \verb+GALFIT+ fits from the cutouts. 

Then, we use \verb+diffima+ \citep{fitstools} to calculate for each cutout image the convolution kernel to convolve the PSF from the PSF of each filter to the target PSF. 

Furthermore, we apply the masks used for the $g'$-band \verb+GALFIT+ fits and combine the \verb+SExtractor+ segmentation maps with only the target being de-masked. 

For the aperture photometry measurements, we use the python package \verb+Photutils+ \citep{photutils}. We measure the flux in elliptical apertures, as this increases the S/N compared to a circular aperture. We use a semi-major axis of $1 R_e$. 
The effective radius, position angle, and axis ratio are obtained from the previous \verb+GALFIT+ fits. Furthermore, we set a minimum aperture area of 100 px to ensure a high enough S/N and a maximum semi-major axis of 15 px to reduce the probability of including nonmasked contamination. For both cases, the axis ratio is kept fixed.

\newpage
\subsubsection{Catalog Processing} \label{sec:catalogprocessing}
Firstly, we reject all objects with $\Delta(u'-g')$ or $\Delta(g'-r')$ larger than 0.2 mag. Here, we aim rather for a clean than for a complete sample.

For the correction of galactic absorption, we use the extinctions from \citet{Schlafly2011} at the center of the galaxy cluster, assuming it to be constant over the field of view. For the K-correction, we use the web tool by \citet{Kcorrection2012} under the assumption that all objects belong to the galaxy cluster. The absolute magnitudes are calculated using the distance modulus and cosmic dimming obtained from the cosmology calculator by
\citet{wrightcosmocalc2006} under the assumption that the objects are at the redshift of the cluster. 

For the comparison of our data with \citet{Binggeli1994} and \citet{Kormendy2009}, we also calculate $B$- and $V$-band magnitudes following \citet{Jester} using $g=g'+0.09$ and $r=r'$ for the sun \citep{Willmer2018}:

\begin{equation}
    V = g' - 0.59\left(g'_{\mathrm{aper}}-r'_{\mathrm{aper}}\right) + 0.03
\end{equation}  
\begin{equation}
    B = g' + 0.39\left(g'_{\mathrm{aper}}-r'_{\mathrm{aper}}\right) + 0.34
\end{equation}

Here, we correct the magnitudes using a photometric zero-point determined in apertures with 10\arcsec~diameter ZP$_{10}$.

\subsubsection{Bicolor Sequence Selection} \label{sec:bi colorsequenceselection}
\begin{figure*}[ht]
    \centering
    \begin{tabular}{c c}
        \textbf{A1656} & \textbf{A262} \\
        \includegraphics[width=0.47\textwidth]{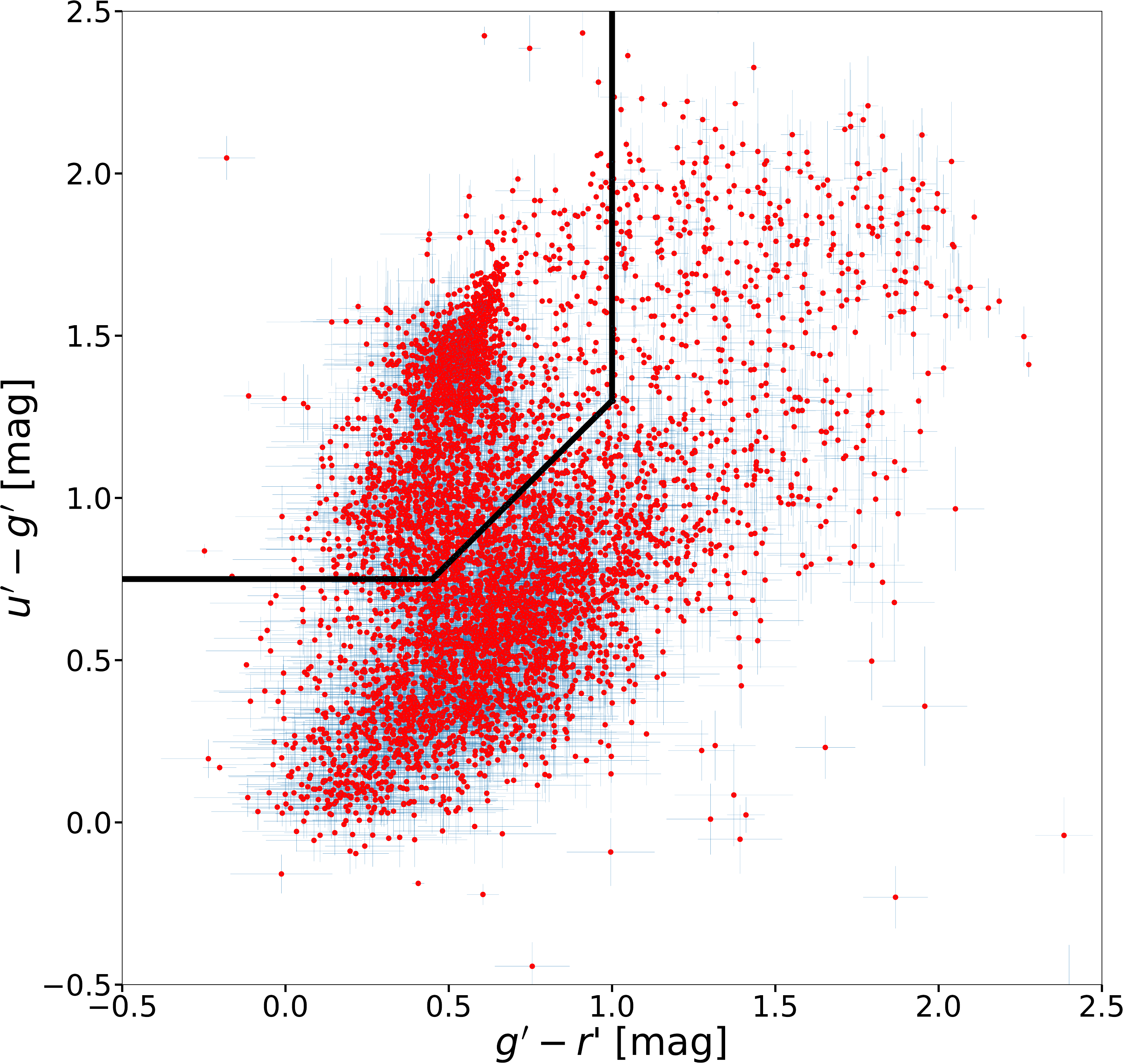} & \includegraphics[width=0.47\textwidth]{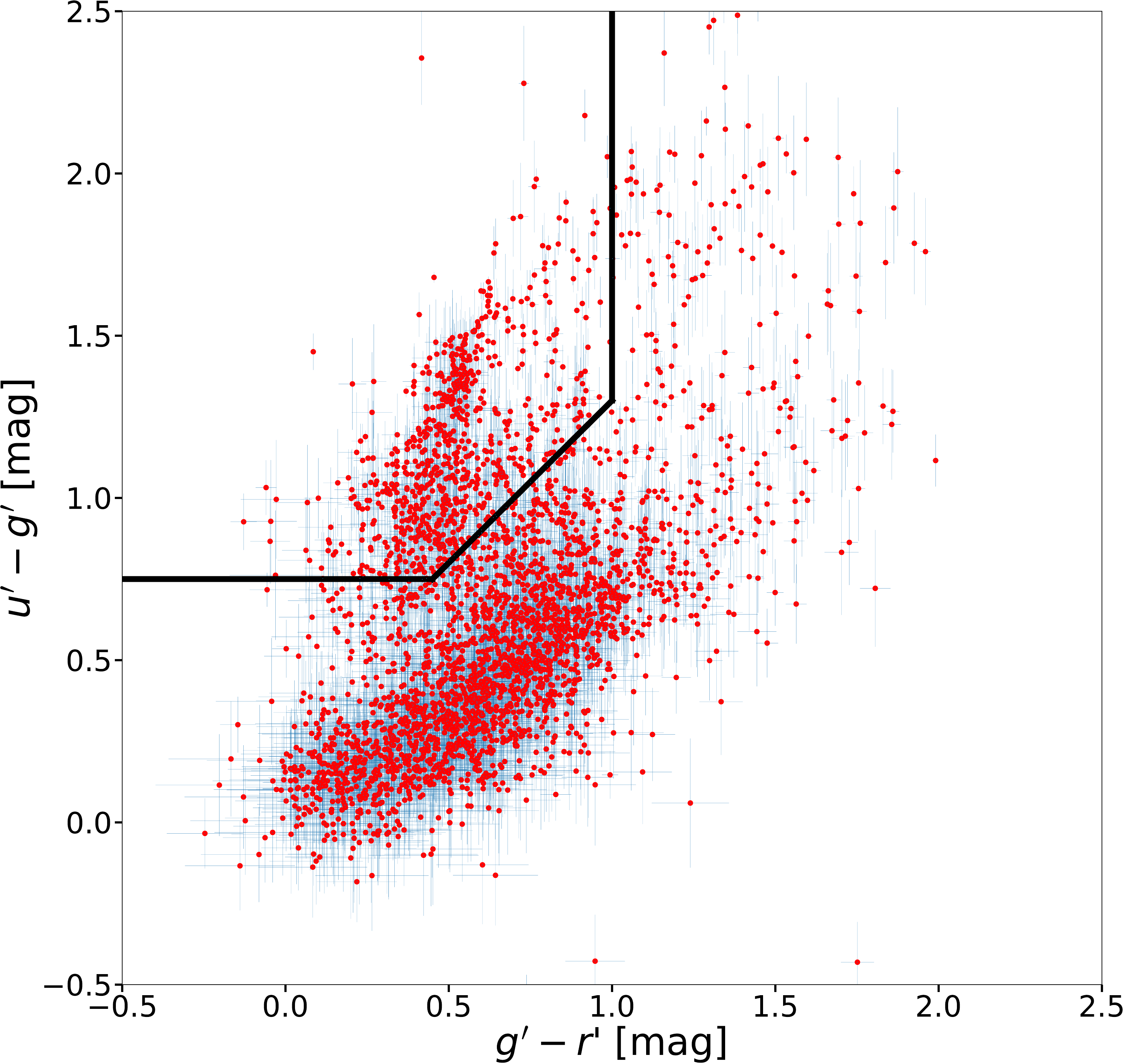} \\
        \\
        \textbf{Reference A1656} & \textbf{Reference A262} \\
        \includegraphics[width=0.47\textwidth]{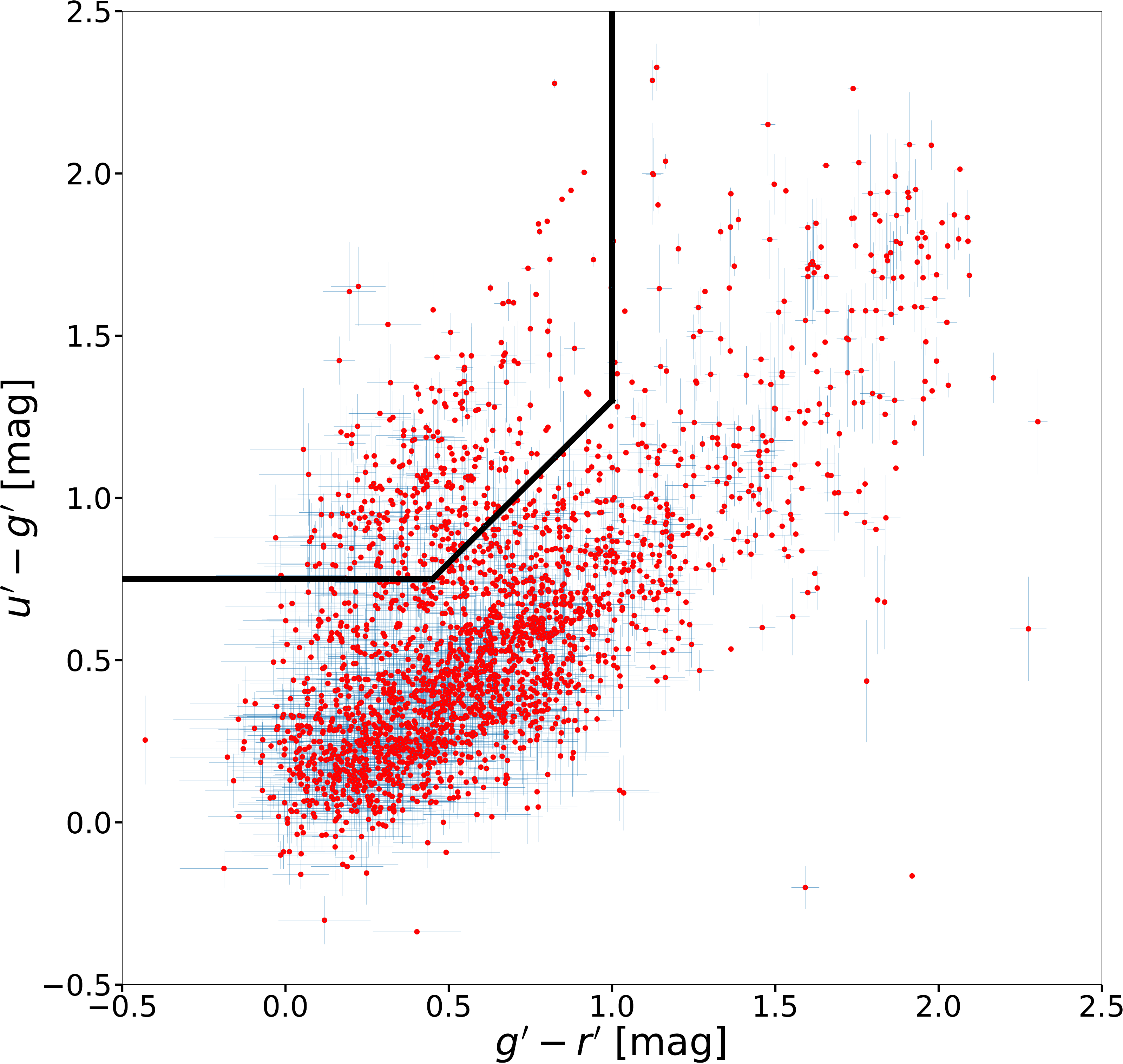} & \includegraphics[width=0.47\textwidth]{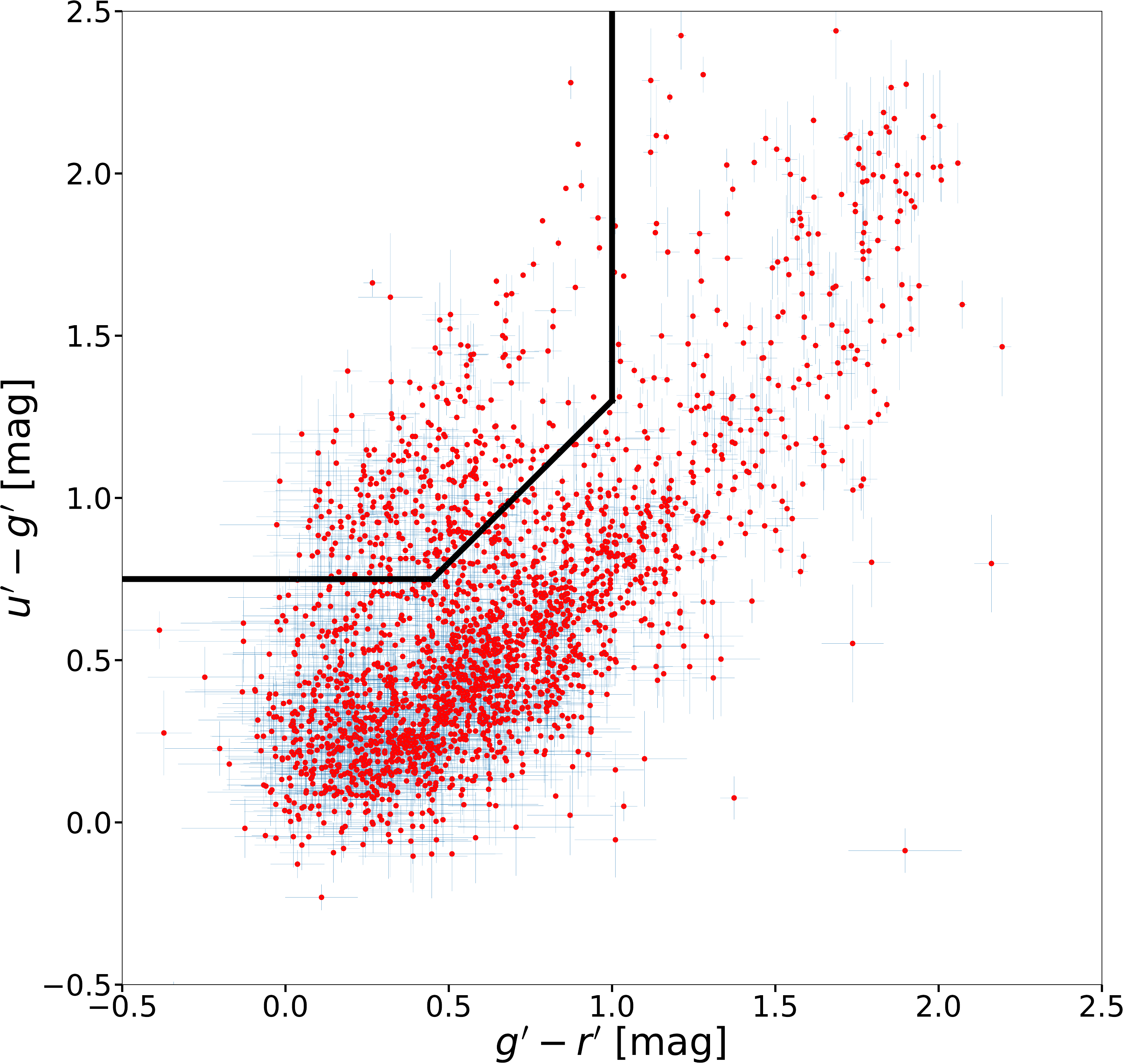}
    \end{tabular}
    \caption{$u'-g'$ vs. $g'-r'$ color--color diagrams for A1656 and A262, as well as for the reference field when alanyzed for the respective cluster. The black line indicates the selection cutoffs. All galaxies in the top-left corner are considered as quiescent.}
    \label{fig:colorcolor}
\end{figure*}

As a preselection of our cluster member sample, we first select quiescent galaxies using the bicolor sequence. Here, we follow \citet{Williams} who found that quiescent and star-forming galaxies form two distinct sequences in color--color diagrams. 
For our selection of quiescent galaxies, we use a $u'-g'$ versus $g'-r'$ color--color diagram. Here, star-forming galaxies that are reddened due to dust move along the diagonal, whereas quiescent galaxies are mainly affected in the $u'$ band by the 4000 $\mathrm{\AA}$ break and, hence, are shifted upward from the diagonal and form a distinct sequence there. For the selection of quiescent galaxies, we use the following criteria:
\begin{align}
    u'-g' &> g'-r + 0.3 \label{eq:diag}\\
    u'-g' &> 0.75\\
    g'-r' &< 1
\end{align}
The diagonal selection criterion in Equation (\ref{eq:diag}) is set to the approximate minimum of the number density between the quiescent and the star-forming sequence. The color--color diagrams are shown in Figure \ref{fig:colorcolor} for A1656 (top left) and A262 (top right). In those diagrams, we consider all objects in the top-left corner to be quiescent. The color--color diagrams for the reference field when analyzed for the respective cluster (bottom) differ from each other because different objects are contained in the sample (due to the different selection criteria), different masks, and different K-correction (corrected under the assumption that they belong to the respective cluster).

\subsubsection{Eyeball Inspection, Remasking, Nucleus Fits, and Rejection} \label{sec:style}
\begin{figure*}[ht]
    \centering
    \includegraphics[width=\textwidth, trim=0cm 0cm 0cm -0.2cm]{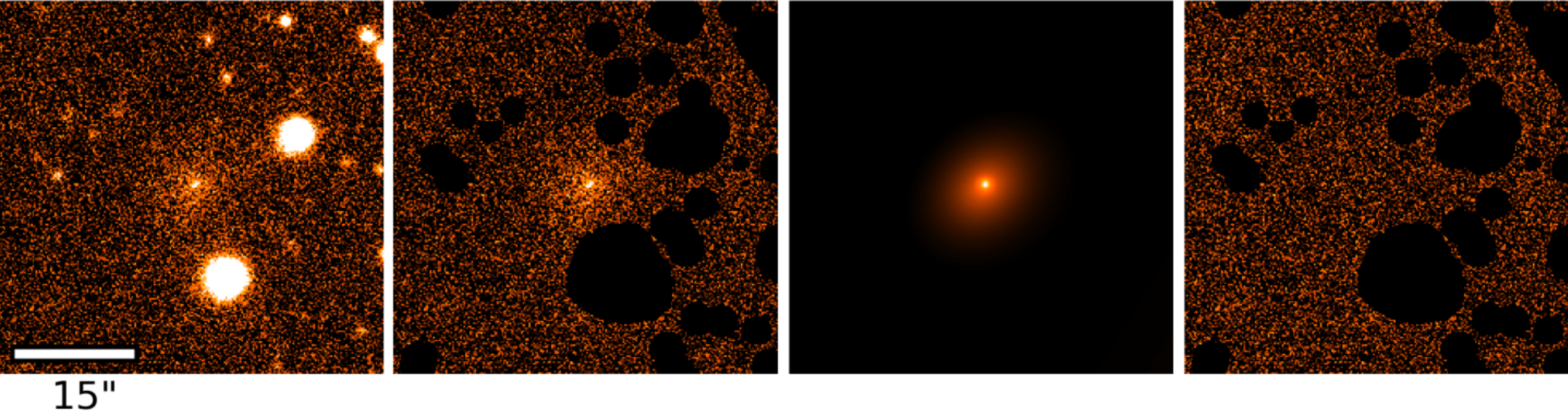}
    \includegraphics[width=0.8\textwidth]{scale0to4g.png}
    \caption{Original cutout image, manually edited masks applied to the cutout image, best-fit single Sérsic plus PSF GALFIT model, and residuum.}
    \label{fig:remaskingnucleus}
\end{figure*}

After the bicolor preselection, all \verb+GALFIT+ fitting results of quiescent galaxies undergo an eyeball inspection. Here we check the masked input images, the best-fit models, and the residuals (see Figure \ref{fig:UDGfirstfit}) for all galaxies and decide whether we have to improve the masks manually or whether a clear nucleus is present that has to be added for the fit. Furthermore, we remove all apparently bad fits that cannot be improved by improving the masks or fitting an additional nucleus. These bad fits are mainly caused by a strong overlap with a bright nearby object, multiple objects being detected as one, or a more complex structure of the galaxy than a simple Sérsic profile.

Then, we rerun \verb+GALFIT+ and redo the color measurements with the improved masks and the nuclei for those galaxies, where this is necessary. Afterward, the fits again undergo an eyeball inspection, and galaxies for which the fits are still not appropriate are removed. 

This remasking and nuclei fitting for our prime UDG in A262 are visualized both before (Figure \ref{fig:UDGfirstfit}) and after these steps have been performed (Figure \ref{fig:remaskingnucleus}).

Afterward, the catalog processing is run again. Finally, we reject all objects with large uncertainties ($\Delta m_{\mathrm{tot}} > 1\,g'\,\mathrm{mag}$, $\mathrm{\Delta \mu_0} > 1\,g'\,\mathrm{mag\,arcsec^{-2}}$, $\mathrm{\Delta \mu_e} > 1\,g'\,\mathrm{mag\,arcsec^{-2}}$, or $\Delta R_e / R_e>0.5$).

For all quiescent galaxies whose $\mu_e$ or $\mu_0$ fits failed or $\mathrm{\Delta \mu_0} > 1\,g'\,\mathrm{mag\,arcsec^{-2}}$ or $\mathrm{\Delta \mu_e} > 1\,g'\,\mathrm{mag\,arcsec^{-2}}$, we calculate $\mu_e$ and $\mu_c$ analytically from the parameters obtained from the $m_{\mathrm{tot}}$ fits:

\begin{align}
        \mu_e&=-2.5log_{10}\left(\frac{10^{-0.4m_{\mathrm{tot}}} (1.999n-0.327)^{2n}}{2\pi n q R_e^2 e^{1.999n-0.327} \Gamma(2n)}\right)\\
        \mu_0&=\mu_e-1.999n+0.327
\end{align}
The respective uncertainties are determined by varying the parameters randomly using a normal distribution around the best-fit value. Note that \verb+GALFIT+ does not provide a covariance matrix, and hence we are overestimating the errors here. Here, we again reject results with $\mathrm{\Delta \mu_0} > 1\,g'\,\mathrm{mag\,arcsec^{-2}}$ or $\mathrm{\Delta \mu_e} > 1\,g'\,\mathrm{mag\,arcsec^{-2}}$. Furthermore, we reject objects with $n>4$ in this procedure, as for those objects, $\mu_0$ is diverging, and $n>4$ is an unrealistically high value for the galaxies we are interested in. Of those analytically determined $\mu_e$ and $\mu_0$, only $\approx 15 \%$ provide acceptable results. 

\subsubsection{Red Sequence Cluster Member Selection} \label{sec:redsequence}
The final cluster member selection is done using a $g'-r'$ red sequence. The fitting routine resembles the one described by \citet{Stott2009}. For the determination of the red sequence, we use all bright galaxies whose parameters were determined in section \ref{sec:EllS0} and all quiescent galaxies with $M_{\mathrm{tot}} <  20\,g'\,\mathrm{mag}$ remaining after the previous selection steps.
Firstly, the median color is determined. Then, an orthogonal distance regression of a linear function is performed on all data points within a color interval with a width of five times the median absolute deviation around the previously determined median color. In the following iteration steps, this width is defined relative to the linear function determined in the previous step. This is iterated five times. As the final selection criterion, we chose that the galaxies must not deviate more from the best-fit red sequence than the quadratically combined width of the intrinsic width and the width introduced by the statistical scatter of the data points. Based on the scatter of the high-S/N data points of the bright elliptical and S0 galaxies, we estimate the intrinsic half-width (hw) of the red sequence to be 0.06\,mag. The statistical scatter is estimated via the mean aperture color error of the apparent magnitude bins, each spanning a range of $1\,g'\,\mathrm{mag}$. Note that this takes only the statistical broadening of the red sequence due to larger errors at the faint end into account, but not a potential real broadening of the red sequence at the faint end. The selection limits ($l_i$) of each such magnitude bin are given by:

\begin{align}
    \begin{aligned}
        l_i&=a\times m_{\mathrm{tot},i}+b \\
        &\pm \sqrt{\mathrm{hw}^2(a^2+1)+\left(3\times \mathrm{mean} \{\Delta (g'-r')\})_i\right)^2}
    \end{aligned}
\end{align}

where $a$ is the slope of the best-fit red sequence, and $b$ is the offset. Then, third-degree polynomials are fitted to the upper and lower limits, which give a smooth selection criterion. 

\begin{figure*}[ht]
    \includegraphics[width=0.49\textwidth, trim=-5cm 0cm 0cm 0cm]{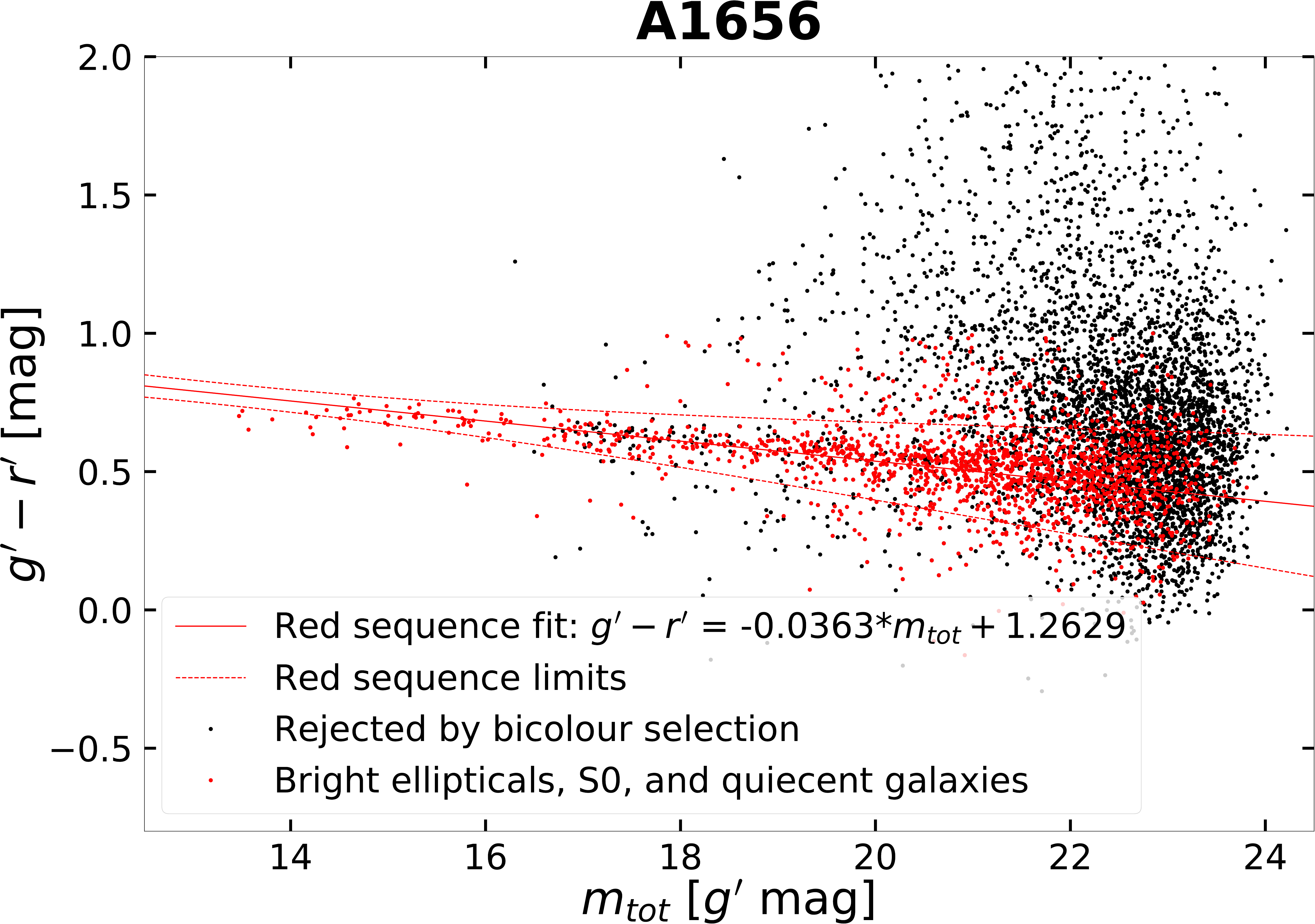}
    \includegraphics[width=0.49\textwidth, trim=-5cm 0cm 0cm -5cm]{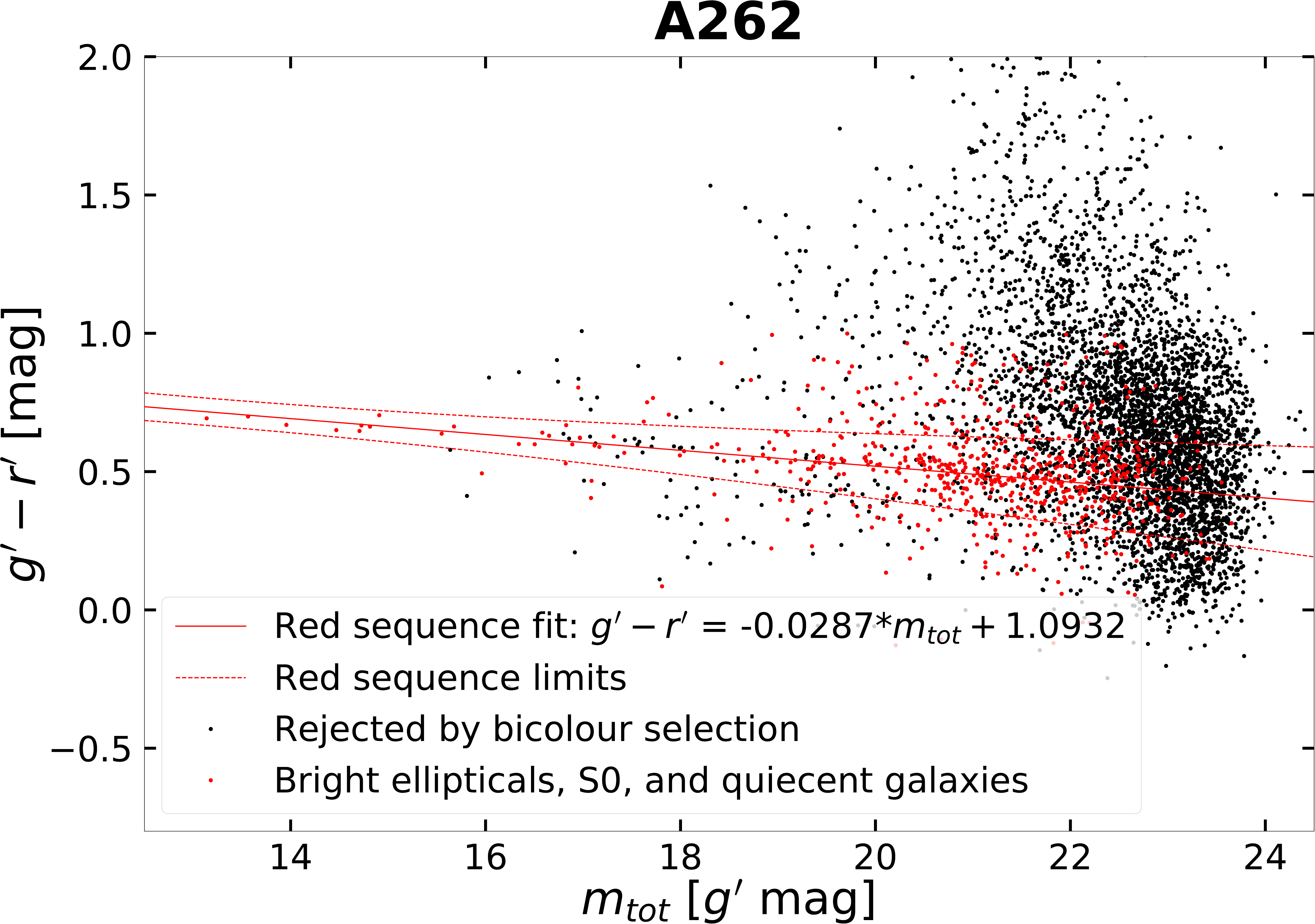}\\
    \includegraphics[width=0.49\textwidth, trim=-5cm 0cm 0cm 0cm]{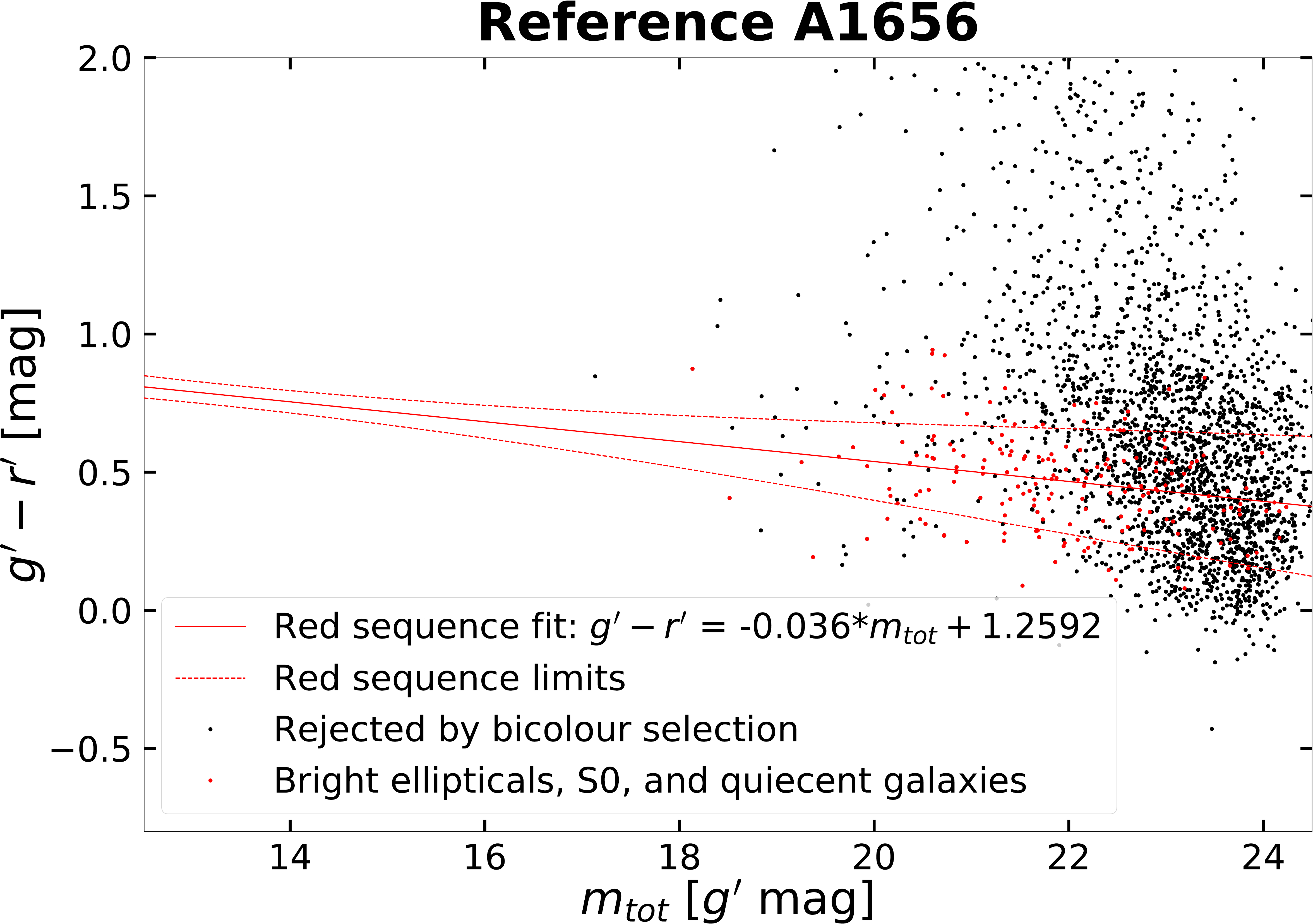}
    \includegraphics[width=0.49\textwidth, trim=-5cm 0cm 0cm -5cm]{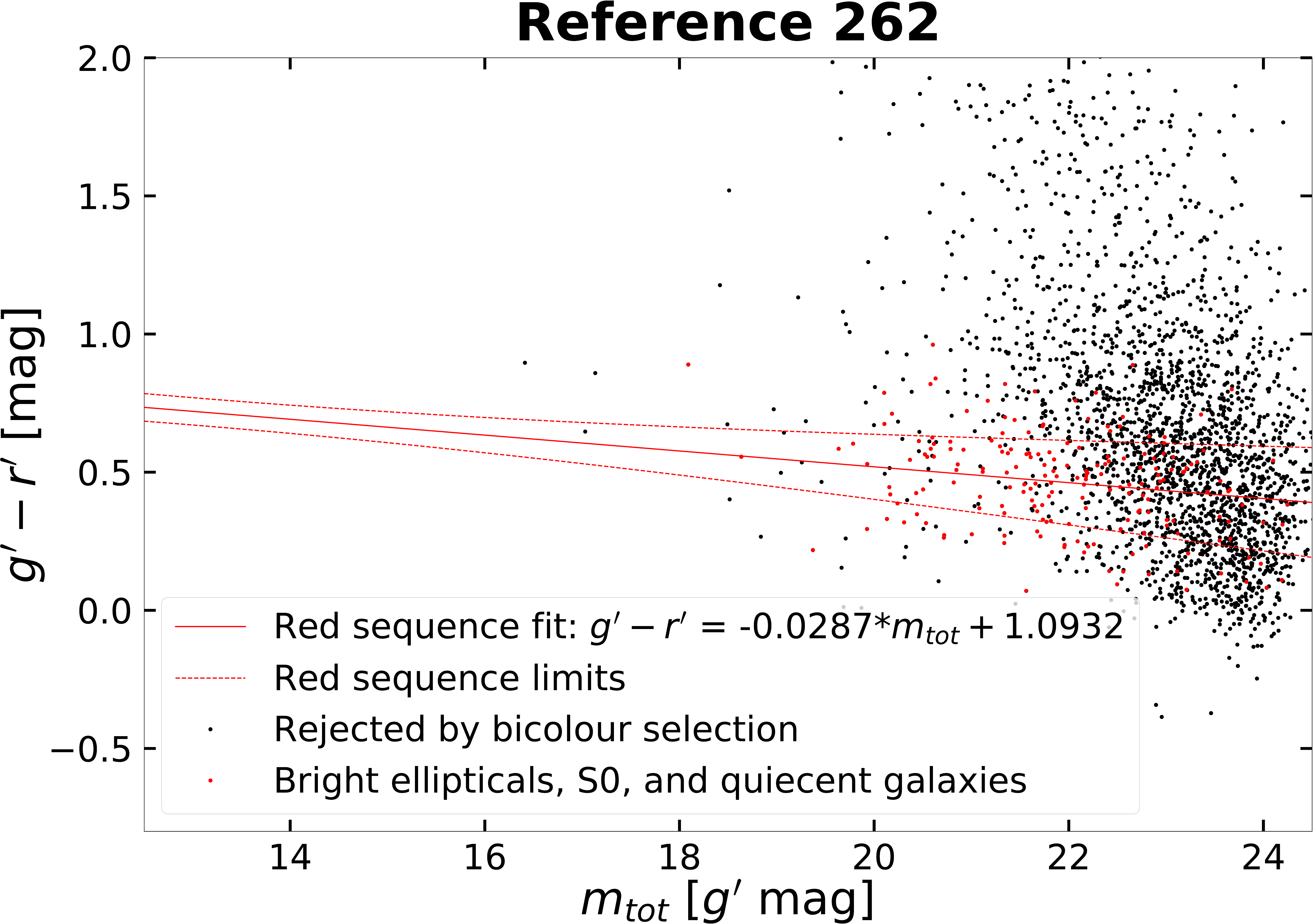}
    \caption{$g'-r'$ color--magnitude diagram of A1656 (top-left panel), A262 (top-right panel), and the reference field analyzed for the respective cluster (bottom). The best-fit red sequence is depicted as a solid red line, and the upper and lower red sequence selection limits are depicted as dashed red lines. For the reference field, the red sequence for the respective cluster is used. Those objects previously rejected in the bicolor selection are marked in black. Those galaxies classified as quiescent, as well as bright ellipticals and S0 are depicted as red points.}
    \label{fig:redsequence2}
\end{figure*}

The data points used for the red sequence fit (red dots), the best-fit red sequence model, and the final selection limits are shown in Figure \ref{fig:redsequence2} for A1656 (top left) and A262 (top right). Black dots indicate likely star-forming galaxies that were previously removed from our sample based on our bicolor sequence. For a detailed discussion about the rejection of star-forming galaxies using the reference field, see section \ref{sec:referencefield}.

\subsection{Reference Field and Further Catalog Cleaning} \label{sec:referencefield}

We have analyzed our reference field pointing twice using our pipeline, once for each cluster assuming their respective distances. The red sequence selection is done using the same selection cutoffs as for the respective galaxy cluster. 

The maximum depths of the A262 and A1656 stacks are comparable to the depth of the reference field. However, the depths of the images of both galaxy clusters decrease significantly toward the outer regions while the depth of the reference field is nearly constant. Therefore, the science and the reference fields are only approximately comparable. Additionally, the reference field is significantly smaller. Hence, we focus not on the absolute or relative numbers but on the parameter regions that objects populate in the reference field  (see section \ref{corellations}). We argue that the higher uncertainties of the colors in regions with lower depth lead to a similar amount of galaxies being scattered into and out of the quiescent sequence and red sequence, that is, no Eddington bias. The bicolor and red sequence plots for the reference pointing are shown in Figures \ref{fig:colorcolor} and \ref{fig:redsequence2} (bottom). On the other hand, we reject more galaxies in the low-S/N regions due to our strict quality cuts of $\Delta(u'-g')$, $\Delta(g'-r')$, $\Delta m_{\mathrm{tot}}$, $\mathrm{\Delta \mu_0}$, $\mathrm{\Delta \mu_e}$, and $\Delta R_e/R_e$. 
Hence, we consider the findings in the reference field as an upper limit except for the absolute number of galaxies, due to its smaller size. 

In the reference field, we do not find a single UDG, neither when analyzed for A262 nor for A1656. There are 111 remaining galaxies found in the reference pointing for A262 and 135 for A1656. Hence, we conclude that contamination by background galaxies mainly affects the more compact dwarfs. In Section \ref{corellations}, we discuss which regions the galaxies found in the reference field pointing populate in multiple parameter spaces.

Furthermore, we use the reference field pointing to test our efficiency in removing background objects using the bicolor sequence and the red sequence. These tests are performed on our catalogs without the manual removal of bad fits, remasking, and nucleus fits of the quiescent galaxies. These steps are only performed for the galaxies selected as quiescent. Otherwise, this would bias our estimate for the efficiency of our background object removal procedure. Here, we still apply our automatic quality cuts. We find that by applying both the bicolor and the red sequence selection, we remove 90\% of the galaxies in the reference pointing when analyzed for A262 and 89\% when analyzed for A1656.
Furthermore, we test the importance of preselecting quiescent galaxies using the bicolor sequence. It removes 71\% more galaxies from the reference field sample analyzed for A262 and 76\% when analyzed for A1656 than if the cluster member selection would have only been done using the red sequence. This demonstrates how crucial the deep $u'$-band data is to properly select cluster members, as it allows us to improve the purity of our sample by about 70\% compared to only using $g'$- and $r'$-band data for the red sequence selection.  
Note, that in Figure \ref{fig:redsequence2} only the $\Delta(u'-g')$, $\Delta(g'-r')$ quality cuts were applied. The other quality cuts are applied afterward. By this quality filter, faint galaxies are more affected than bright ones and for those faint galaxies, the fraction of non-star-forming galaxies is higher, as shown in Figure \ref{fig:redsequence2}.

\begin{figure*}
    \includegraphics[width=0.98\textwidth]{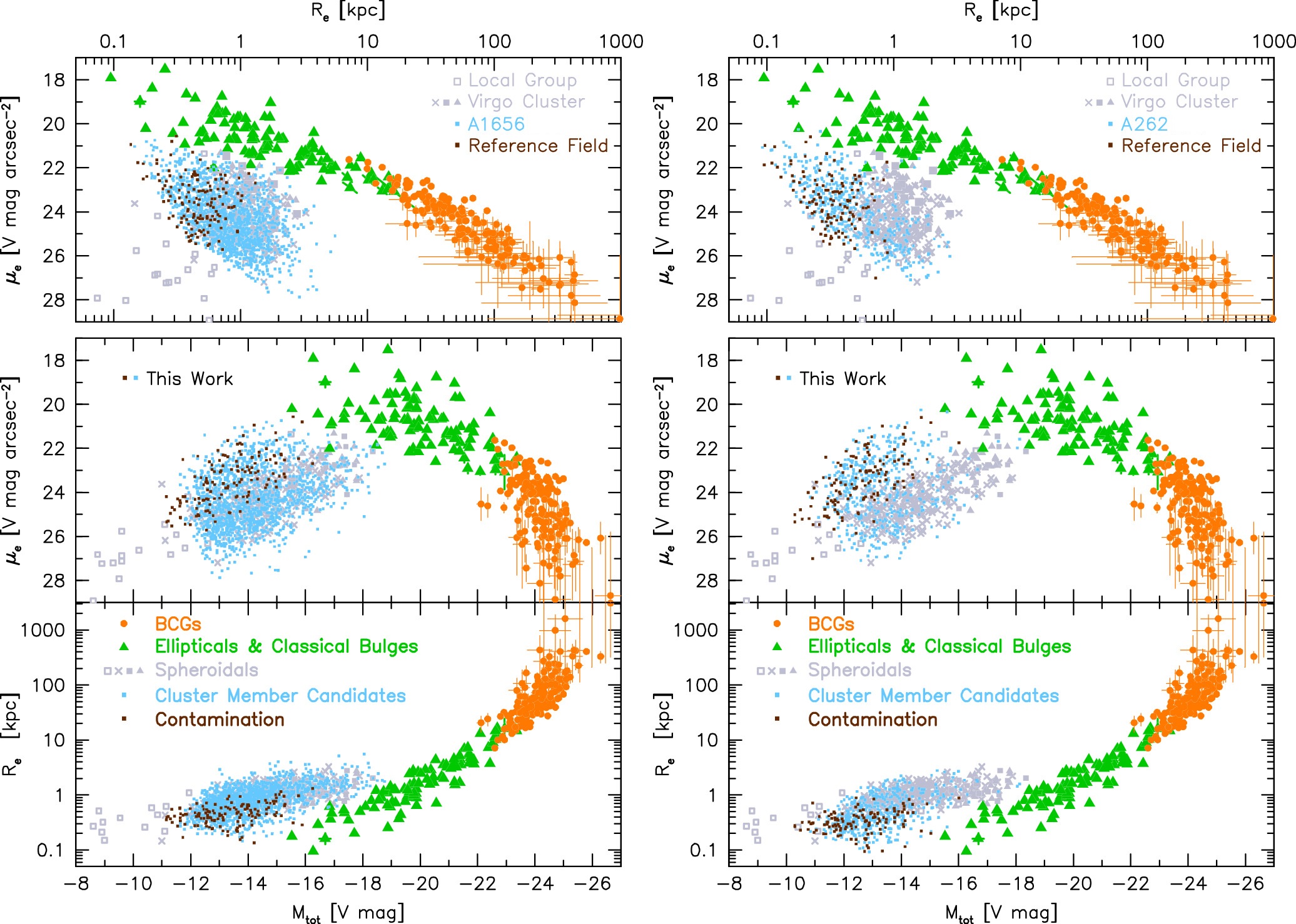}
    \caption{Comparison between $M_{\mathrm{tot}}$, $R_e$, and $\mu_e$ of dwarf cluster member candidates (light blue) from our A1656 (right) and A262 (right) sample, as well as galaxies from the reference field analyzed for the respective cluster (brown). The basis for this plot is Figure 37 in \citet{Kormendy2009} with updates in Figure 2 in \citet{KormendyBender2012}, Figure 14 in \citet{Bender2015}, and Figure 16 in \citet{kluge} including BCGs (orange), ellipticals (green), classical bulges (green), and spheroidals (gray).}
    \label{fig:simplefinal}
\end{figure*}

\begin{figure*}
    \centering 
    \includegraphics[width=0.78\textwidth]{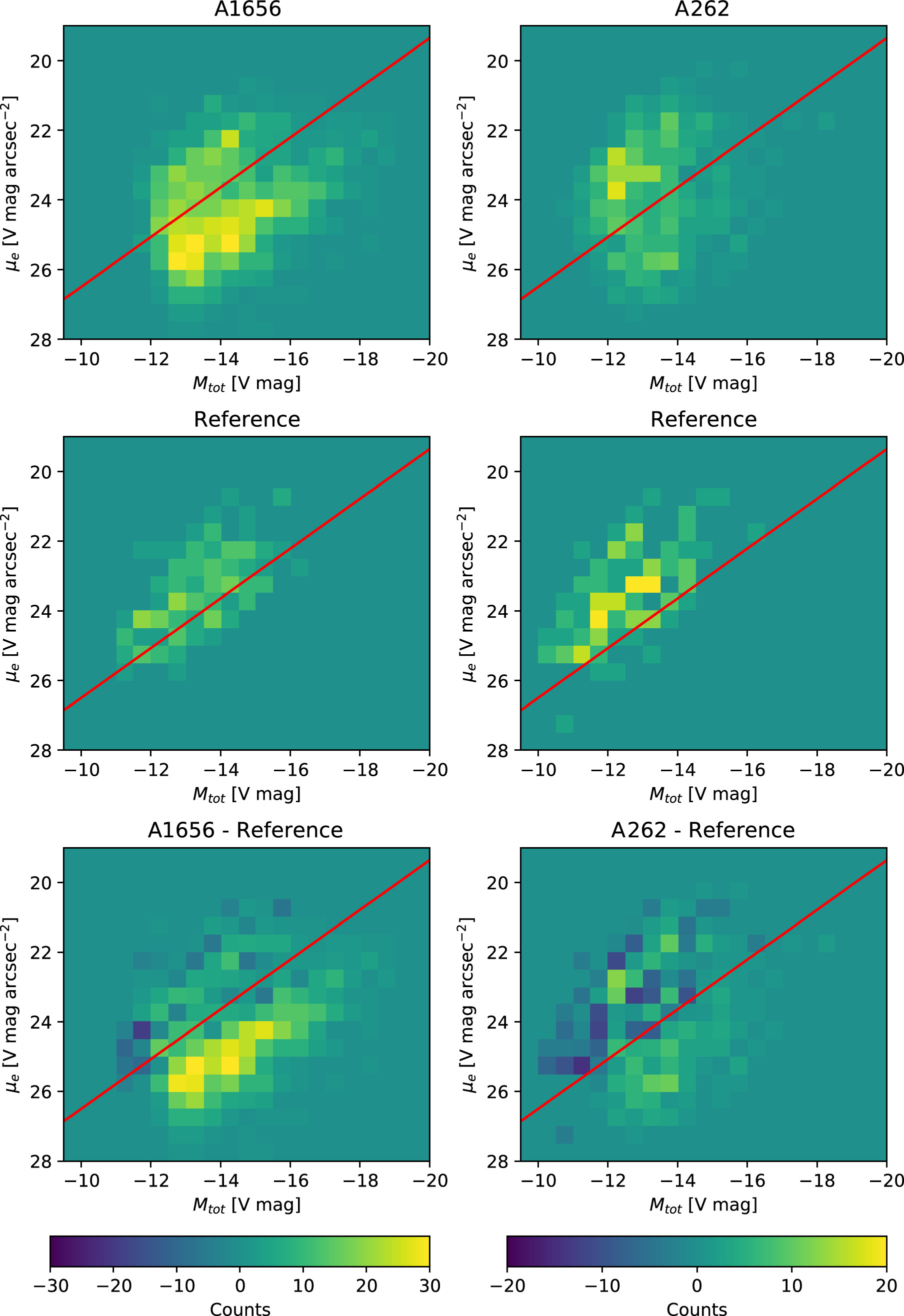}
    \caption{Number density of our A1656 (top left) and A262 (top right) dwarf cluster member candidates in the $M_{\mathrm{tot}}-\mu_e$ parameter space. The middle panel shows the number density of the galaxies found in the reference field when analyzed for the respective cluster and scaled to match the size of the cluster images. The bottom panels show the number density in the $M_{\mathrm{tot}}-\mu_e$ parameter space of the dwarf cluster member candidates after subtracting the number density from the reference field. The red line indicates the chosen cutoff in between the two sequences.}
    \label{fig:contaminationtest}
\end{figure*}

In Figure \ref{fig:simplefinal} we plot the $M_{\mathrm{tot}}-R_e$, $M_{\mathrm{tot}}-\mu_e$, and $R_e-\mu_e$ scaling relations of our dwarf cluster member candidates (light blue) in A1656 (left) and A262 (right), as well as of the galaxies found in the reference field when analyzed for the respective cluster (brown). The basis for these plots is Figure 37 in \citet{Kormendy2009} with updates in Figure 2 in \citet{KormendyBender2012}, Figure 14 in \citet{Bender2015}, and Figure 16 in \citet{kluge} including BCGs (orange), ellipticals (green), classical bulges (green), and spheroidals (gray). The galaxies found in the reference field are compact (bright $\mu_e$ and small $R_e$), forming a cloud that is significantly separated from the region in those parameter spaces where UDGs are expected to be. In the $M_{\mathrm{tot}}-\mu_e$ parameter space, we recognize that most of the galaxies from the reference field representing the contamination of our cluster member sample lie above the spheroidal sequence from \citet{Kormendy2009}. Indeed, plotting the number density of the dwarf cluster member candidates in the $M_{\mathrm{tot}}-\mu_e$ parameter space in Figure \ref{fig:contaminationtest} (top) for A1656 (left) and A262 (right), we find two sequences. Note here that we use a different scaling for both clusters due to the significantly different richness. For the reference field (middle panel), we do only find the upper sequence. Here, the number density is scaled with the nonmasked area to match approximately the number density expected for the contamination in the cluster sample. However, the scaling is just a rough proxy. The scaled reference field number density gives an upper limit for the contamination due to the higher depth and assumed higher completeness of detected objects due to less overlap with other galaxies in the dense cluster environment.
Subtracting the scaled number density of the reference field in the $M_{\mathrm{tot}}-\mu_e$ parameter space from the number density of the galaxies found in the galaxy clusters basically eliminates the upper sequence (Figure \ref{fig:contaminationtest} bottom). For A262, the subtraction actually leads to a negative number density where the upper sequence was due to the subtracted number density being an upper limit of the contamination and the generally low richness of A262. 

This indicates that the upper sequence visible for both clusters is actually dominated by interloping galaxies. Hence, we remove the galaxies from the upper sequence from our sample. For this, we set a cutoff in between the two sequences of A1656 (indicated by the red line in Figure \ref{fig:contaminationtest}). For A262, we use the same cutoff despite the smaller distance modulus of A262 due to its lower richness and hence stronger relative contamination. This final selection cutoff removes 91.2\% of the galaxies in the reference field pointing analyzed for A262 and 74.4\% when analyzed for A1656. For A262, this cutoff removes 297 of 472 galaxies, and for A1656, removes 406 of 1305 galaxies. We want to stress that this cutoff does not imply that there are no galaxies in the respective clusters above this threshold but just that the sample is strongly contaminated by interloping background galaxies there. Furthermore, we want to make clear that we still expect some contamination for the compact galaxies of our final sample. Using the fraction of galaxies rejected by the $M_{\mathrm{tot}}-\mu_e$ cutoff in the reference field and the number of rejected galaxies in the cluster member sample (conservatively assuming that they are all interloping galaxies), we estimate a conservative upper limit for the contamination of our final cluster member sample of 15.6\% for both clusters.



\section{Results and Discussion} \label{sec:resultsdiscussion}
Overall, we are left with 185 dwarf galaxy cluster members in A262 and 899 in A1656. Of those, we find 11 galaxies fulfilling the \citet{vanDokkum2015} UDG definition in A262 and 48 such UDGs in A1656, compared to eight UDGs in A1656 which \citet{vanDokkum2015} found within our common field of view.

\subsection{Comparison to Literature} \label{literature}

\begin{figure*}[ht]
    \centering
    \includegraphics[width=0.4\textwidth]{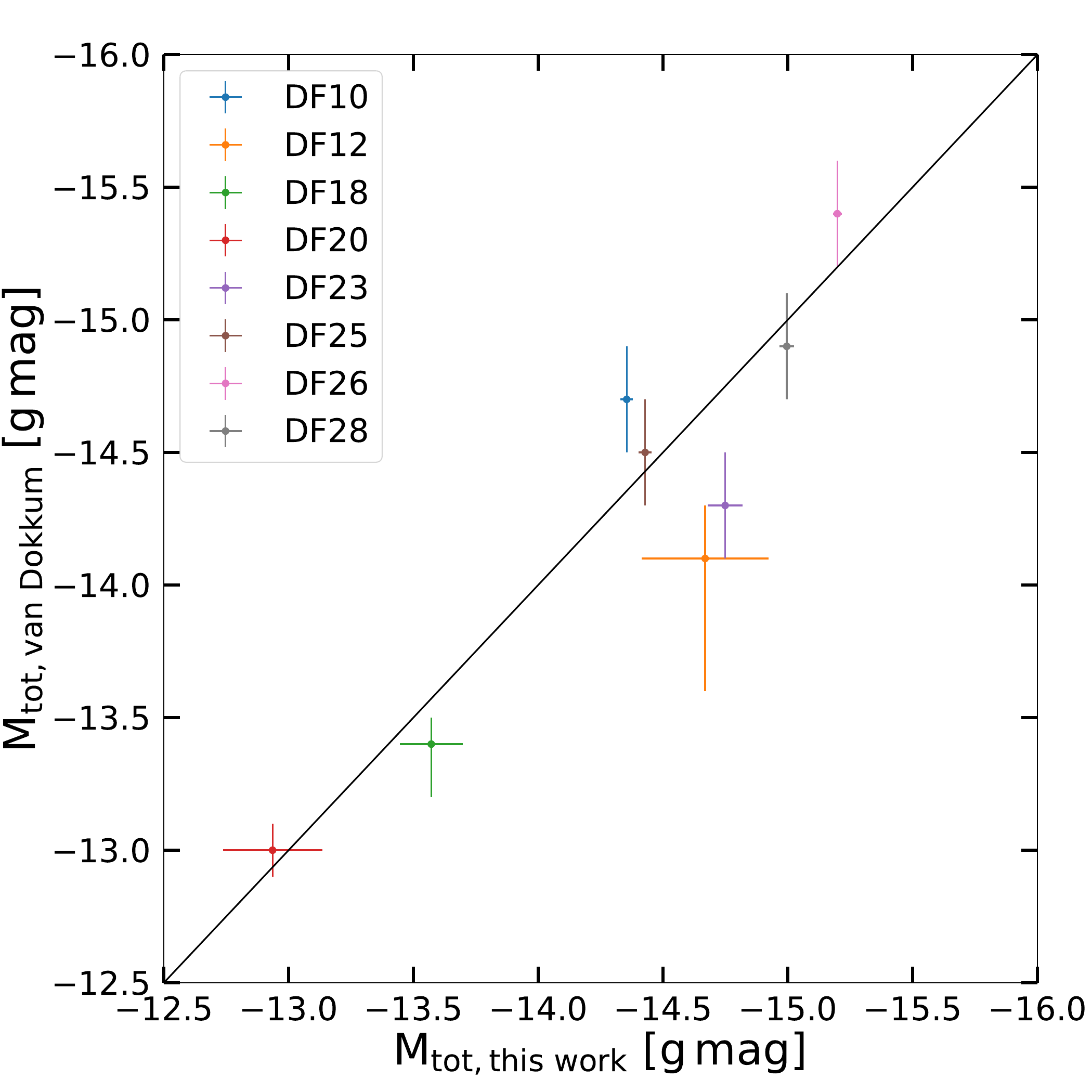}
    \includegraphics[width=0.4\textwidth]{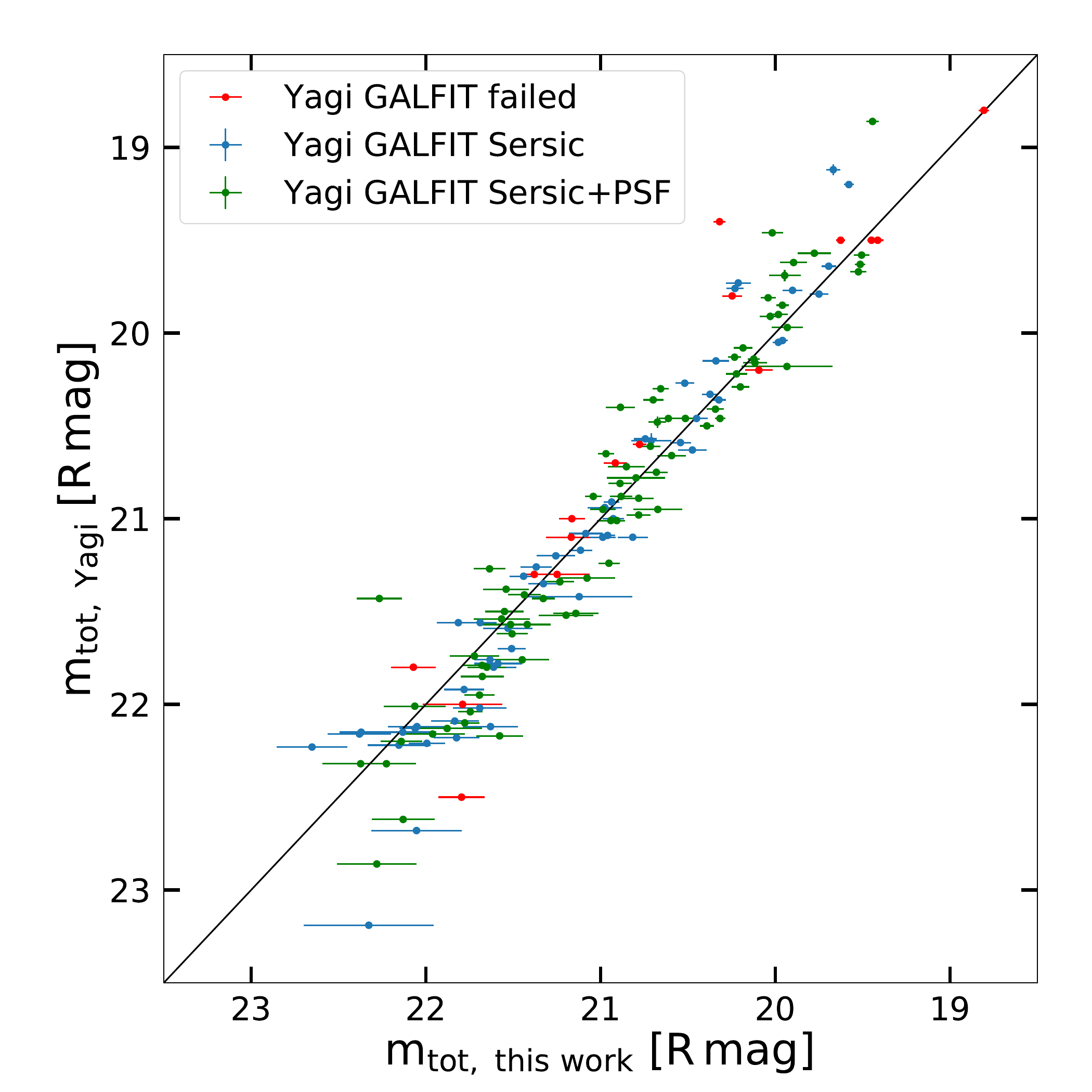}
    \includegraphics[width=0.4\textwidth]{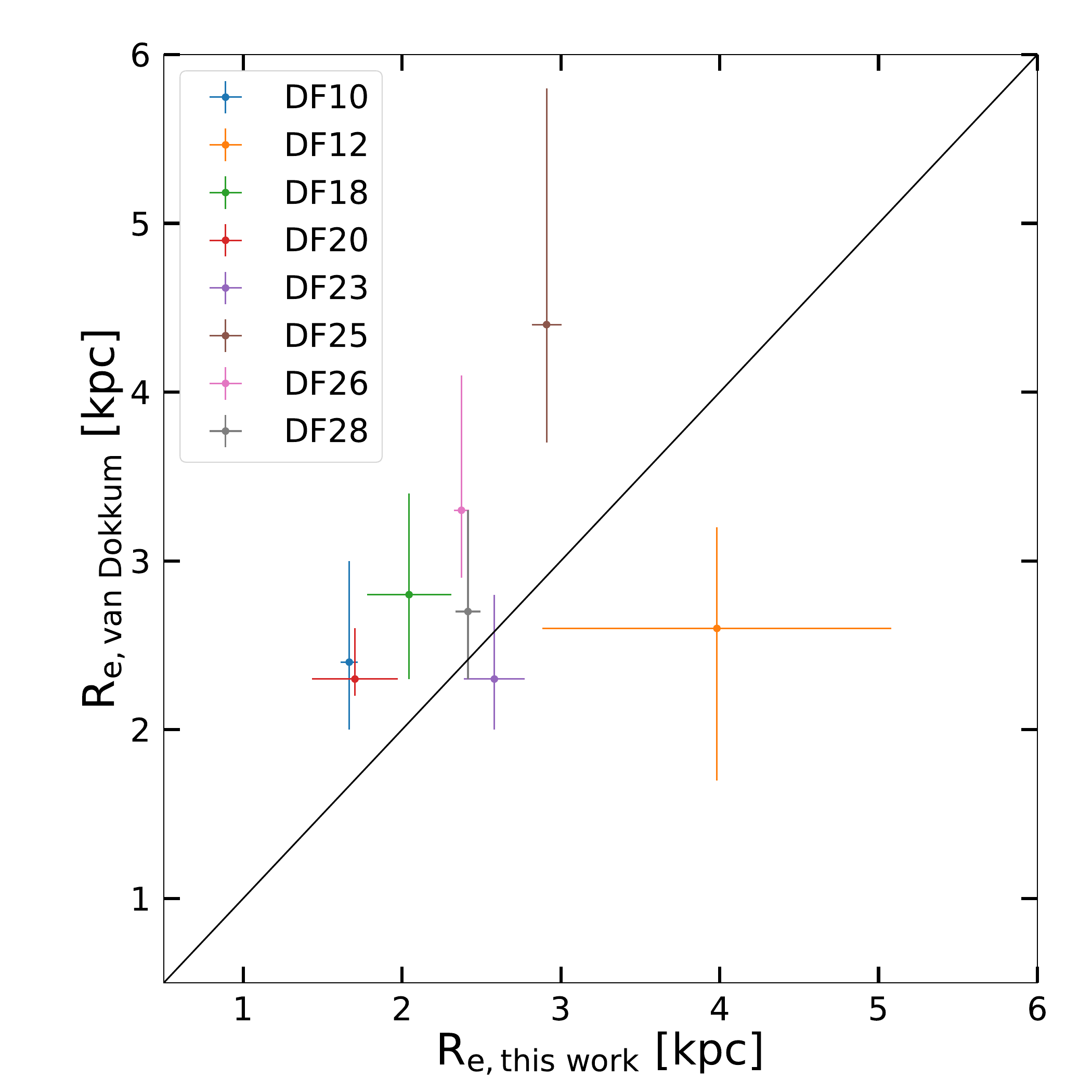}
    \includegraphics[width=0.4\textwidth]{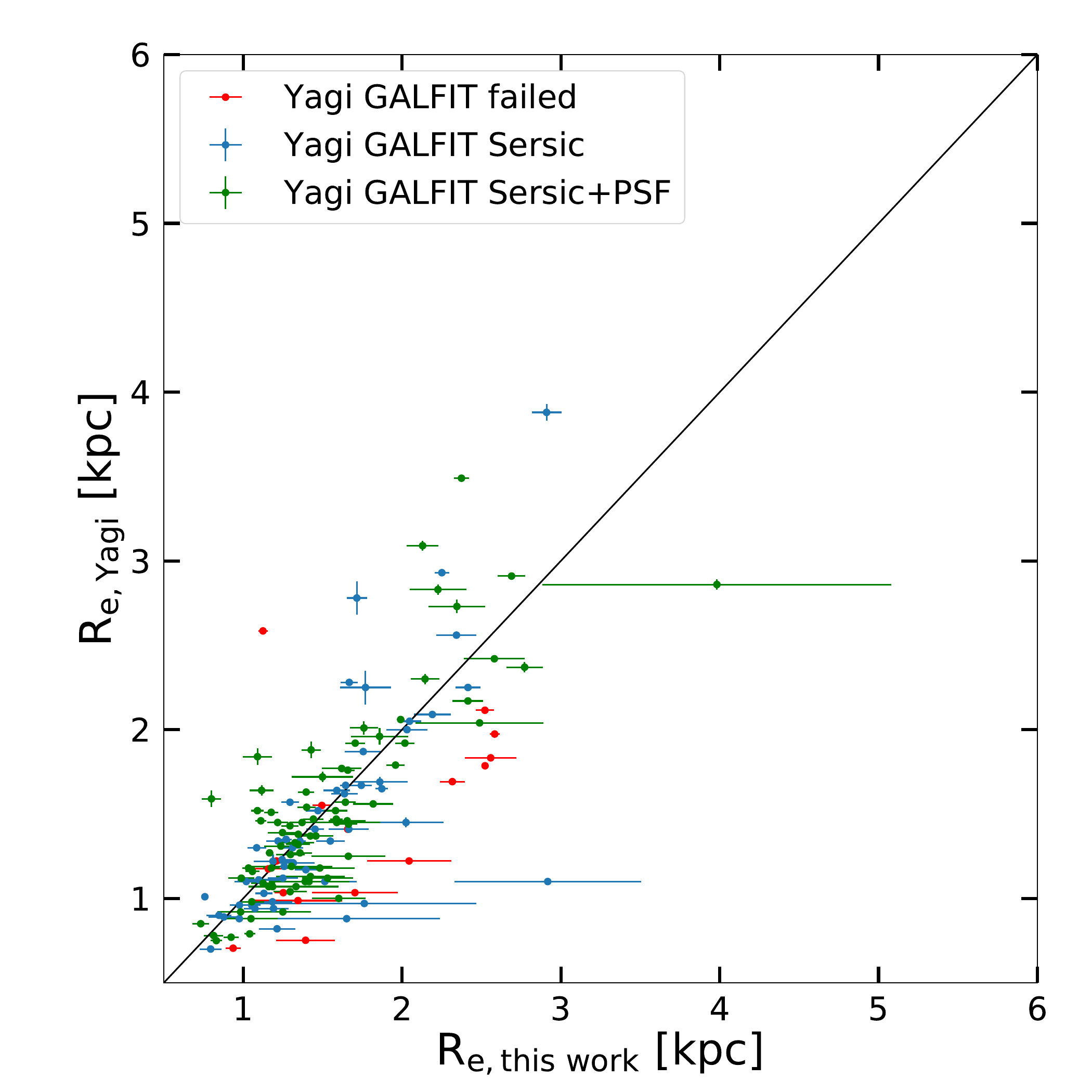}
    \includegraphics[width=0.4\textwidth]{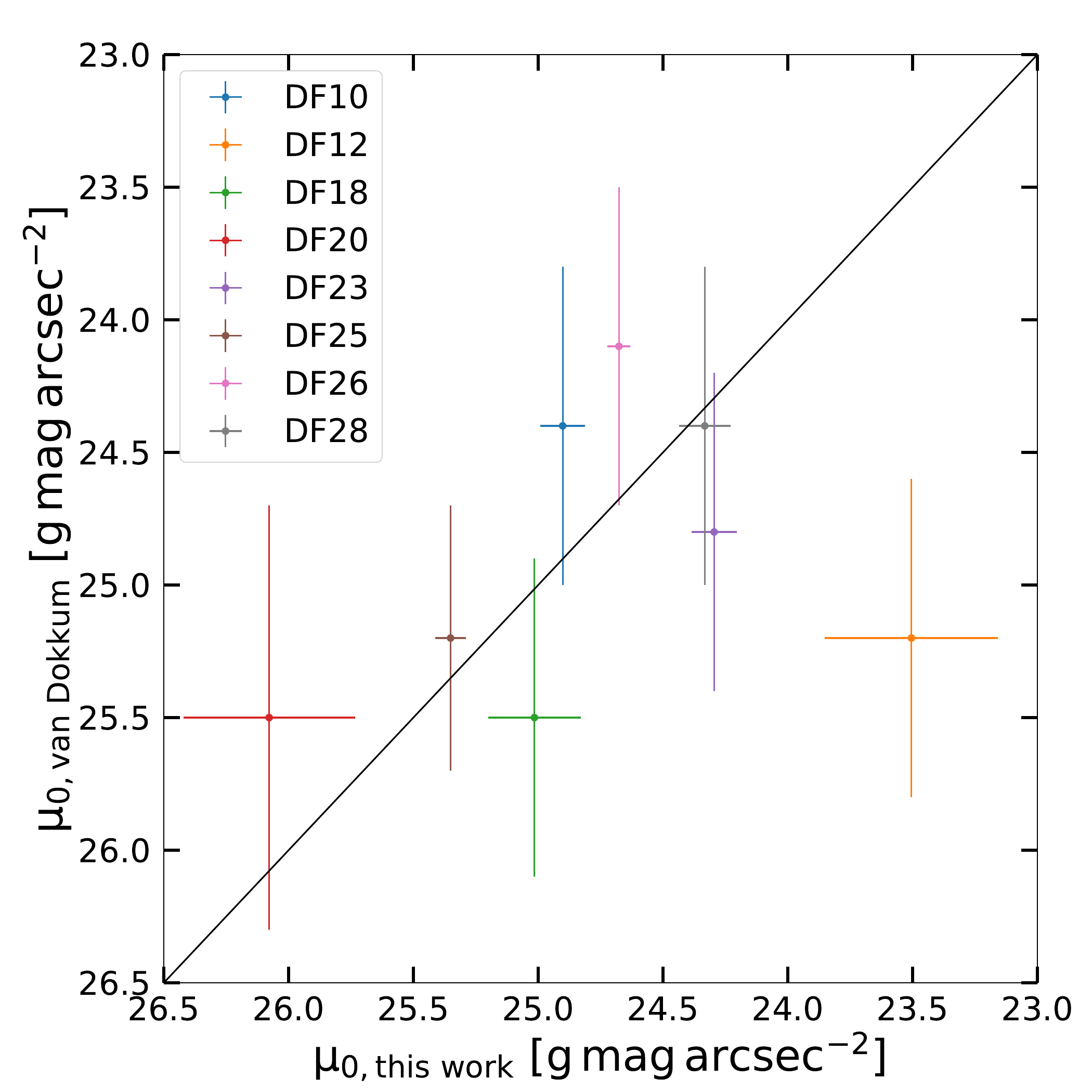}
    \includegraphics[width=0.4\textwidth]{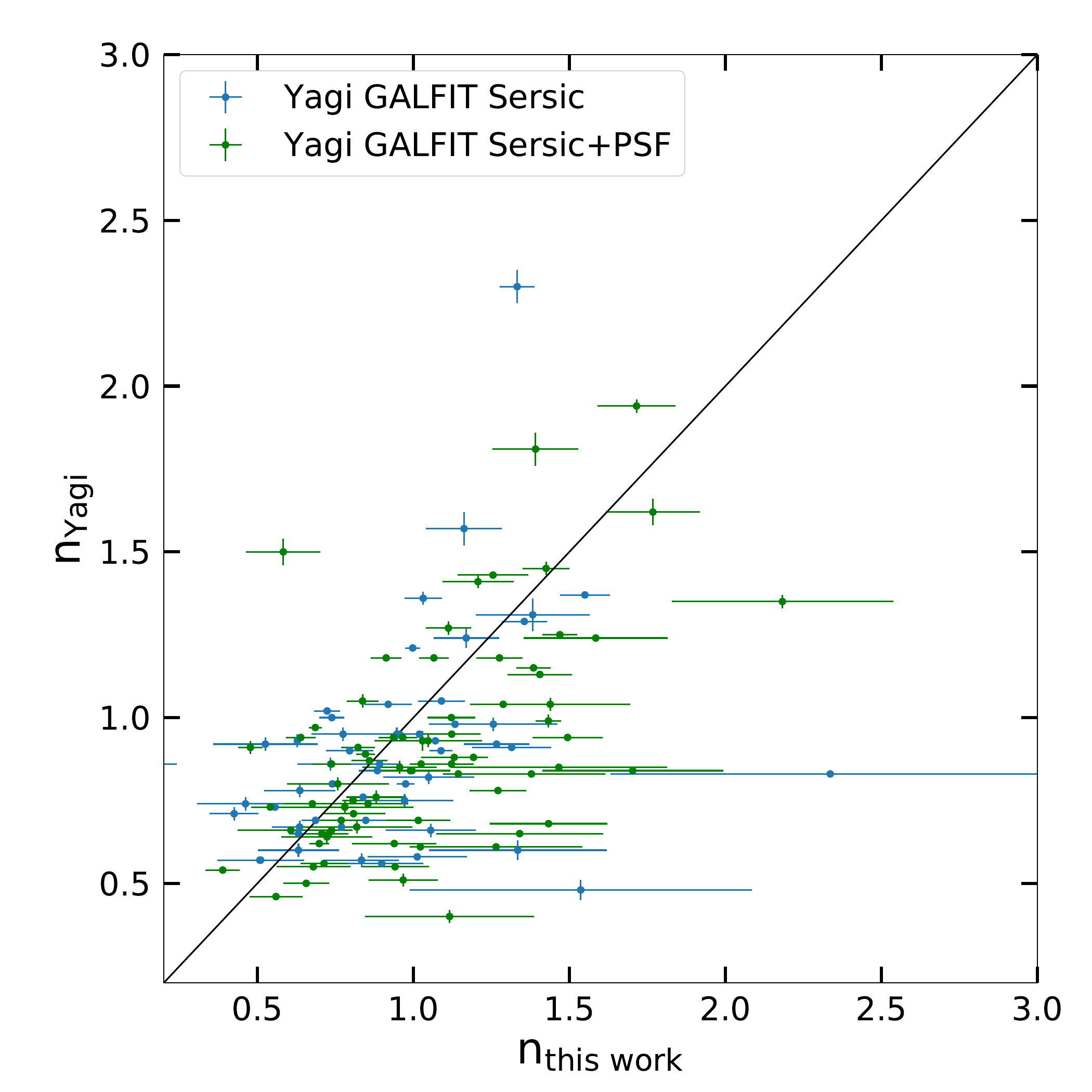}
    \caption{Comparison between our measured structural parameters and those obtained by \citet[][left]{vanDokkum2015} and \citet[][right]{Yagi2016}.}
    \label{fig:vDokkumcomp}
\end{figure*}

\begin{figure*}[ht]
    \centering
    \includegraphics[width=\textwidth]{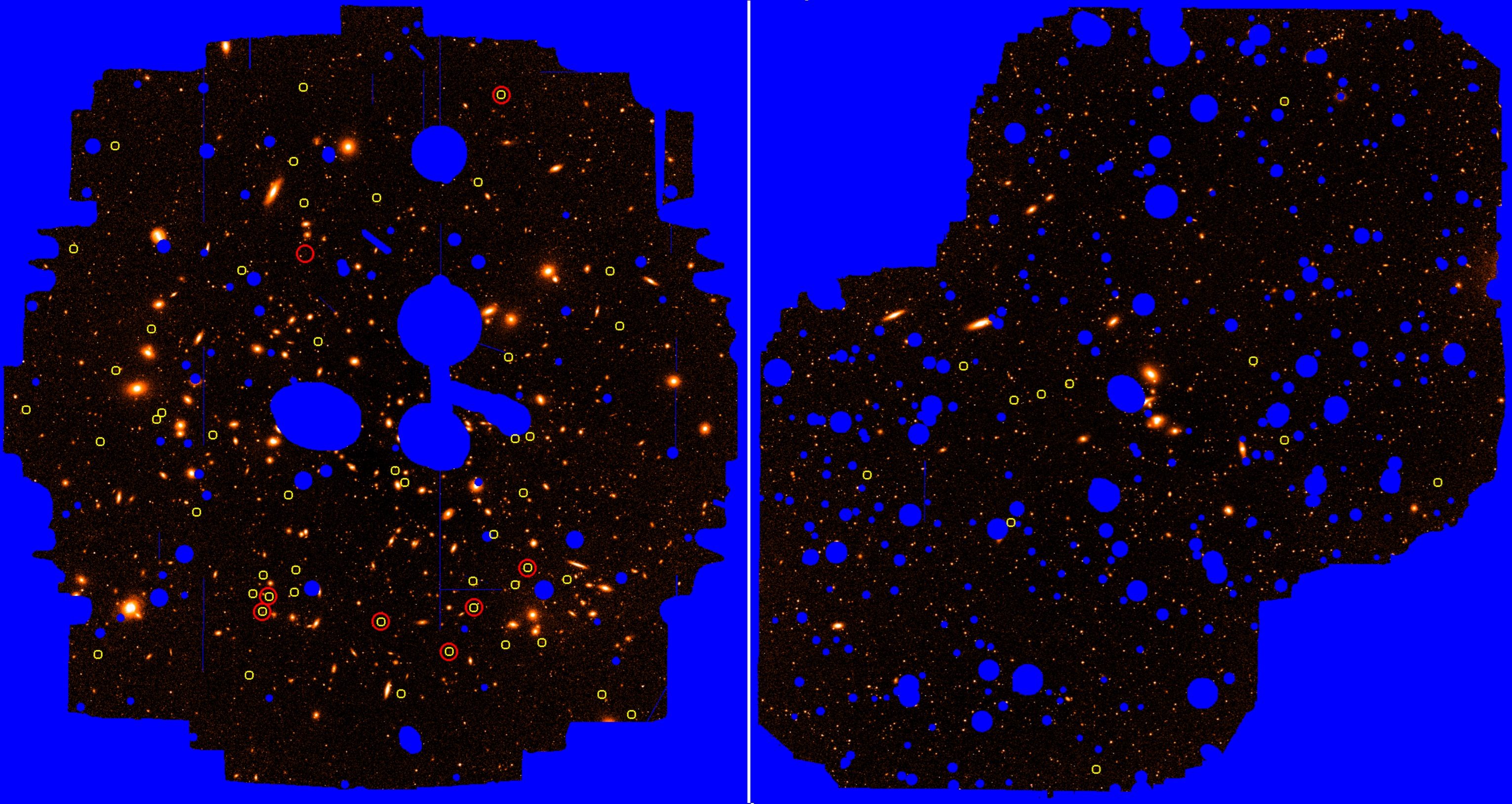}
    \caption{Overview of our A1656 (left) and A262 (right) UDG samples (yellow circles) and DF UDGs (red circles).}
    \label{fig:UDGoverview}
\end{figure*}

We detected all eight DF UDGs (DF10, DF12, DF18, DF20, DF23, DF25, DF26, and DF28) found by \citet{vanDokkum2015} in the region covered by our survey and successfully measured their structural parameters. In Figure \ref{fig:vDokkumcomp} (left), we compare our measured structural parameters for those eight UDGs with the results of \citet{vanDokkum2015}. 
Here, our $g'$-band measurements are converted to $g$ magnitudes by $g-g'=0.09$ \citep{Willmer2018}. For DF12, we find $\mu_{0}=23.48\pm0.34\,g'\,\mathrm{mag\,arcsec^{-2}}$ which is not fulfilling the UDG definition. Furthermore, we detected DF20 as two objects. 
However, our measured structural parameters for DF20 still agree reasonably well with those measured by \citet{vanDokkum2015}. Generally, our measured $M_{\mathrm{tot}}$ and $\mu_0$ agree well with the measurements by \citet{vanDokkum2015}, but we tend to find smaller $R_e$. This might be due to us varying $n$, whereas \citet{vanDokkum2015} fixed $n=1$. In addition to those eight, we find further 42 UDGs that were not detected by \citet{vanDokkum2015}. In total, we end up with 48 UDGs (excluding DF12), which is six times more than \citet{vanDokkum2015} found in our common region. Furthermore, \citet{vanDokkum2015} noted that they did not find UDGs close to the cluster core due to crowding and ICL. We visualize the spatial distribution of our UDG sample in Figure \ref{fig:UDGoverview} with yellow circles for A1656 (left) and A262 (right). As shown here, we do find UDGs significantly closer to the cluster center. The DF UDGs are depicted as red circles. 
We trace this back to our BCG and ICL subtraction, as well as our higher resolution.

By comparing our results to \citet{Yagi2016}, we find in our final sample 145 galaxies that they classified as UDGs. Here, we consider every galaxy with a counterpart in the catalog of \citet{Yagi2016}, which is separated by  $<1\arcsec$. 

The sheer difference in numbers between those 145 galaxies and the 48 UDGs in our sample fulfilling the original UDG definition by \citet{vanDokkum2015} indicates that the majority of the UDGs found by \citet{Yagi2016} are only due to them using their less-strict UDG definition. Actually, out of these 145 UDGs found by \citet{Yagi2016}, only 41 fulfill the UDG definition by \citet{vanDokkum2015} using our measured parameters. Eight of our UDGs were not found by \citet{Yagi2016}. We  note here that \citet{Yagi2016} include all galaxies in their sample that are in their \verb+SExtrator+ catalog even if their \verb+GALFIT+ fits did not converge. In our sample, we are much more restrictive by requiring that a galaxy has to be fitted well by \verb+GALFIT+ (not only converged but also relatively small uncertainties of all structural parameters and accepted in the eyeballing procedure), has a small color uncertainty ($\mathrm{<0.2\,mag}$), has to be on the quiescent sequence in the color--color diagram, and has to be on the red sequence. Furthermore, we note here that for 16 out of these 145 galaxies, the \verb+GALFIT+ fits of \citet{Yagi2016} did not converge, whereas ours did. This might be due to our more elaborate masking procedure.

In Figure \ref{fig:vDokkumcomp} (right), we compare our measured structural parameters to those of \citet{Yagi2016}. Here, we compare our best fits to the best model parameters of \citet{Yagi2016}, obtained from either a single Sérsic or single-Sérsic+PSF \verb+GALFIT+ model. If their \verb+GALFIT+ fits failed, then we use their single-Sérsic fits obtained with \verb+SExtractor+. To convert our $g'$-band magnitudes to the $R$ band, we use $r\approx R+0.09$ \citep{Yamanoi2012} and $r\approx r'$ \citep{Willmer2018}, as well as our measured $g'-r'$ colors. Our measured $m_{\mathrm{tot}}$ and $n$, as well as this time also $R_e$ agree well with the findings of \citet{Yagi2016}. Worth noting is that for the galaxies for which the \verb+GALFIT+ fits of \citet{Yagi2016} failed, we tend to find larger $R_e$ than their \verb+SExtractor+ measurements. Also note that we cannot compare the Sérsic indices for those galaxies for which the \verb+GALFIT+ fits of \citet{Yagi2016} failed, as they were not published. 

\citet{Yagi2016} found nuclei in 50\% of their UDGs. For our UDGs in A1656 we only found 4\% to host a nucleus. This cannot be due to the different UDG definition, as we found a nucleus only for 0.7\% of those galaxies for which we have counterparts in the \citet{Yagi2016} sample. We attribute this difference in the fraction of nucleated UDGs to the higher resolution ($\mathrm{FWHM=0.\arcsec7}$) of the data used by \citet{Koda} and \citet{Yagi2016}. In A262, where the apparent size of the UDGs is larger and, hence, the separation between the nucleus and stellar body is easier, we find 2 out of 11 UDGs to host a nucleus.

\subsection{Morphology from Hubble Space Telescope Data}
\label{sec:HSTmorph}
As a check for the purity of our dwarf galaxy sample, we inspect their morphology. Massive galaxy clusters should mainly contain elliptical galaxies, S0's, and spheroidals, and nearly no spiral galaxies or irregular galaxies, whereas the population of background field galaxies should mainly consist of the last two galaxy types \citep{Dressler1980}. In order to check the morphology of the dwarf galaxies from our final sample, we require a higher resolution than our ground-based WWFI data. For this, we use archival reduced Hubble Space Telescope (HST) data obtained from the Mikulski Archive for Space Telescopes. For A262, there is only data for the BCG available, and hence, we do not consider A262 in this discussion. We aligned the background of the F475W images to our A1656 $g'$ band stack using constant offsets. For aligning the background of the F814W images, we used $i$-band data from the Legacy Survey DR10 taken for the Dark Energy Camera Legacy Survey \citep{dey+2019}. The images were resampled and stacked using SWarp \citep{swarp}.

All galaxies from our sample before applying the $M_{\mathrm{tot}}-\mu_e$ selection cutoff that are covered by those archival HST data are visually inspected in color images. We report that only 14 galaxies out of 421 galaxies covered by the HST data show a spiral morphology (none of them was previously identified as a UDG). These 14 galaxies are shown in Appendix \ref{app:SpiralsHST}. Six of these 14 galaxies were rejected by the $M_{\mathrm{tot}}-\mu_e$ selection cutoff, and eight remain in our final sample. Of these 421 galaxies covered by HST data, 90 are rejected by the  $M_{\mathrm{tot}}-\mu_e$ selection cutoff and 331 remain in our final sample. This results in a fraction of galaxies with a spiral morphology of 6.7\% for the galaxies removed by the $M_{\mathrm{tot}}-\mu_e$ cutoff and only 2.4\% for our final sample. We conclude that from a morphological view, our galaxy sample is consistent with the expected morphology of typical galaxies in massive galaxy clusters. Also note that these spiral galaxies are not necessarily background galaxies, as a few spiral galaxies can still exist even in massive galaxy clusters.

\begin{figure*}[ht]
    \centering
    \includegraphics[width=0.43\textwidth]{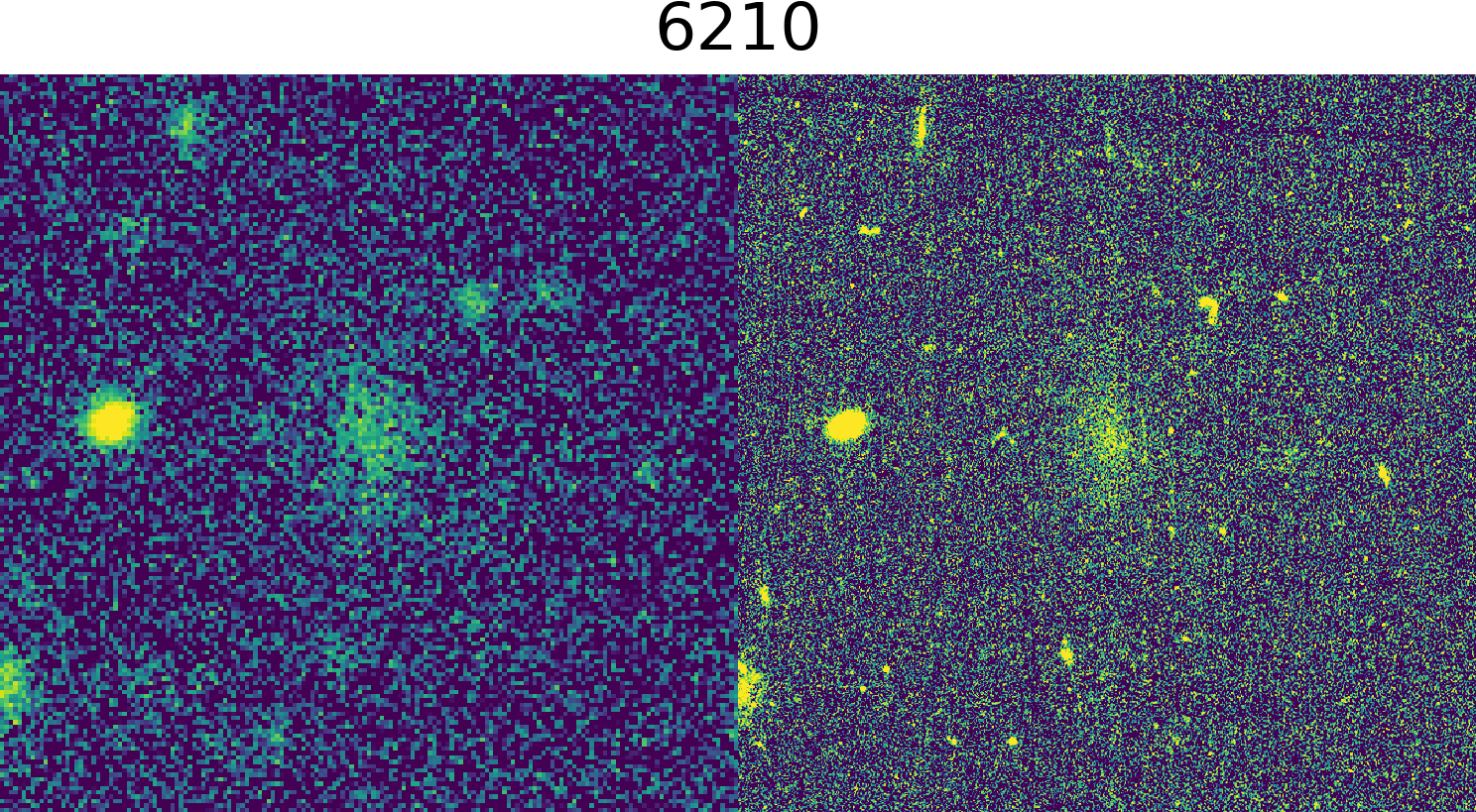}
    \includegraphics[width=0.43\textwidth]{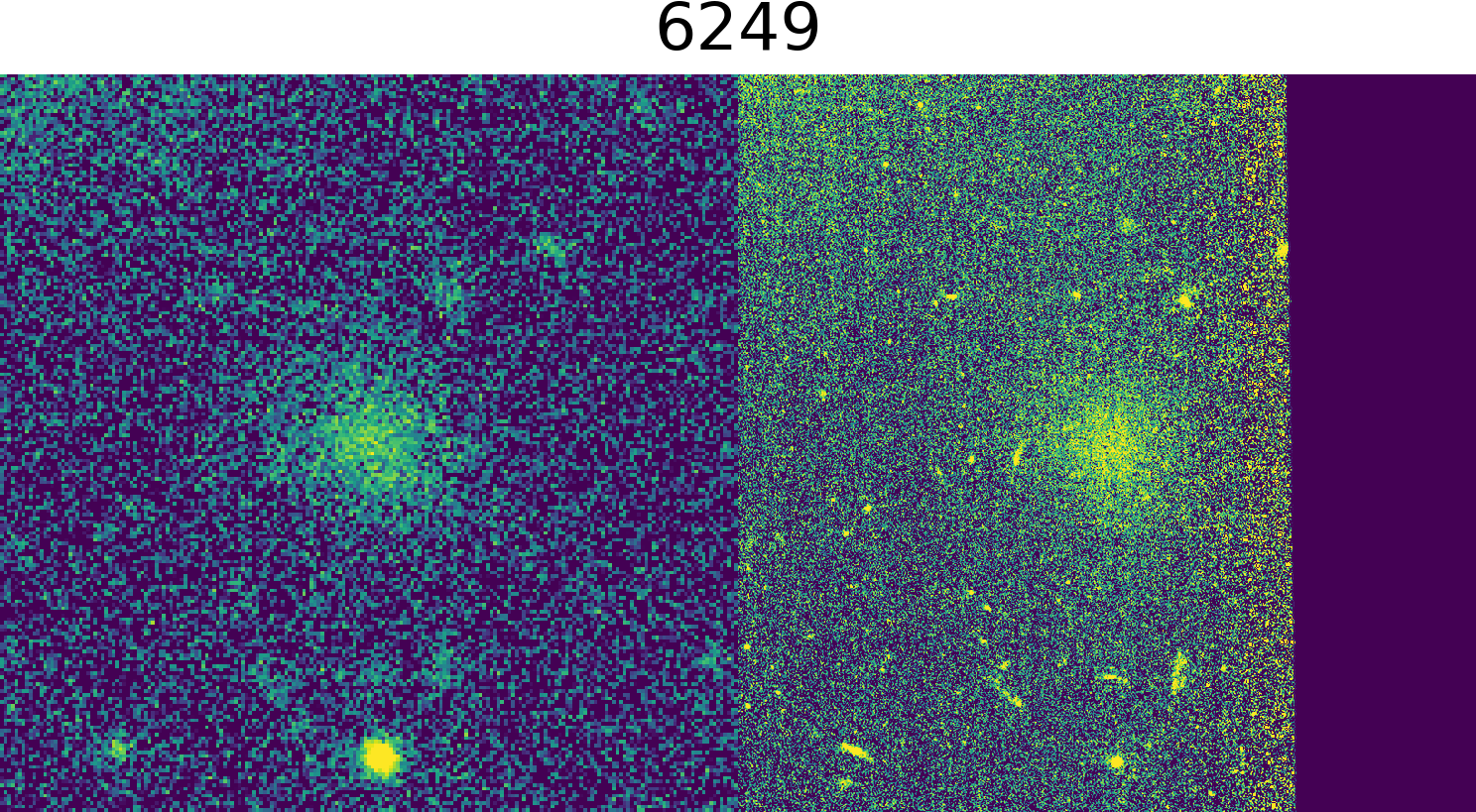}
    \includegraphics[width=0.43\textwidth,trim=0cm 0cm 0cm -0.3cm]{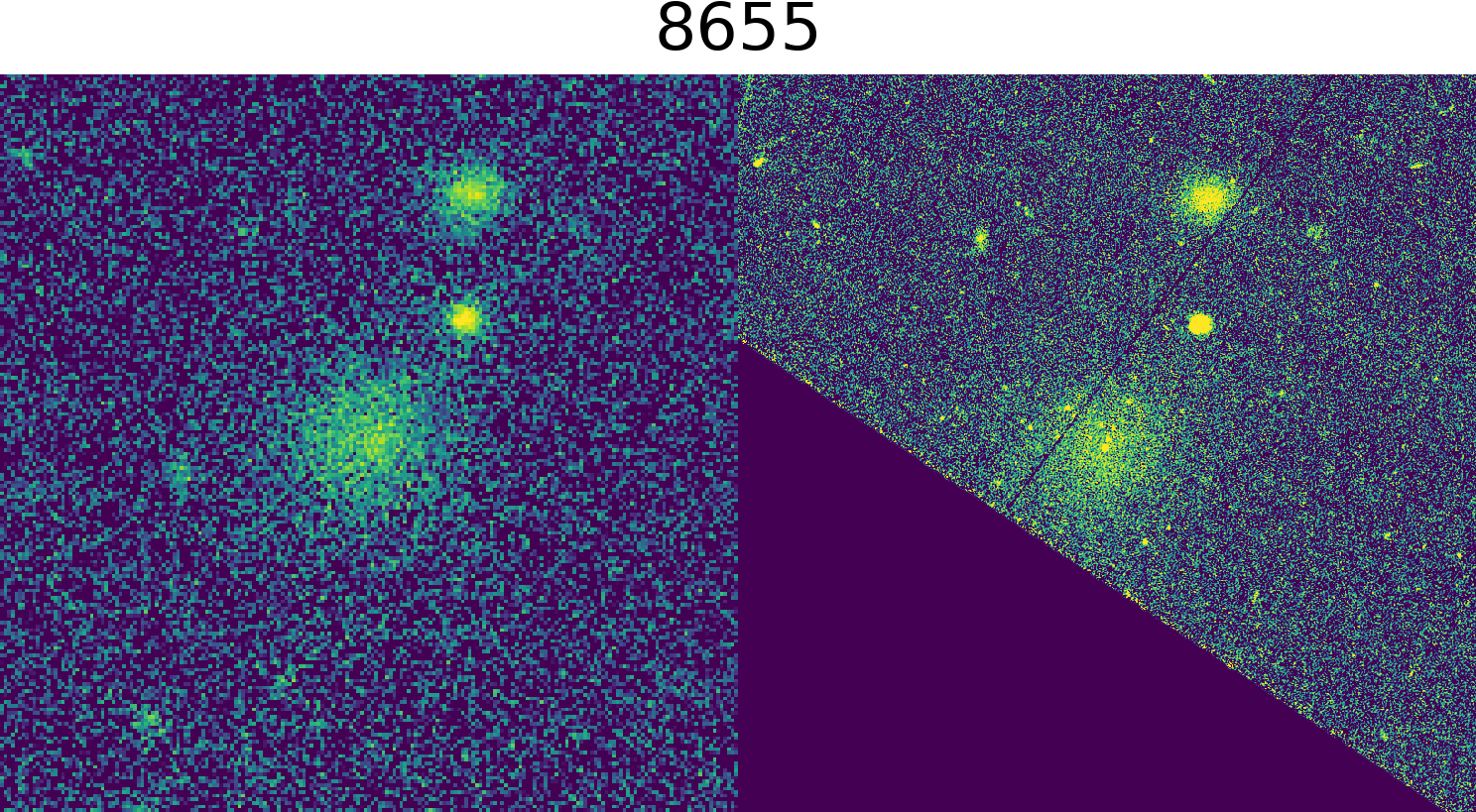}
    \includegraphics[width=0.43\textwidth,trim=0cm 0cm 0cm -0.3cm]{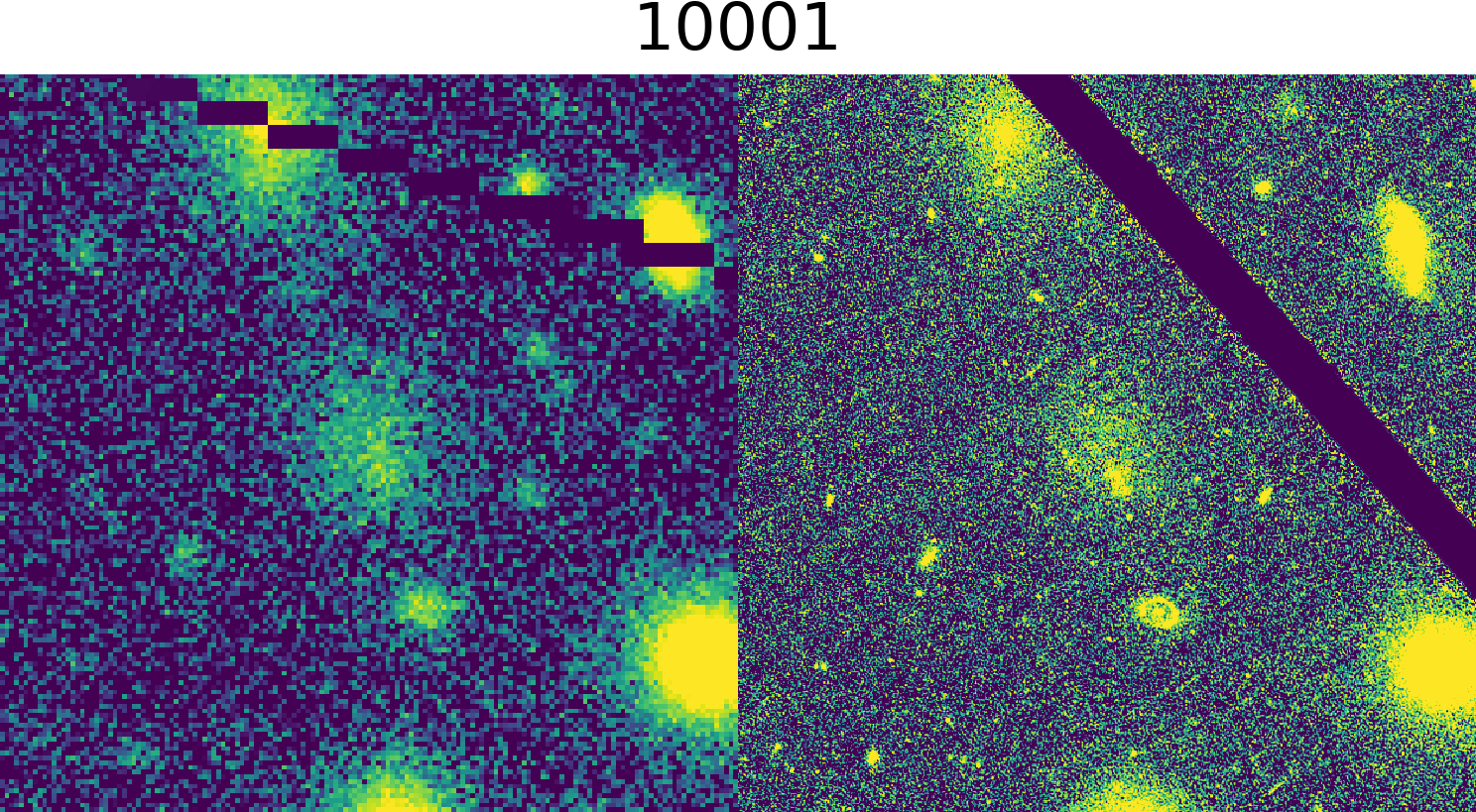}
    \includegraphics[width=0.43\textwidth,trim=0cm 0cm 0cm -0.3cm]{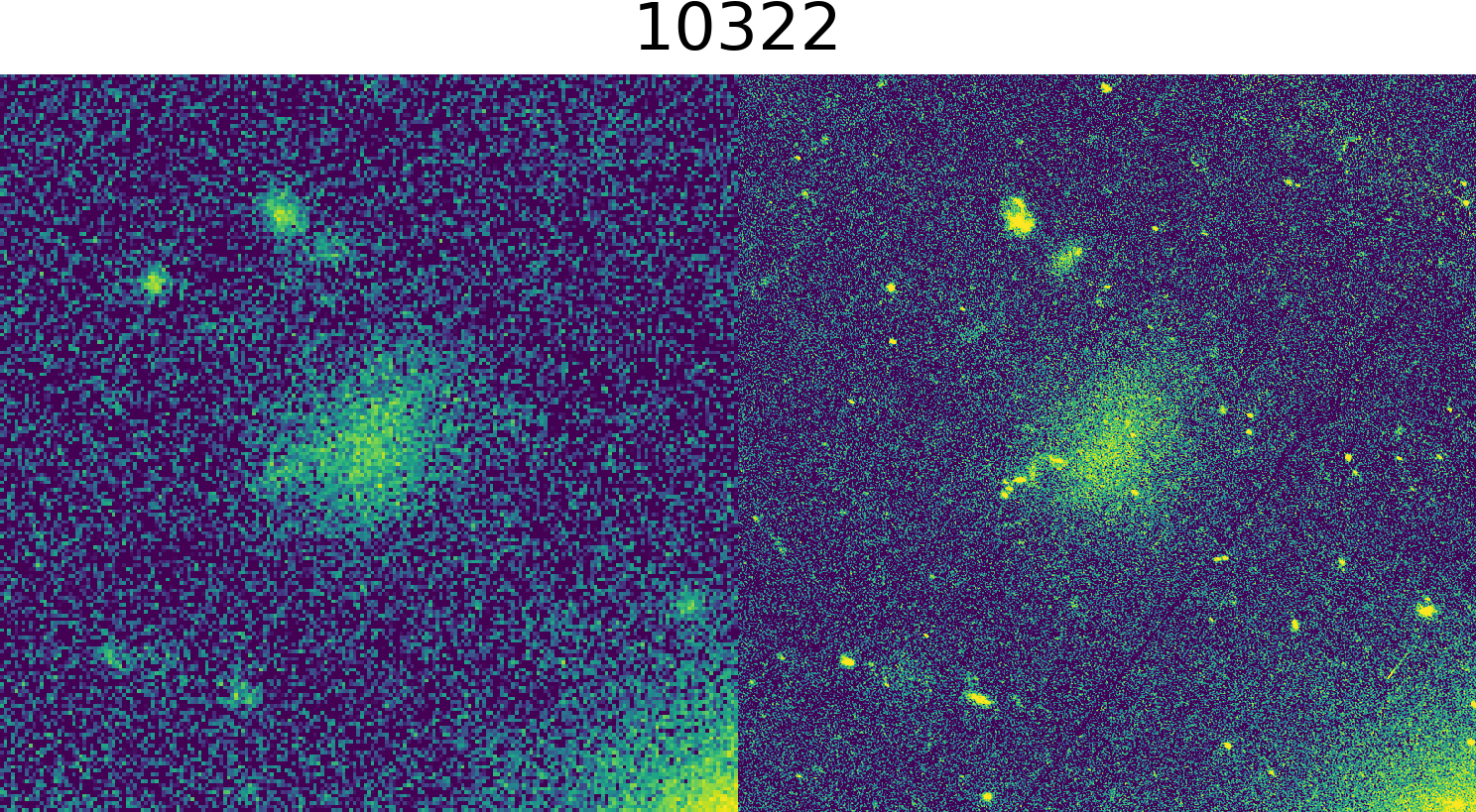}
    \includegraphics[width=0.43\textwidth,trim=0cm 0cm 0cm -0.3cm]{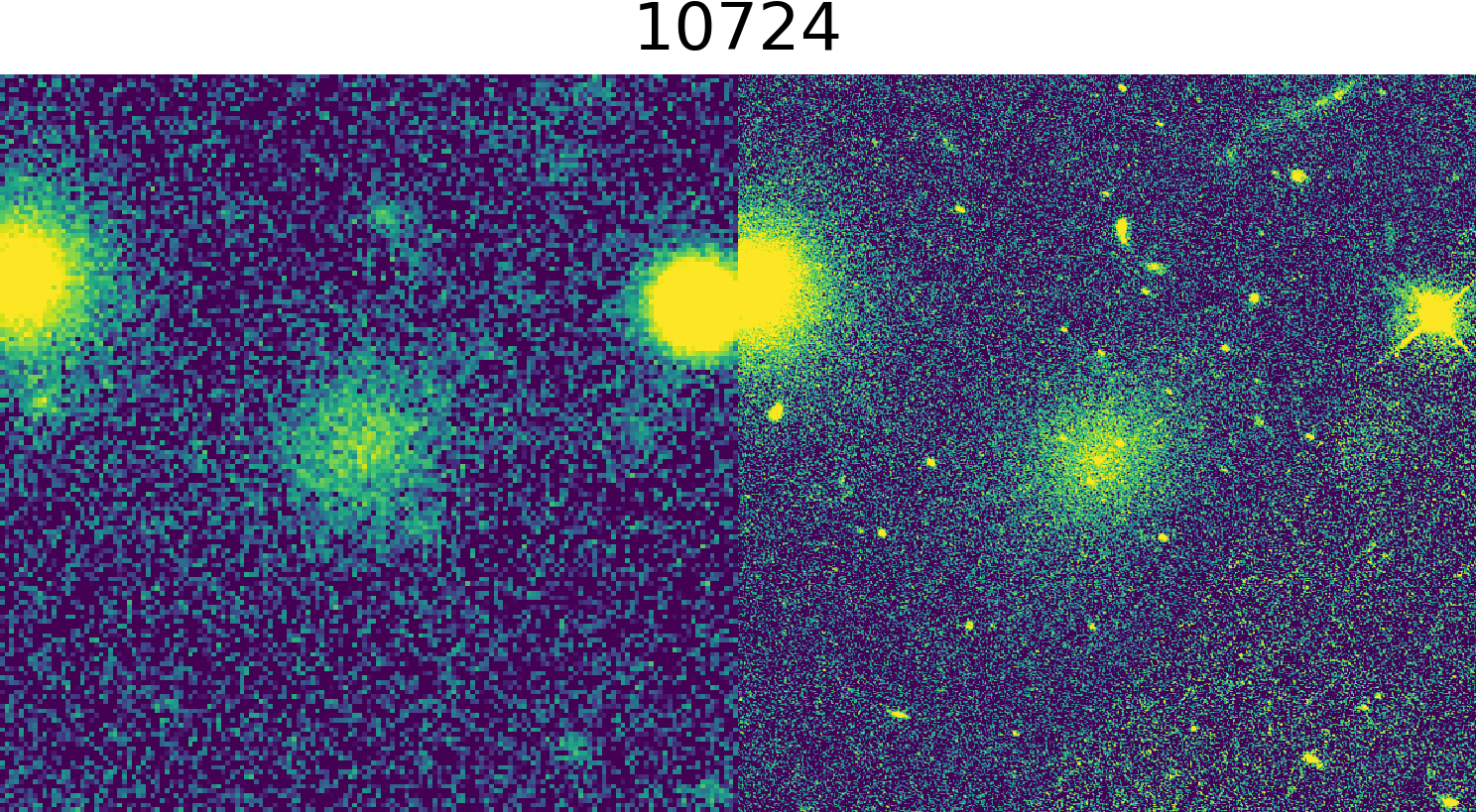}
    \includegraphics[width=0.43\textwidth,trim=0cm 0cm 0cm -0.3cm]{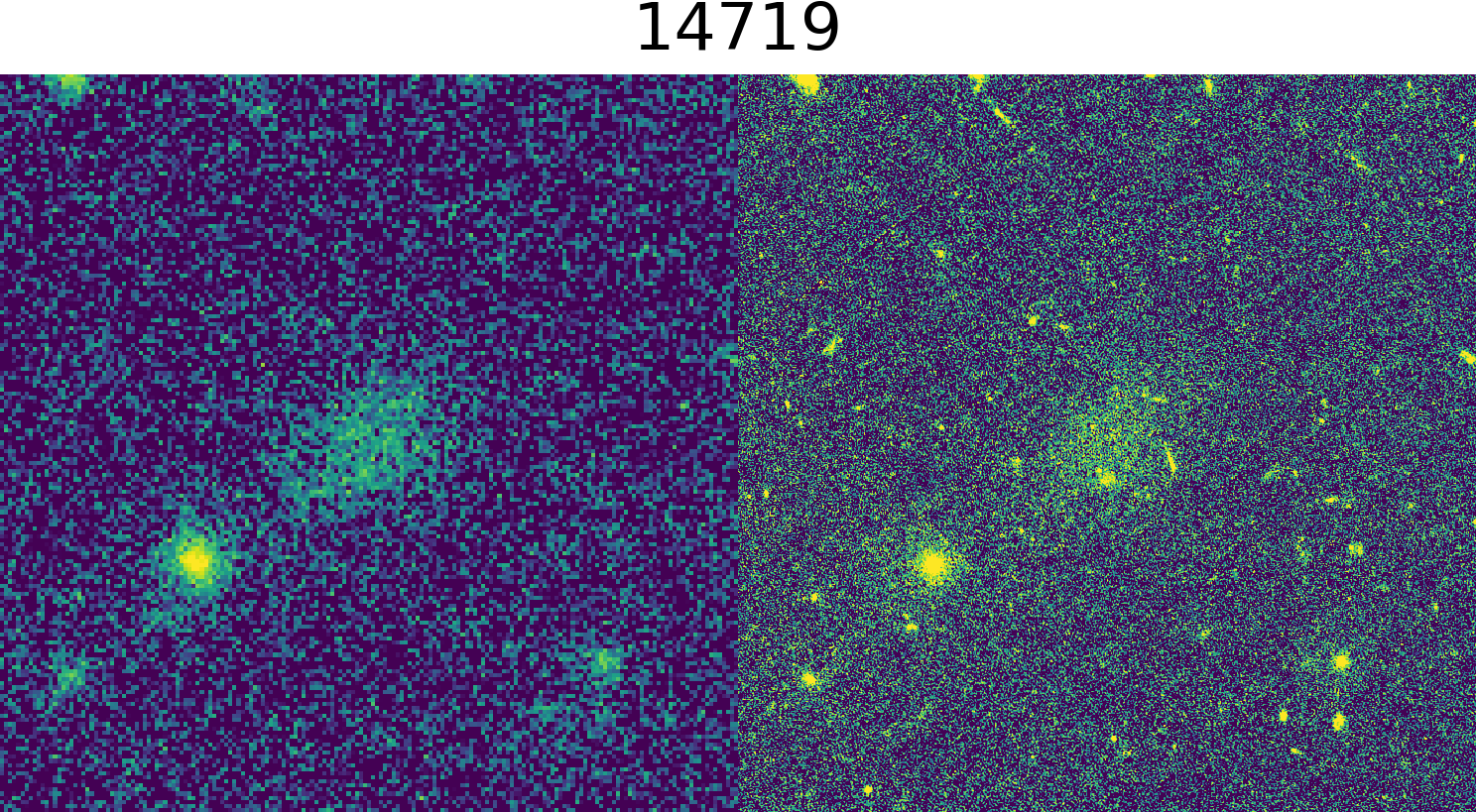}
    \includegraphics[width=0.43\textwidth,trim=0cm 0cm 0cm -0.3cm]{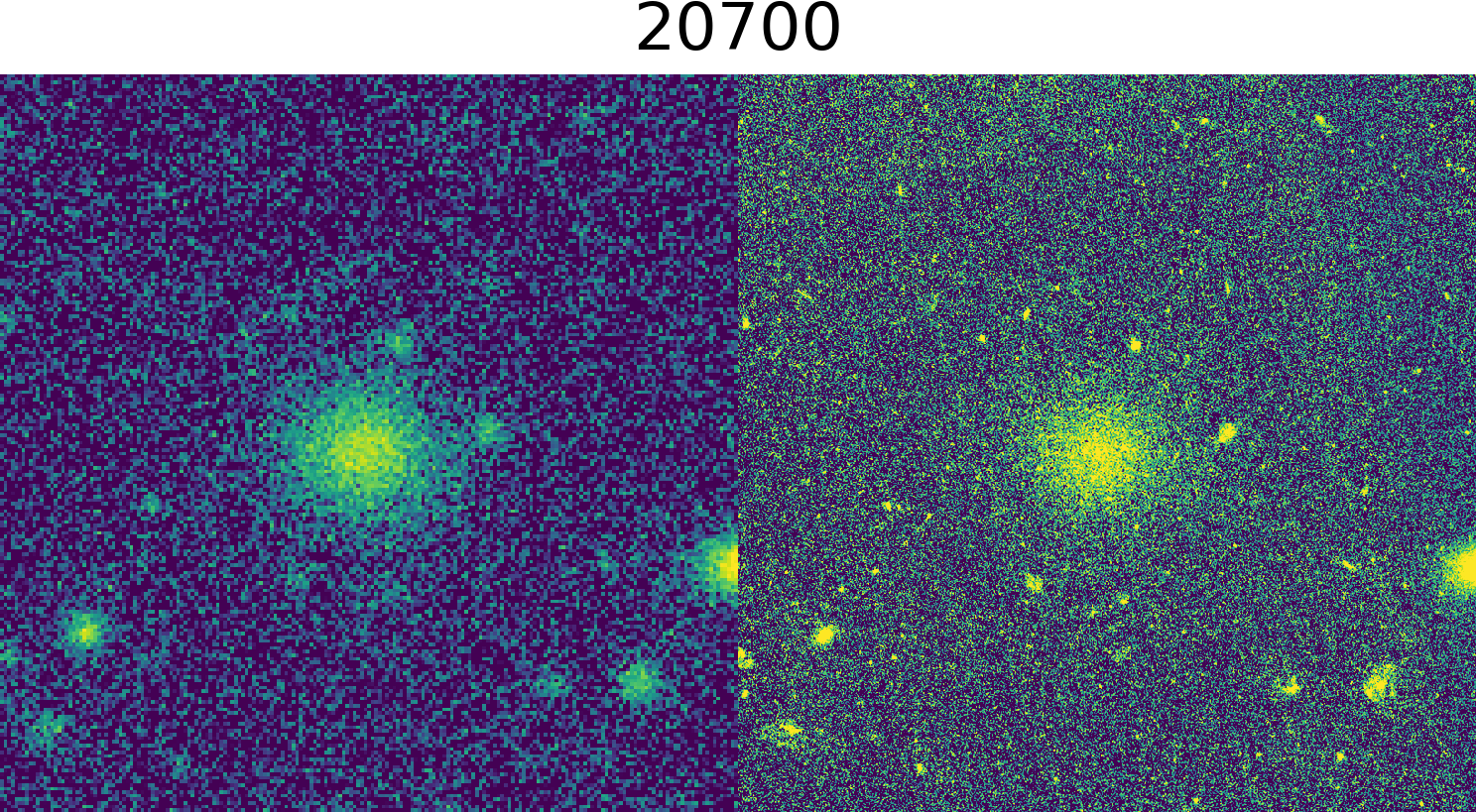}
    \includegraphics[width=0.43\textwidth,trim=0cm 0cm 0cm -0.3cm]{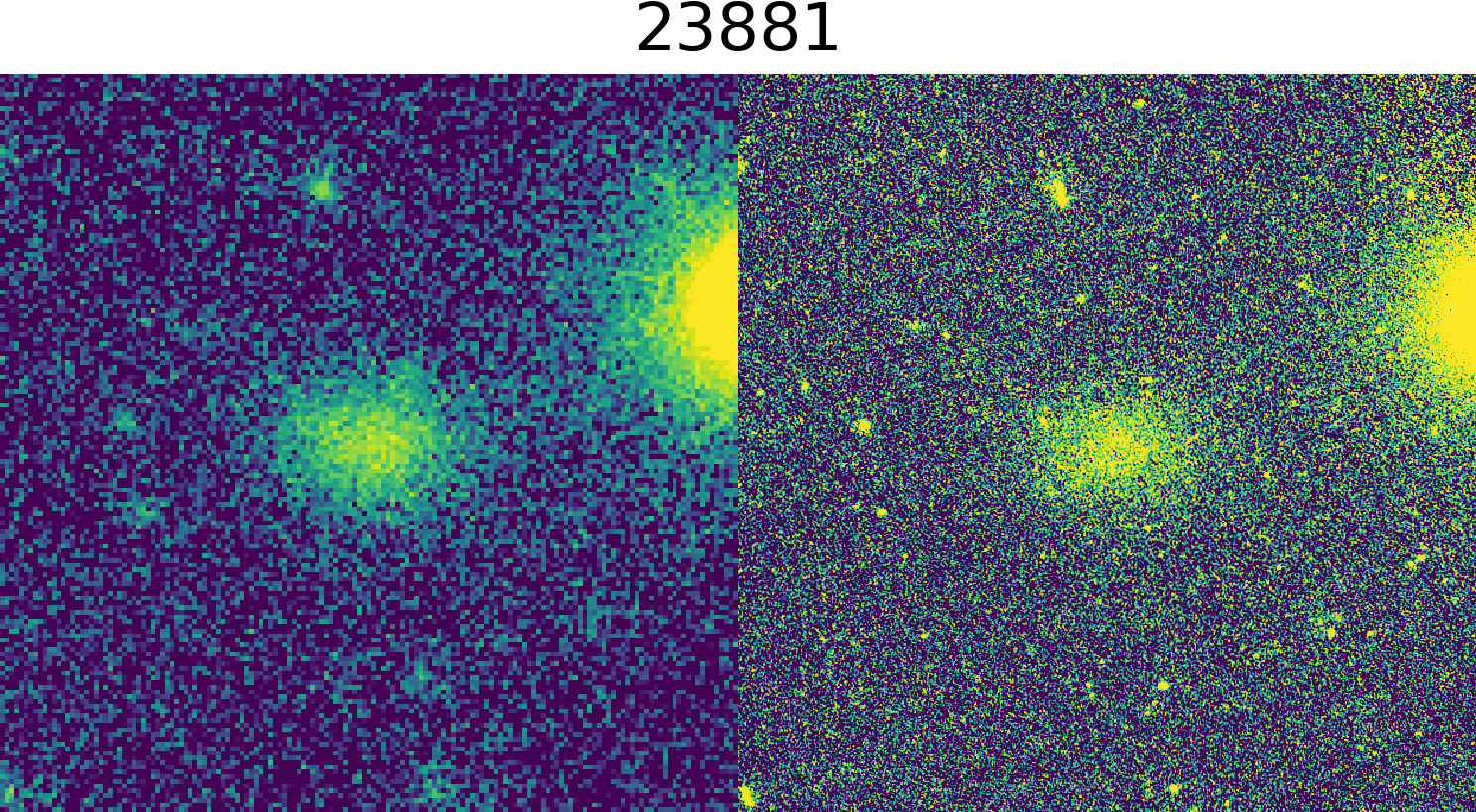}
    \includegraphics[width=0.43\textwidth,trim=0cm 0cm 0cm -0.3cm]{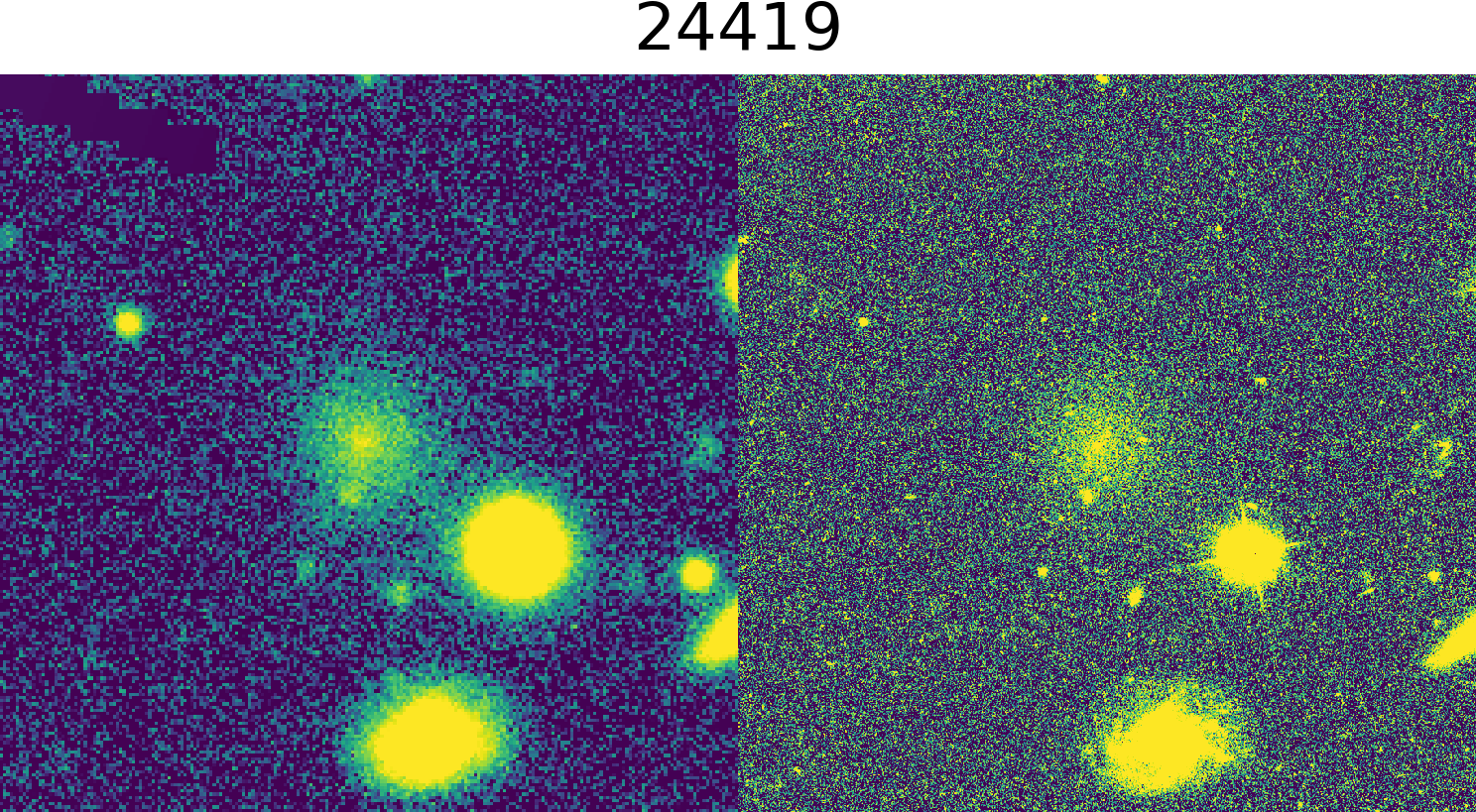}
    \caption{Comparison between WWFI $g'$ band (left) and HST F475W (right) data of selected UDGs.}
    \label{fig:UDGsWSTHST}
\end{figure*}

Furthermore, we report that 22 UDGs are covered by the HST data (catalog numbers: 6210, 6249, 8655, 8989, 9298, 10001, 10322, 10724, 13248, 14719, 15786, 16014, 16886, 17332,  20108, 20700, 21359, 23881, 24419, 26679, 30472, 38476). A few of them are depicted in Figure \ref{fig:UDGsWSTHST} showing the image cutout used for the \verb+GALFIT+ fits of our $g'$-band WWFI data (left) and the corresponding F475W HST image (right). Note that due to the significantly varying depth of the HST images and the diffuseness of UDGs, not all regions covered by HST data are deep enough to inspect even the most diffuse UDGs visually. All non-UDGs are bright enough to visually inspect them. We do not see a sign of spiral arms for any UDG. Their morphology is smooth, some appear to be GC rich, some GC poor as already reported in other studies \citep[see, e.g.][]{Amorisco2018,Gannon2022}.

\subsection{Parameter Correlations} \label{corellations}
\begin{figure*}
    \centering 
    \includegraphics[width=0.8\textwidth]{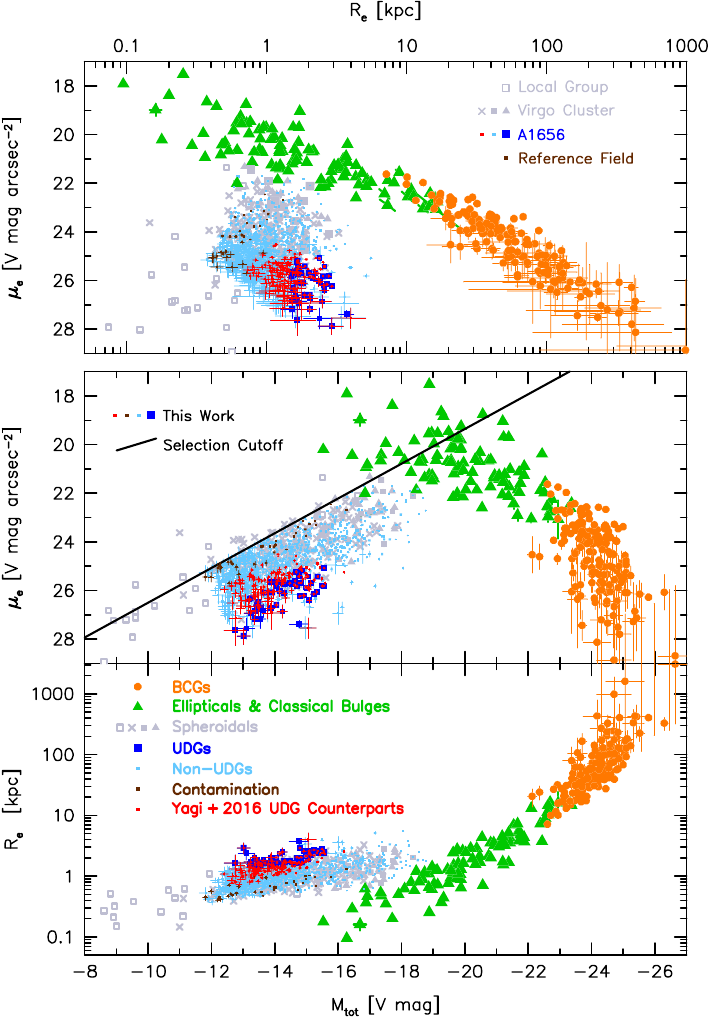}
    \caption{Comparison between $M_{\mathrm{tot}}$, $R_e$, and $\mu_e$ of UDGs (dark blue) and non-UDG cluster members (light blue) from our A1656 sample, as well as galaxies from the reference field analyzed for A1656 (brown). Galaxies with a counterpart in \citet{Yagi2016} are depicted in red. The $M_{\mathrm{tot}}-\mu_e$ cutoff is indicated by the black line. The basis for this plot is Figure 37 in \citet{Kormendy2009} with updates in Figure 2 in \citet{KormendyBender2012}, Figure 14 in \citet{Bender2015}, and Figure 16 in \citet{kluge} including BCGs (orange), ellipticals (green), classical bulges (green), and spheroidals (gray).}
    \label{fig:finalplotA1656}
\end{figure*}

\begin{figure*}
    \centering 
    \includegraphics[width=0.8\textwidth]{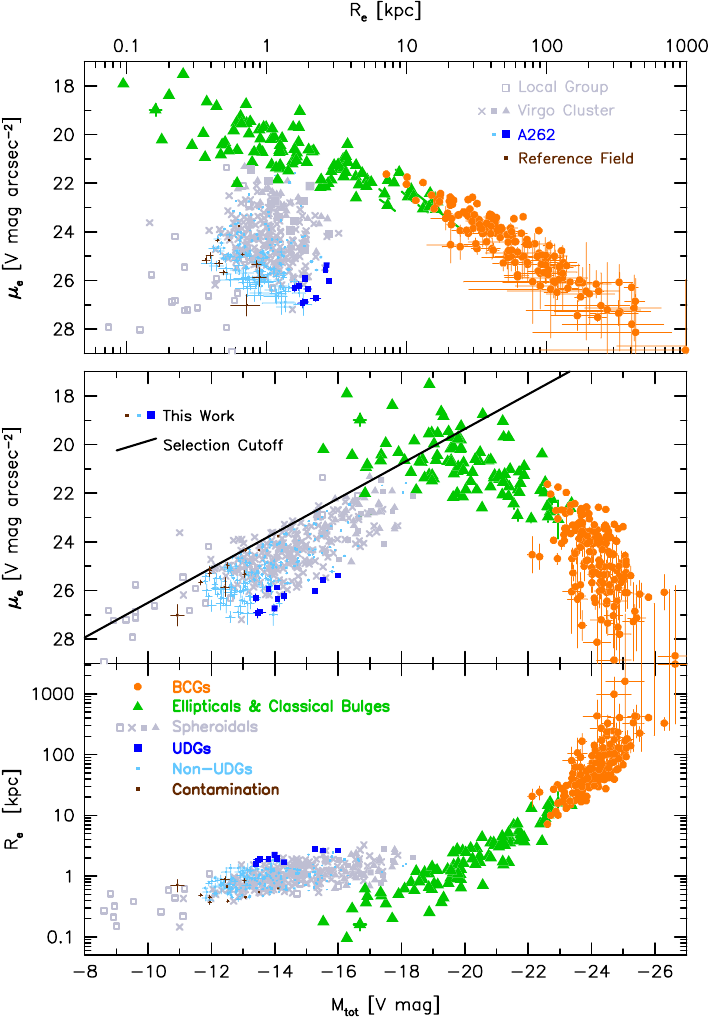}
    \caption{Comparison between $M_{\mathrm{tot}}$, $R_e$, and $\mu_e$ of UDGs (dark blue) and non-UDG cluster members (light blue) from our A262 sample, as well as galaxies from the reference field analyzed for A262 (brown). The $M_{\mathrm{tot}}-\mu_e$ cutoff is indicated by the black line. The basis for this plot is Figure 37 in \citet{Kormendy2009} with updates in Figure 2 in \citet{KormendyBender2012}, Figure 14 in \citet{Bender2015}, and Figure 16 in \citet{kluge} including BCGs (orange), ellipticals (green), classical bulges (green), and spheroidals (gray).}
    \label{fig:finalplotA262}
\end{figure*}
We investigate which regions our UDGs and non-UDG cluster members populate in the $M_{\mathrm{tot}}-R_e$, $M_{\mathrm{tot}}-\mu_e$, and $R_e-\mu_e$ parameter spaces, and where they lie relative to other galaxy populations. Furthermore, we study which regions of these parameter spaces are still affected by interloping background objects using our reference field. These parameter relations are shown for A1656 in Figure \ref{fig:finalplotA1656} and for A262 in Figure \ref{fig:finalplotA262}. The basis for these plots is Figure 37 in \citet{Kormendy2009} with updates from \citet{KormendyBender2012}, \citet{Bender2015}, and \citet{kluge}. The structural parameters of ellipticals are taken from \citet{Bender1992} and \citet{Kormendy2009}, those of classical bulges are from \citet{Fisher2008}, \citet{Kormendy2009}, and \citet{KormendyBender2012}. Here, we do not distinguish between cored ellipticals, cuspy ellipticals, and classical bulges, as they follow the same parameter relations. The structural parameters of BCGs are from \citet{kluge}. The data points of Local Group spheroidals are from \citet{Mateo1998} and \citet{McConnachieIrwin2006} and those of Virgo spheroidals are from  \citet{Ferasse2006}, \citet{Gavazzi2005}, and \citet{Kormendy2009}. Of those literature data, spheroidals are depicted in gray, ellipticals and classical bulges in green, and BCGs in orange. Our UDG sample is depicted in dark blue and non-UDG cluster members in light blue. For A1656, we depict all galaxies in our final sample for which we find a counterpart in the catalog of \citet{Yagi2016} as small red dots. Depicted in brown are all galaxies that remain in the sample for the reference field when analyzed for the respective cluster representing the contamination for our cluster member sample. 

In all three parameter relations and for both clusters, UDGs lie on the diffuse end of the spheroidal population and slightly extend it. UDGs are well separated from the elliptical and BCG populations. 
There is no dichotomy between UDGs and spheroidals from our sample nor from the literature in any of these parameter spaces. We even find a few galaxies that are more extreme than most UDGs without fulfilling the UDG definition because their central surface brightness is too bright. Most of those extreme non-UDGs have a high Sérsic index $n>2$. These galaxies might have an undetected nucleus that increases the central light profile, leading to a higher $n$. 

Comparing now our UDG sample (original \citealt{vanDokkum2015} definition) to our \citet{Yagi2016} UDG counterparts in all of these three parameter spaces, we find that the \citet{Yagi2016} UDG definition predominantly extends the original \citet{vanDokkum2015} UDG definition toward ordinary spheroidals and only adds a few galaxies in the regions of the parameter spaces that are populated by UDGs. 

In addition to the fact that we do not find a single UDG in the reference field, UDGs and the galaxies from the reference field are very well separated in all three parameter relations. Hence, we can conclude that our UDG sample should not be affected by interloping background galaxies.

By observing A262 in addition to A1656, we can check whether we can find more extreme UDGs in terms of both fainter surface brightness and larger size because of its smaller distance modulus, larger apparent size, and less crowding. However, we do not find more diffuse UDGs in A262 than in A1656. Instead, we find more diffuse galaxies in A1656. This might hint at the cluster environment playing a key role in forming the most diffuse UDGs (but also galaxies not fulfilling the UDG definition), e.g., by the higher gravitational potential or the higher richness and, hence, more interactions between the galaxies \citep[see, e.g.,][]{Dressler1980,KormendyBender2012,Duc,Poggianti2019,Sales,Shin2020,Tremmel2020}.
However, those galaxies are very rare and with relatively large uncertainties in their structural parameters and, hence, the absence of such galaxies in A262 could just be due to low number statistics in this cluster. 

We further note here that we detect one UDG in A1656 with extreme structural parameters that we did not include in our final sample despite being quiescent, a red sequence member, and having a converged \verb+GALFIT+ fit due to its too high uncertainties of the best-fit parameters ($\mu_e=29.6\pm1.5\,g'\,\mathrm{mag\,arcsec^{-2}}$, $\mu_0=24.80\pm0.96\,g'\,\mathrm{mag\,arcsec^{-2}}$,  $m_{\mathrm{tot}}=20.8\pm1.1\,g'\,\mathrm{mag}$, $n=2.3\pm1.2$, $R_e=7.0\pm7.8\,\mathrm{kpc}$, $u'-g'=1.37\pm0.14$, $g'-r'=0.57\pm0.12$, R.A. = 194.8011234, decl. = 27.92548284, \verb+SExtractor+ catalog number: 17748). Furthermore, we rejected a few apparently very diffuse and large galaxies in the eyeballing due to unreliable fitting results. This indicates that there might exist even more diffuse galaxies than those contained in our final sample of which the structural parameters could be measured with even deeper data. 

\begin{figure}
    \centering 
    \includegraphics[width=0.47\textwidth]{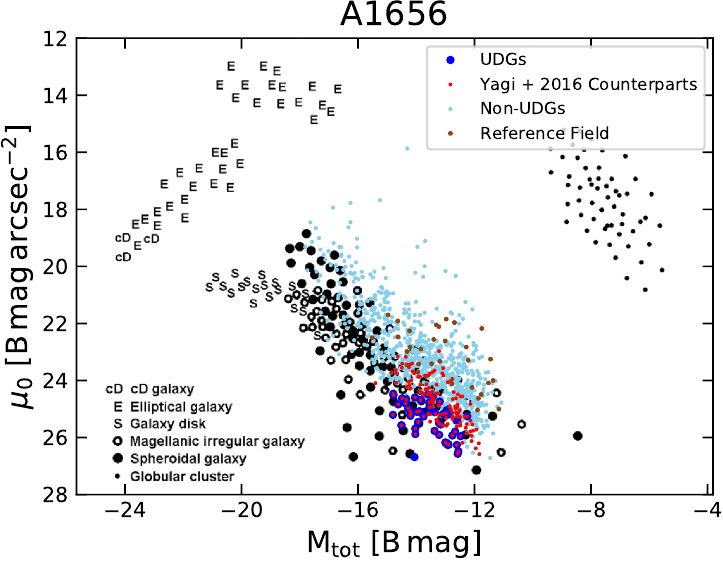}
    \includegraphics[width=0.47\textwidth,trim=0cm 0cm 0cm -0.5cm]{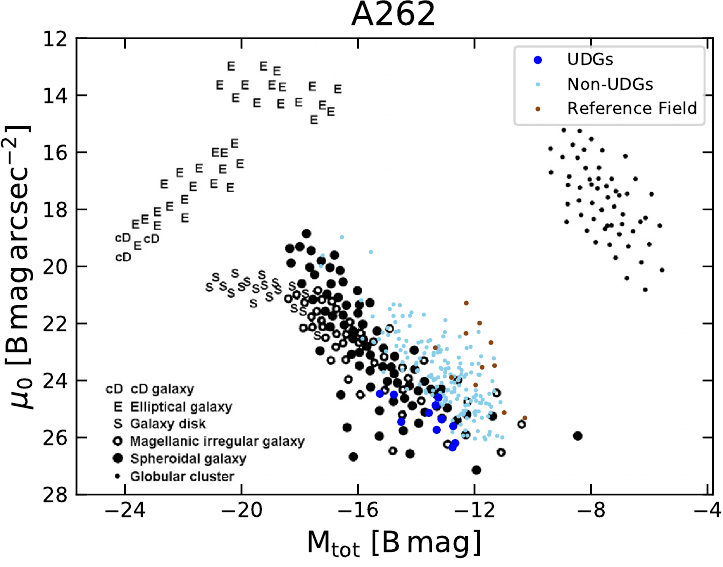}
    \caption{ Correlations between $M_{\mathrm{tot}}$ and $\mu_0$ for UDGs (dark blue) and non-UDGs (light blue) in A1656 (top) and A262 (bottom), as well as galaxies from the reference field analyzed for the respective cluster (brown). Galaxies with a counterpart in \citet{Yagi2016} are depicted in red. The basis for this plot is Figure 3 in \citet{Kormendy1985} with updates in Figure 1 in \citet{Binggeli1994} and in Figure 1 in \citet{Kormendy2009}. For elliptical and cD galaxies, $\mu_0$ corresponds to the highest surface brightness resolved by HST.}
    \label{fig:BinggeliComp}
\end{figure}
The next comparison is in the $\mu_0-M_{\mathrm{tot}}$ parameter space. In Figure \ref{fig:BinggeliComp}, we plot our measured parameters of the galaxies in A1656 (top) and A262 (bottom) over Figure 1 from \citet{Kormendy2009} which is based on Figure 3 of \citet{Kormendy1985} with updates in Figure 1 of \citet{Binggeli1994}. The literature data points are from \citet{Kormendy1985}, \citet{Bothun1987}, \citet{vanderKruit1987}, \citet{BinggeliCameron1991,BinggeliCameron1993}, \citet{Caldwell1992}, and \citet{Faber1997}. In this parameter space, UDGs populate the same region as the spheroidals in \citet{Binggeli1994}. Actually, the sample used by \citet{Binggeli1994} contains even more extreme galaxies than UDGs with similar $\mu_0$ and brighter $M_{\mathrm{tot}}$. Parts of our non-UDG sample extend the spheroidal population in this parameter space at the faint $M_{\mathrm{tot}}$ end. Moreover, UDGs are separated from the galaxies remaining in the sample of the reference field. The contamination mainly affects the bright $\mu_0$ and faint $M_{\mathrm{tot}}$ region above the spheroidal sequence of \citet{Binggeli1994}.

\begin{figure}[t]
    \centering
    \includegraphics[width=0.47\textwidth]{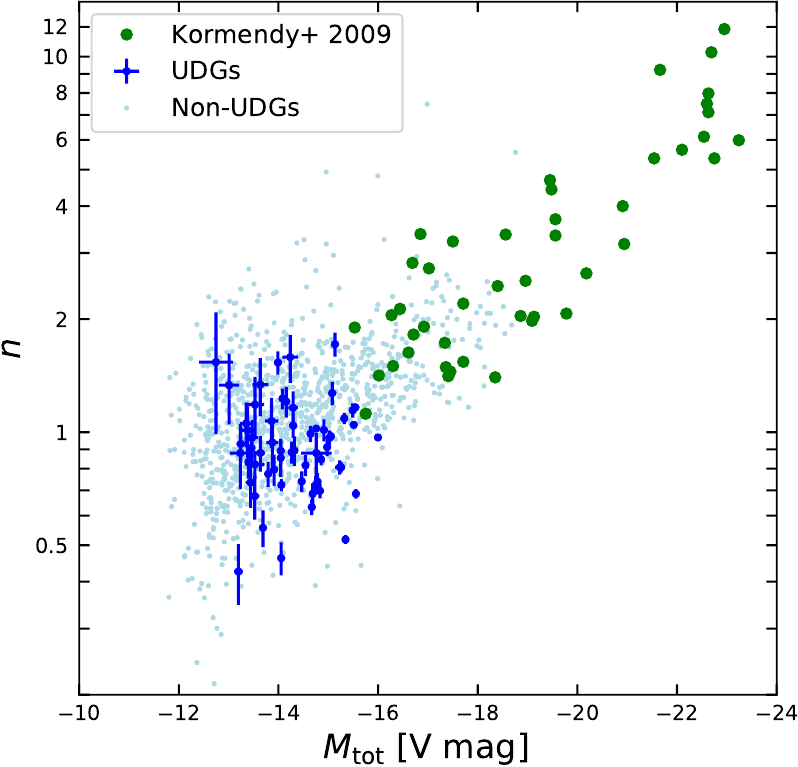}
    \caption{Total $V$-band magnitude vs. Sérsic index $n$ of UDGs (dark blue) and non-UDGs (light blue) in our sample, as well as ellipticals, S0 bulges, and (bright) spheroidals from \citet[][green]{Kormendy2009}. Uncertainties are only shown for UDGs for better clarity.}
    \label{fig:nvsMtot}
\end{figure}
Furthermore, in the $M_{\mathrm{tot}}-n$ parameter space, UDGs as well as non-UDGs in our sample follow the same scaling relation as spheroidals and ellipticals from \citet[][see Figure \ref{fig:nvsMtot}]{Kormendy2009}. In this scaling relation, there is no trend that UDGs have smaller or larger $n$ than non-UDGs of the same $M_{\mathrm{tot}}$.

\begin{figure}[t]
    \centering
    \includegraphics[width=0.45\textwidth]{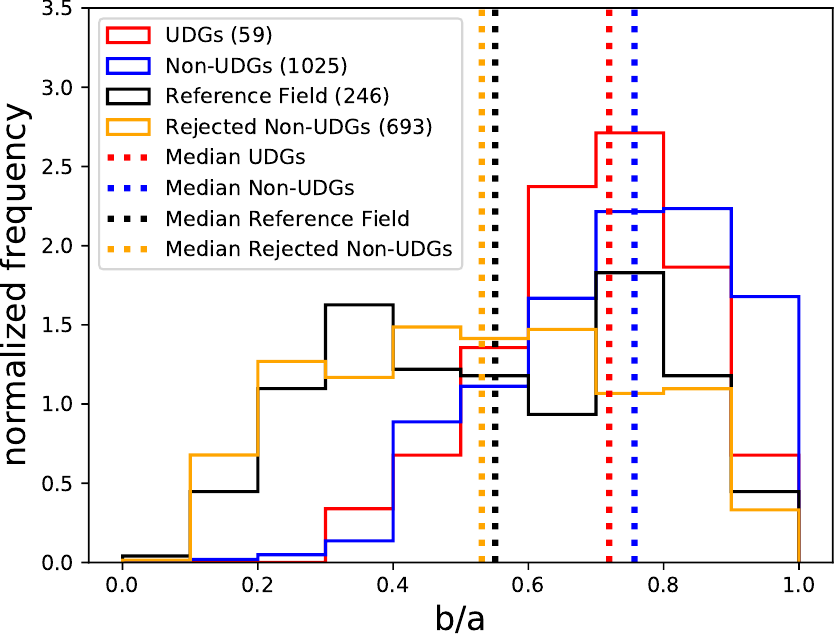}
    \caption{Axis ratio distribution of all UDGs (red) and non-UDGs (blue), as well as of all galaxies from the reference field (black) and all galaxies rejected by the $M_{\mathrm{tot}}-\mu_e$ cutoff (orange). The number of objects is given in brackets. The median axis ratio of each distribution is depicted as a vertical dotted line.}
    \label{fig:histos}
\end{figure}
The axis ratio distributions of UDGs, non-UDGs, as well as all galaxies in the reference field before applying the $M_{\mathrm{tot}}-\mu_e$ cutoff and all galaxies removed from the cluster sample by the $M_{\mathrm{tot}}-\mu_e$ cutoff are shown in Figure \ref{fig:histos}. Both, the UDG, as well as the non-UDG populations in our sample show a peaked axis ratio distribution. They tend to be relatively round with a median $b/a=0.72$ for UDGs and $b/a=0.76$ for non-UDGs. This agrees well with the findings of \citet{vanDokkum2015}. On the other hand, the axis ratio distribution of the galaxies found in the reference field before applying the $M_{\mathrm{tot}}-\mu_e$ cutoff is flat 
with a median $b/a=0.55$. For the galaxies removed from the cluster sample by the $M_{\mathrm{tot}}-\mu_e$ cutoff we also find a similar flat distribution with a median $b/a=0.53$. The flat axis ratio distributions of the galaxies in the reference field and the galaxies removed from the cluster sample by the $M_{\mathrm{tot}}-\mu_e$ cutoff resemble the distribution of randomly oriented thin disks, whereas the peaked distribution of UDGs and non-UDGs resembles the expected distribution of spheroid-shaped galaxies \citep[e.g.,][]{Ryden1996,PadillaStrauss2008}. 

The $b/a$ distributions of UDGs and non-UDGs are quite similar -- a Kolmogorov–-Smirnov test gives a p-value of 0.18. Furthermore, the inferred spheroidal shape of the UDGs indicates that they cannot be formed simply by quenching star formation at high redshift of a spiral galaxy. There must be a violent process involved in the formation history that turned a disk galaxy into a spheroid.

The axis ratio distribution of galaxies from the reference field and of the galaxies removed from the cluster sample by the $M_{\mathrm{tot}}-\mu_e$ cutoff appear to be quite similarly flat, indicating that we indeed mainly rejected interloping background spirals from the sample by the $M_{\mathrm{tot}}-\mu_e$ cutoff. Despite these  axis ratio distributions appearing at first glance quite similar, they are not drawn from exactly the same distribution. A Kolmogorov-Smirnov test gives a p-value of only 0.02. We argue that this difference could arise from galaxies that are actually in the cluster being removed by the $M_{\mathrm{tot}}-\mu_e$ cutoff, cosmic variance, and the higher depth of the reference field.

\section{Summary and Conclusion} \label{sec:summaryconclusion}

We have developed a pipeline to measure the structural parameters as well as $u'-g'$ and $g'-r'$ colors of tens of thousands of potential UDGs and other dwarf galaxies in A262 and A1656. In order to measure accurate structural parameters with \verb+GALFIT+, we have developed a sophisticated masking procedure. We have identified and separated dwarf galaxy cluster  member candidates in A262 and A1656 from diffuse background galaxies based on their location in the $u'-g'$ vs. $g'-r'$ color--color diagram and red sequence membership. Furthermore, we found that the remaining contamination of our sample forms a distinct sequence in the $M_{\mathrm{tot}}-\mu_e$ parameter space, and hence, we rejected the majority of the remaining interloping background galaxies by an $M_{\mathrm{tot}}-\mu_e$ cutoff. Overall, we found and successfully measured the structural parameters of 185 dwarf galaxy cluster members in A262 and 899 in A1656. Among these, we found 11 UDGs in A262 and 48 UDGs in A1656. The latter is six times more than the eight UDGs found by \citet{vanDokkum2015} within our common field of view. Furthermore, we found multiple UDGs that are much closer to the cluster center than the UDGs found by \citet{vanDokkum2015}.

Moreover, we detected a few very diffuse galaxies with colors consistent to be cluster members but excluded them from our final catalog due to unreliable \verb+GALFIT+ fits. This indicates that we did not yet reach the limit of measuring the structural parameters of the most diffuse galaxies.

With the analysis of the reference field, we showed that the color information is crucial to drastically improve the  purity of our sample. The $u'-g'$ vs. $g'-r'$ color--color preselection and the following red sequence selection remove about $90\%$ of interloping galaxies. By using the $u'-g'$ vs. $g'-r'$ color--color diagram to preselect quiescent galaxies additionally to the more traditional red sequence selection, we improved the purity of our sample by $\sim 70\%$ compared to using only the latter. Our final $M_{\mathrm{tot}}-\mu_e$ selection cutoff removes $\sim 90\%$ of the remaining contamination in A262 and about 75\% in A1656. We give a conservative upper limit for the contamination of our final
cluster member sample of 15.6\% for both clusters. In our reference field, we did not find a single UDG, and we found that UDGs in both galaxy clusters populate distinctly separated regions in the $M_{\mathrm{tot}}-R_e$, $M_{\mathrm{tot}}-\mu_e$, $R_e-\mu_e$, and $M_{\mathrm{tot}}-\mu_0$ parameter spaces. Hence, we consider our UDG sample to be free of interloping galaxies. However, for the compact end of our non-UDG sample, we expect significant contamination. 

We found that UDGs populate in the $M_{\mathrm{tot}}-R_e$, $M_{\mathrm{tot}}-\mu_e$, and $R_e-\mu_e$ parameter spaces the same region as the most diffuse Virgo spheroidals analyzed by \citet{Ferasse2006}, \citet{Kormendy2009}, and \citet{Gavazzi2005} and slightly extend this population. We even find a few non-UDGs that are more diffuse in terms of these structural parameters. In the $M_{\mathrm{tot}}-\mu_0$ parameter space, we find that UDGs populate the same region as the spheroidals in \citet{Binggeli1994}. Furthermore, we found that UDGs, as well as non-UDGs follow the same $M_{\mathrm{tot}}-n$ scaling relation as spheroidals, ellipticals, and classical bulges. Lastly, we confirmed that UDGs have a spheroidal shape based on the axis ratio distribution and that the axis ratio distribution of UDGs is similar to that of non-UDGs in our sample. Overall, we conclude that UDGs do not form a distinct population but form the diffuse end of the spheroidal population (also frequently referred to as dSph or dE). 

Furthermore, we found that the UDG definition used by \citet{Koda} and \citet{Yagi2016} extends the definition by \citet{vanDokkum2015} toward ordinary spheroidals. About $70\%$ of their sample for which we have a counterpart in our sample do not fulfill the original UDG definition by \citet{vanDokkum2015}.  

Generally, the classification of galaxies should rely on physical differences between the different populations. The dichotomy of ellipticals and spheroidals relies on distinct formation processes: mergers for ellipticals and conversion from spiral galaxies by environmental effects and by energy feedback for spheroidals \citep{Kormendy2009} like the proposed formation scenarios for UDGs \citep[e.g.,][]{amorisco,diCintio,Sales,Shin2020,Tremmel2020}.

In the $M_{\mathrm{tot}}-R_e$, $M_{\mathrm{tot}}-\mu_e$, and $R_e-\mu_e$ parameter spaces, these galaxy populations follow different scaling relations \citep{Kormendy2009}. However, the spheroidal and elliptical populations overlap slightly in these parameter spaces and, hence, cannot be perfectly separated here. In the $M_{\mathrm{tot}}-\mu_0$ parameter space, spheroidals are well separated from ellipticals of the same $M_{\mathrm{tot}}$ by the extra-light component caused by their formation in wet mergers \citep[see Figure \ref{fig:BinggeliComp}, and e.g., ][]{Kormendy2009}. The significant difference in this parameter space can be used to discriminate between these two populations.

Using different names for essentially the same galaxy population does not clarify the discussion. However, subclassifying the most diffuse spheroidals as UDGs makes sense to ensure that every study is discussing about the same galaxies when studying how the most diffuse spheroidals can be formed, how they can survive in the centers of massive galaxy clusters, and their dark matter content. Despite the definition limits for UDGs introduced by \citet{vanDokkum2015} being arbitrary, one has to stick to a clear definition and not significantly extend the sample toward ordinary spheroidals by using a different definition but the same name \citep[e.g.,][]{Koda,Yagi2016,Sales}. Significantly extending the studied subpopulation dilutes the inferred formation mechanisms and can lead to misinterpretations.

We have shown that UDGs are not a distinct population but are only the diffuse end of the already well-known spheroidal population. However, investigating the properties of UDGs can still be a fruitful endeavor. Despite spheroidals from the Local Group providing an even more dark matter-dominated probe \citep[e.g.,][]{BattagliaNipoti2022}, UDGs still provide an excellent probe to study the nature of dark matter beyond the Local Group in a much denser cluster environment. So far, only the extreme cases of UDGs with either low or high GC counts were studied, suggesting either over- or undermassive halos with respect to the $M_{\mathrm{stellar}}-M_{\mathrm{halo}}$ relation \citep[e.g.,][]{Gannon2023}. Furthermore, future studies should not only focus on UDGs but also probe "normal" spheroidals to get a representative sample to obtain a full understanding of structure formation in the dwarf galaxy regime.

\section{Acknowledgements}
We thank the anonymous referee for providing comments and suggestions that allowed us to significantly improve the paper.

The Wendelstein 2.1m telescope project was funded by the Bavarian government and by the German Federal government through a common funding process. Part of the 2.1m instrumentation including some of the upgrades for the infrastructure were funded by the Cluster of Excellence “Origin of the Universe” of the German Science foundation DFG.

This work would not have been practical without extensive use of NASA's Astrophysics Data System Bibliographic Services and the SIMBAD database, operated at CDS, Strasbourg, France.

This research is based on observations made with the NASA/ESA Hubble Space Telescope obtained from the Space Telescope Science Institute, which is operated by the Association of Universities for Research in Astronomy, Inc., under NASA contract NAS 5–26555. These observations are associated with programs 10397, 10861, 11711, 12918, 13777, 14182, 14361. The HST data presented in this paper were obtained from the Mikulski Archive for Space Telescopes (MAST).

We also used the image display tool SAOImage DS9 developed by Smithsonian Astrophysical Observatory and the image display tool Fitsedit, developed by Johannes Koppenhoefer.

Software: Astropy \citep{astropy2018}, Photutils \citep{photutils}, numpy \citep{numpy}, scipy \citep{scipy}, matplotlib \citep{matplotlib}, SExtractor \citep{Sextractor}, SCAMP \citep{scamp}, SWarp \citep{swarp}.

\bibliography{UDGpaper}{}
\bibliographystyle{aasjournal}

\appendix
\section{Zero-point consistency check}
As a consistency check of the $u'$ band zero-point, we test whether the colors of the stars in the three different fields are consistent with each other. This is especially important for A262, as it is not directly calibrated relative to SDSS because it is not covered by SDSS. For this test, we use all stars used for the PSF measurements (see Section \ref{sec:psfmeasurements}). The density contours in the $u'-g'$ versus $g'-r'$ parameter space agree well with each other. Furthermore, we fit third-order polynomials to the distribution of the stars the A1656 and the reference field in the $u'-g'$ versus $g'-r'$ parameter space. Then we fit the resulting polynomials to the A262 data, with only the $u'$-band magnitude as a free parameter. We find a $u'$ band offset between A262 and A1656 of +0.06 mag and between A262 and the reference field an offset of $-0.03$ mag. Using fifth-order polynomials for this test, we find an offset between A262 and A1656 of 0.00 mag and between A262 and the reference field an offset of 0.02 mag. Hence, we consider the $u'$ band zero-point to be consistent, within the uncertainty of this method.
\section{Spiral Galaxies}
\label{app:SpiralsHST}
\begin{figure}[ht]
    \centering
    \includegraphics[width=0.47\textwidth]{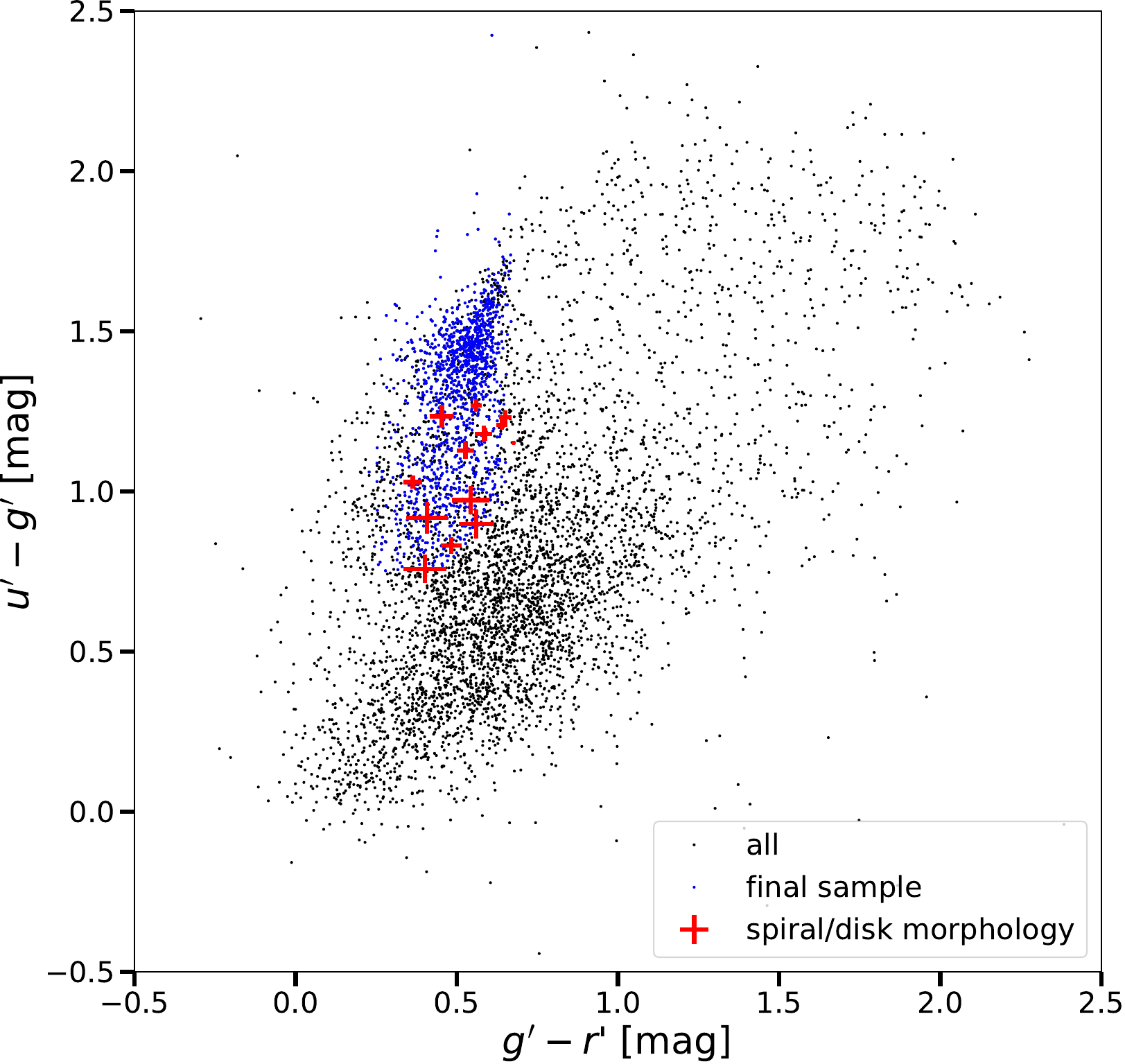}
    \caption{$u'-g'$ vs. $g'-r'$ color--color diagrams for A1656. Red points represent the 14 galaxies with a spiral morphology in the HST observations, blue points represent our final A1656 galaxy sample, and black points represent all galaxies that were rejected by the bicolor and red sequence selection.}
    \label{fig:colorcolorspirals}
\end{figure}
\begin{figure*}[ht!]
    \centering
    \includegraphics[width=0.22\textwidth]{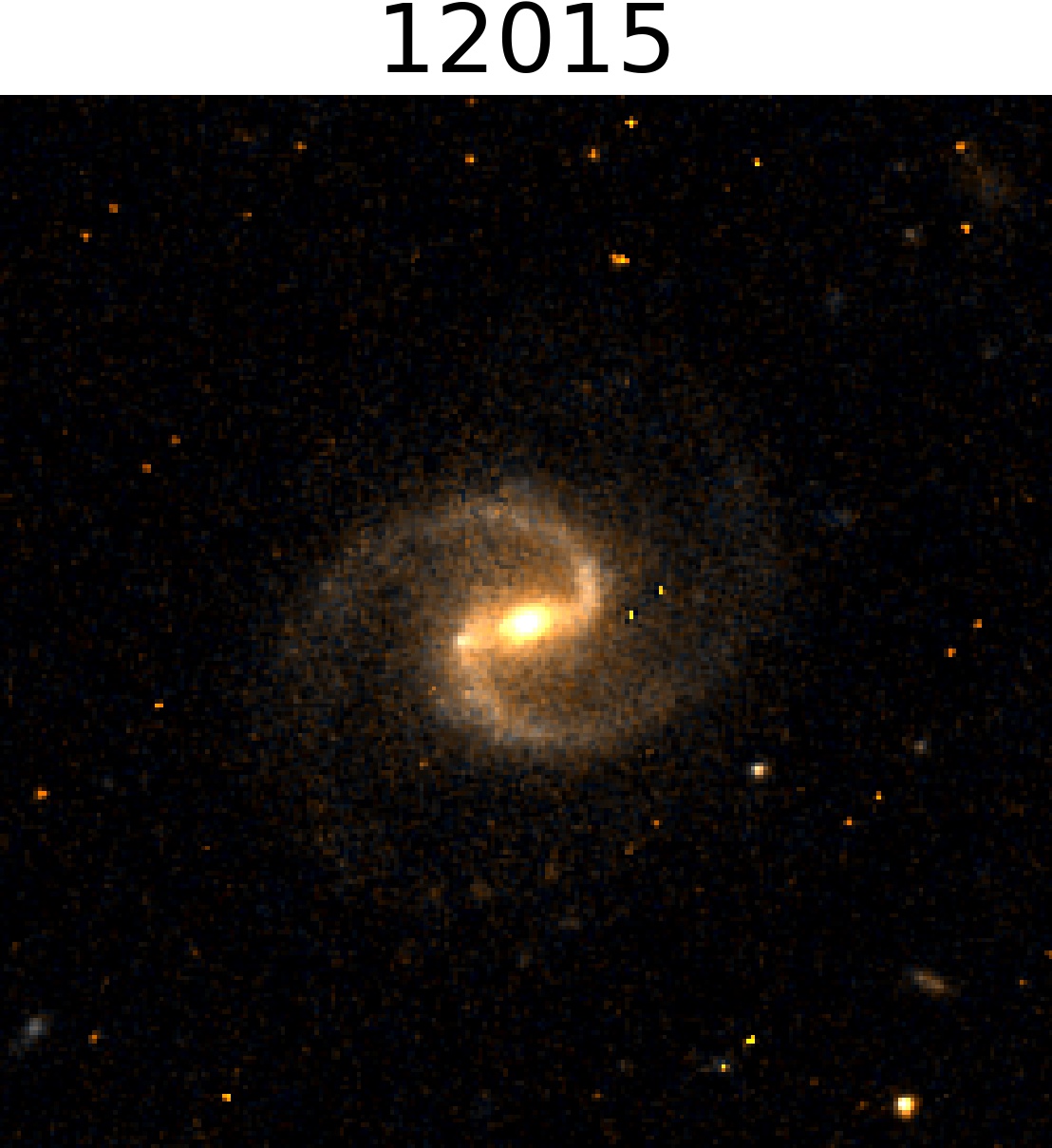}
    \includegraphics[width=0.22\textwidth]{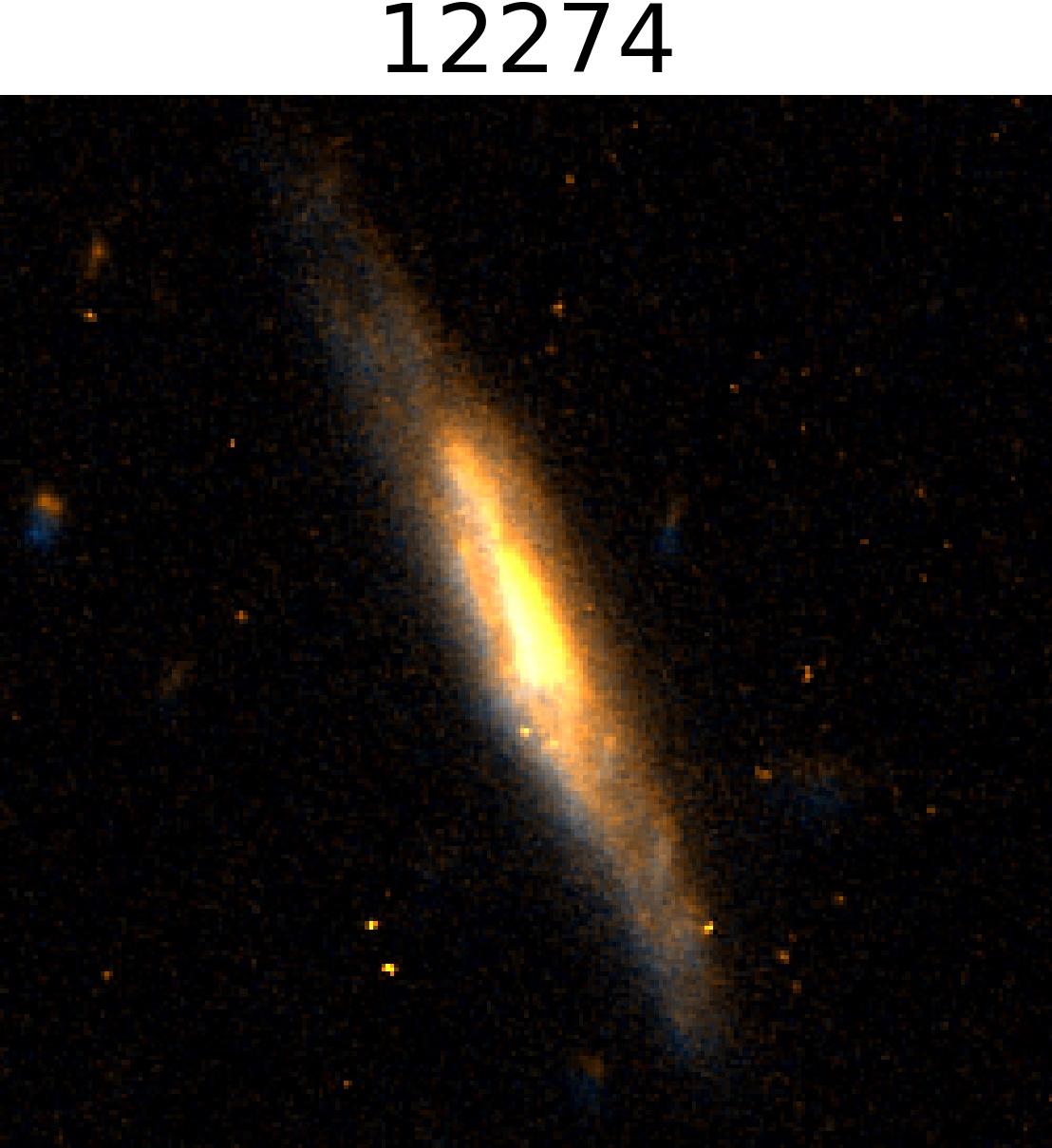}
    \includegraphics[width=0.22\textwidth,trim=0cm 0cm 0cm -0.3cm]{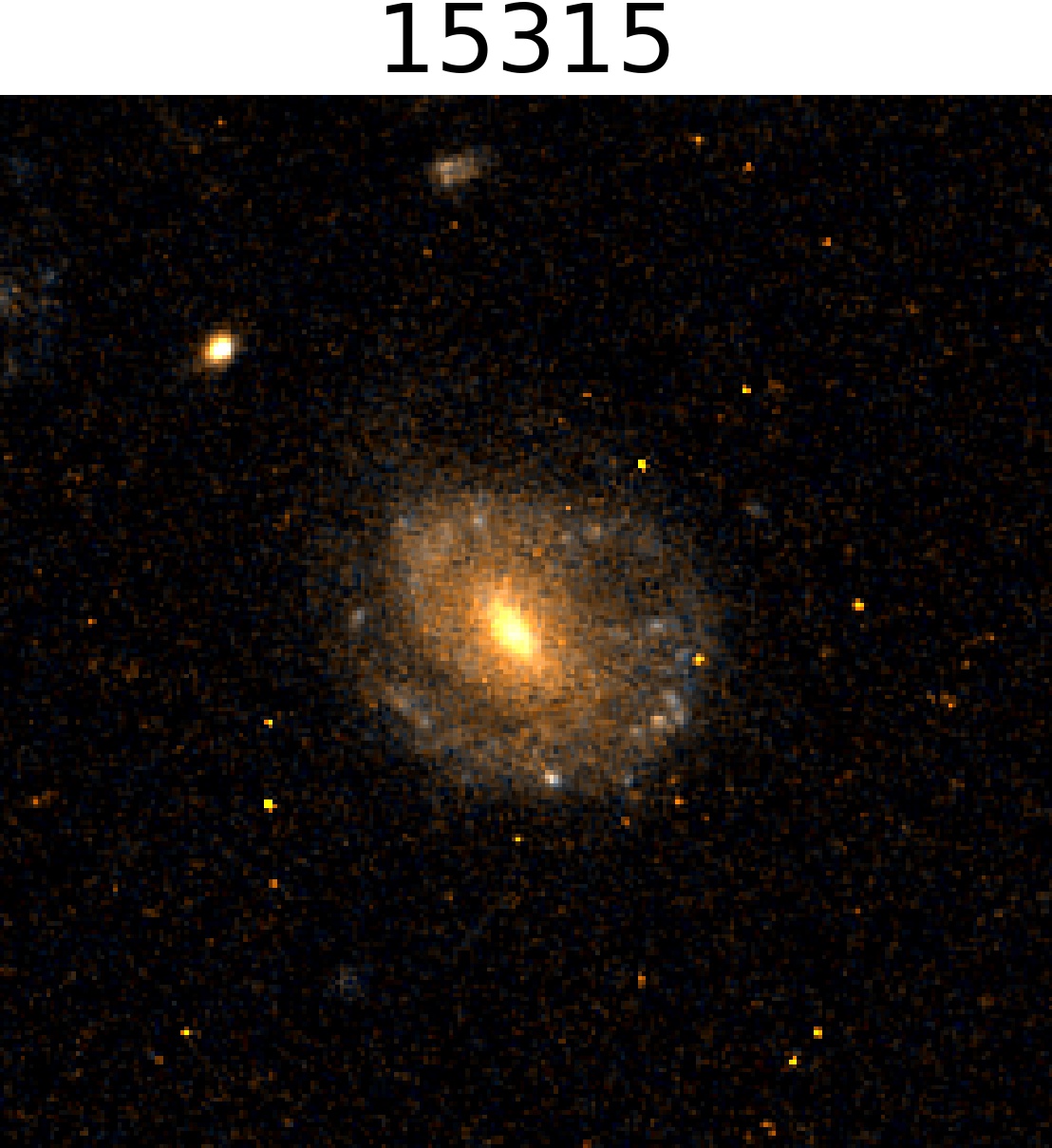}
    \includegraphics[width=0.22\textwidth,trim=0cm 0cm 0cm -0.3cm]{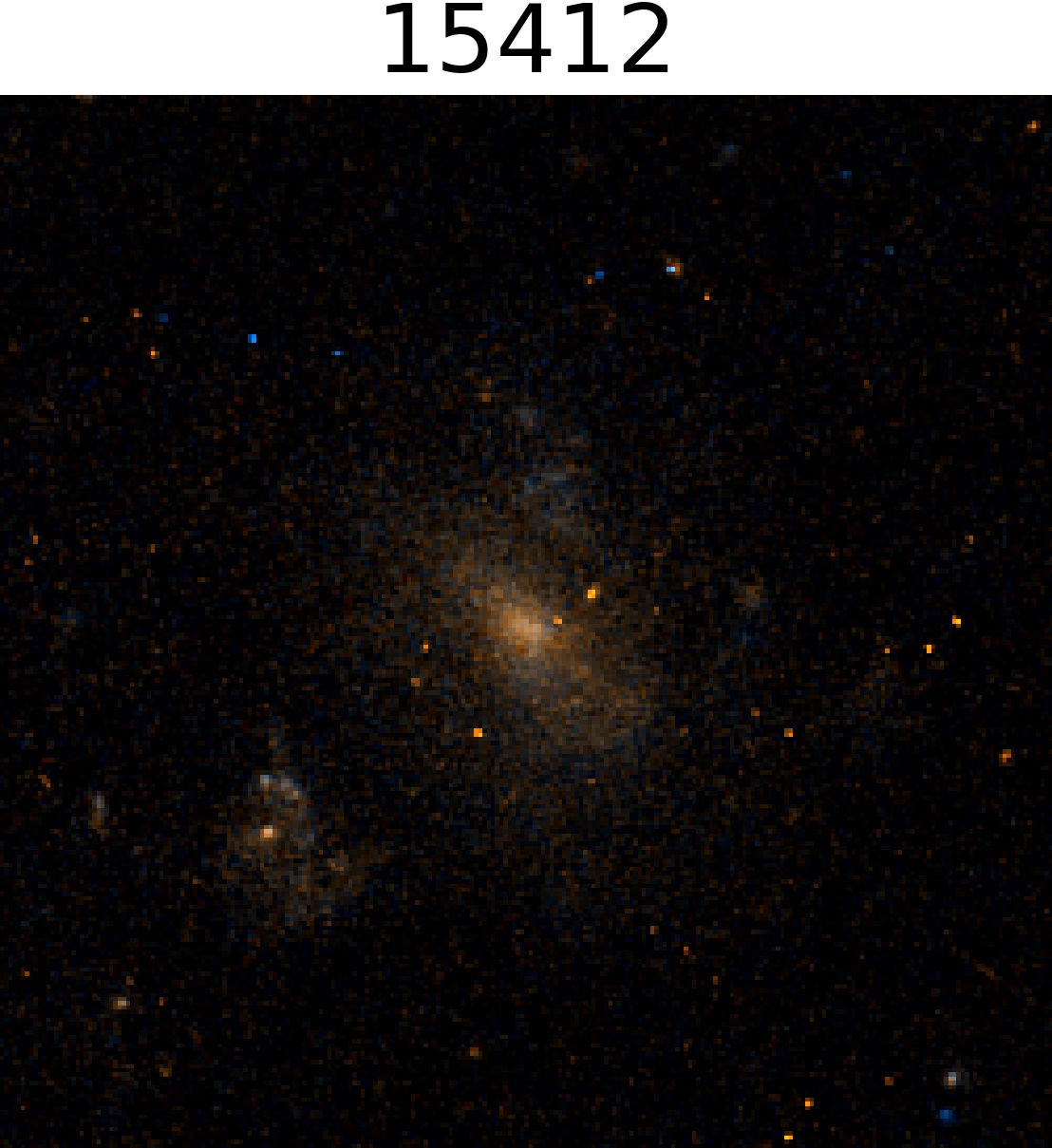}
    \includegraphics[width=0.22\textwidth,trim=0cm 0cm 0cm -0.3cm]{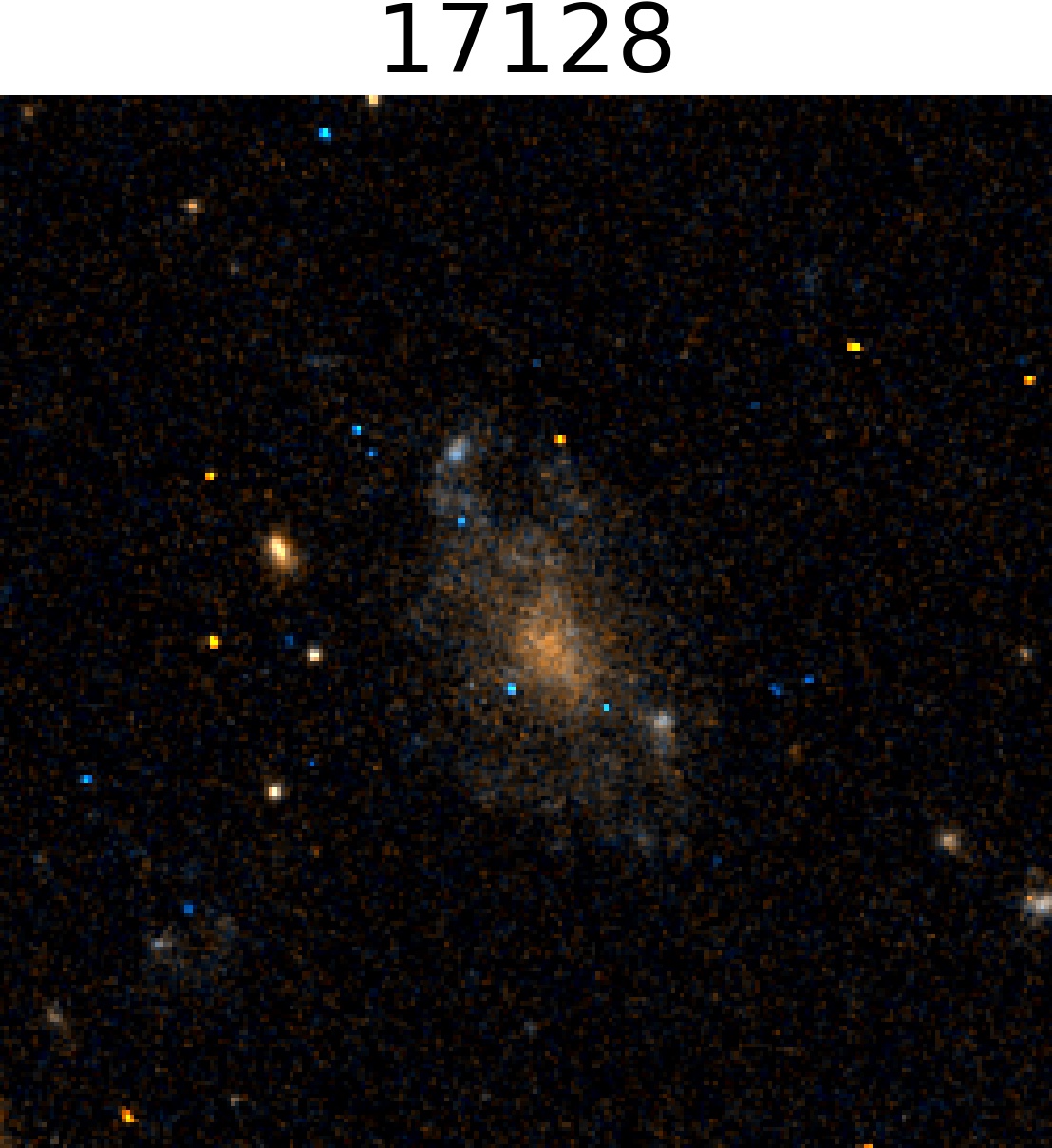}
    \includegraphics[width=0.22\textwidth,trim=0cm 0cm 0cm -0.3cm]{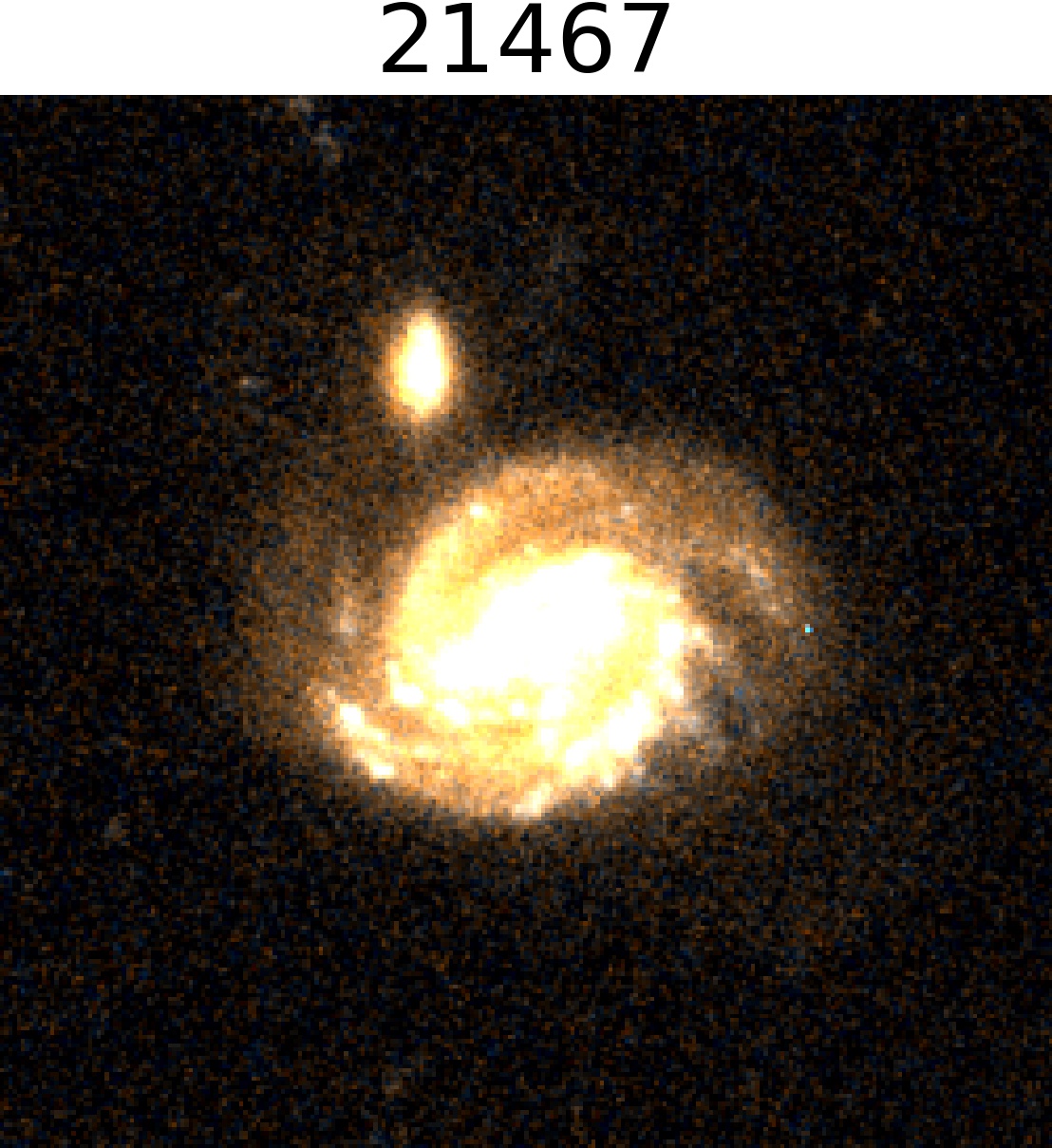}
    \includegraphics[width=0.22\textwidth,trim=0cm 0cm 0cm -0.3cm]{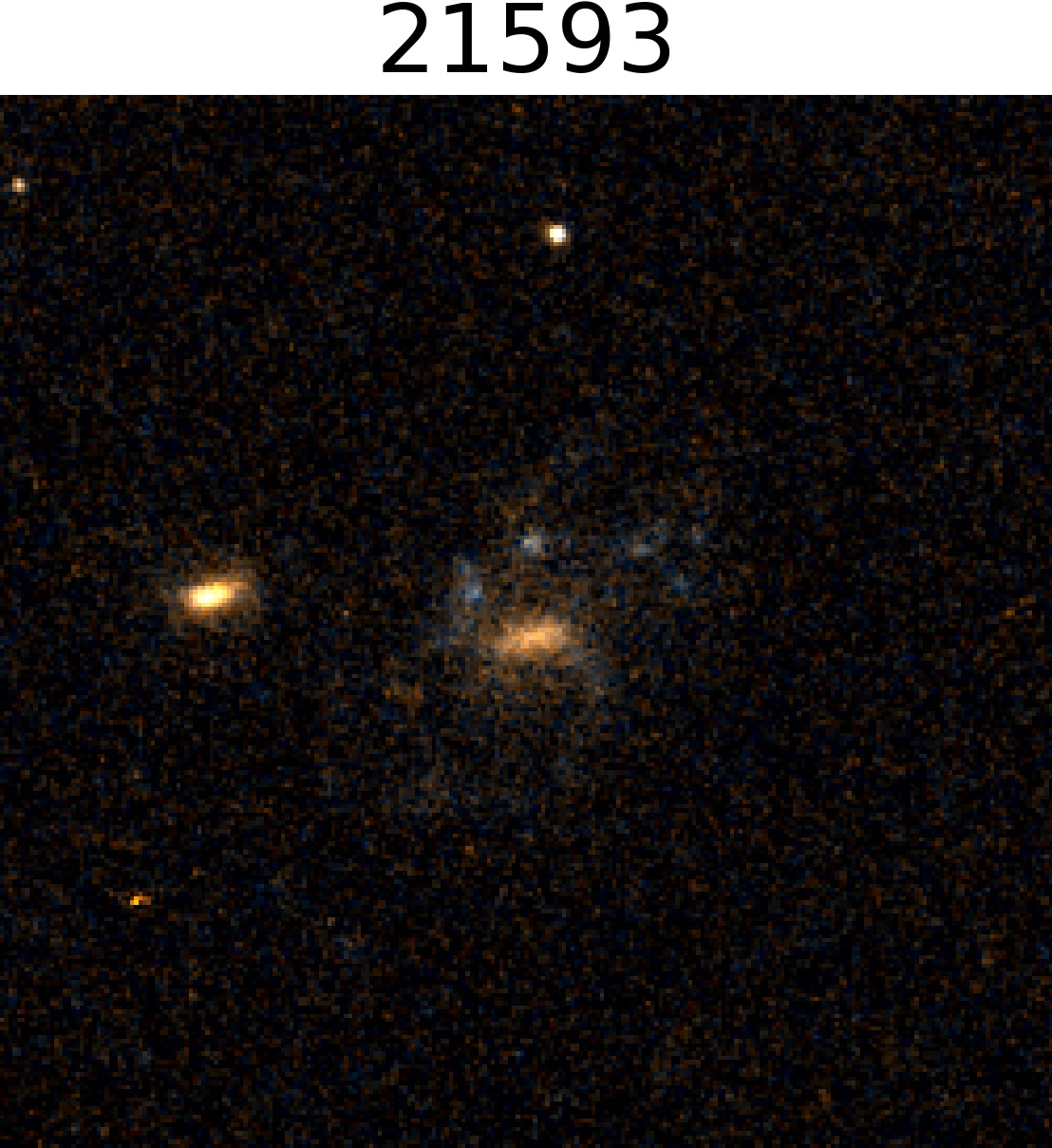}
    \includegraphics[width=0.22\textwidth,trim=0cm 0cm 0cm -0.3cm]{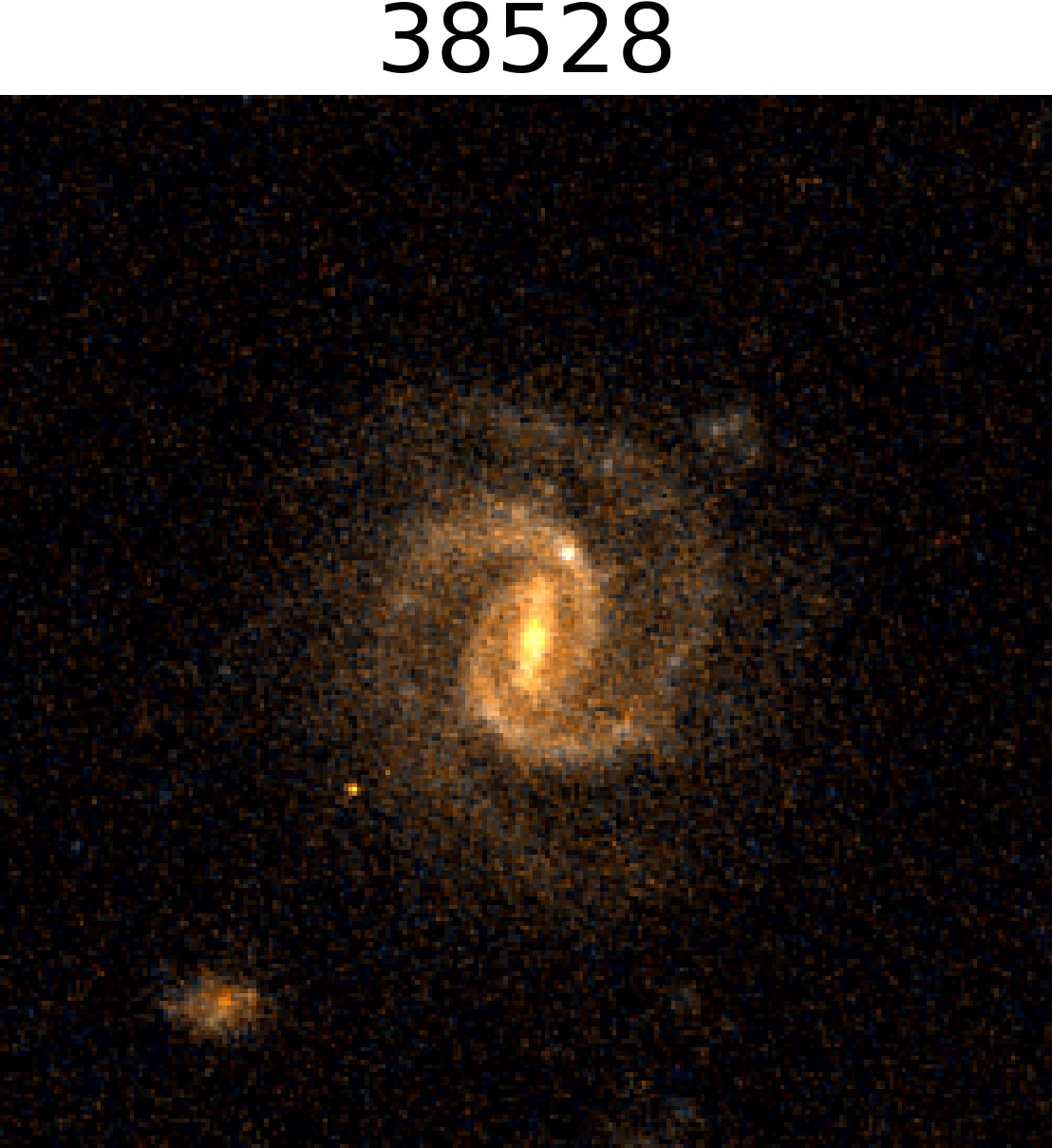}
    \caption{HST color images of galaxies in our final sample with a spiral morphology covered by HST archival images.}
    \label{fig:spirals}
\end{figure*} 
\begin{figure*}[ht!]
    \centering
    \includegraphics[width=0.22\textwidth]{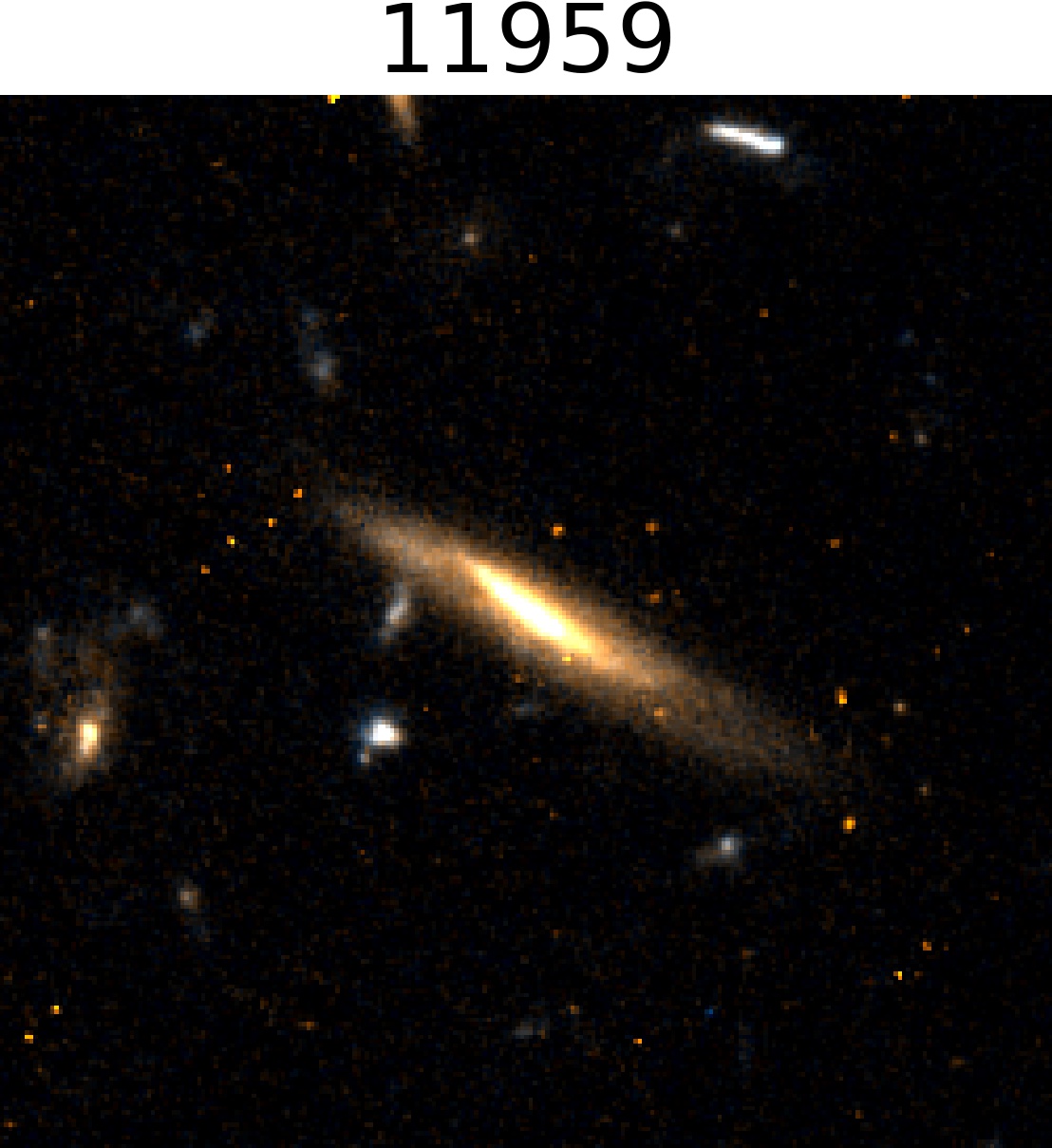}
    \includegraphics[width=0.22\textwidth]{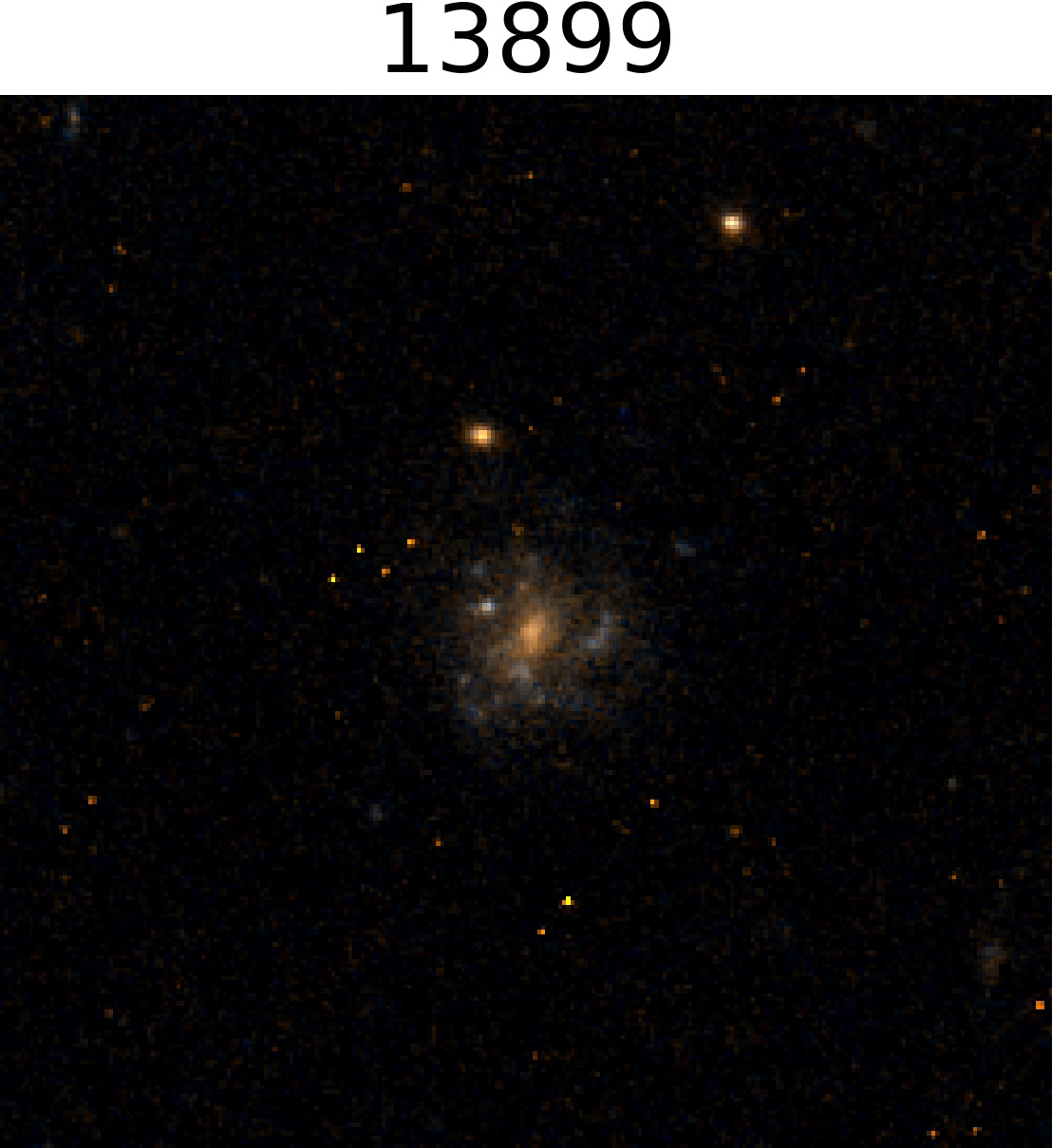}
    \includegraphics[width=0.22\textwidth,trim=0cm 0cm 0cm -0.3cm]{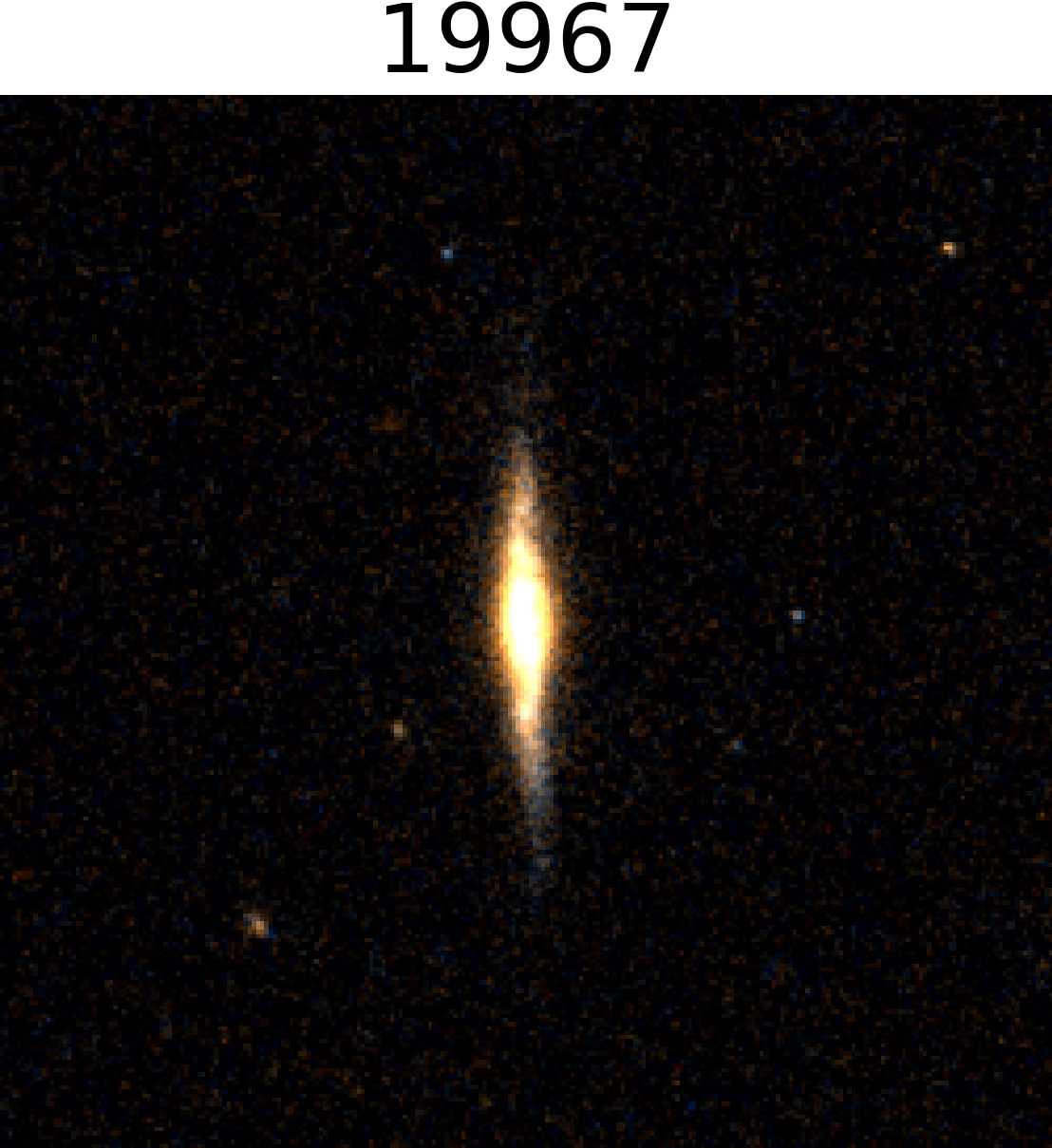}
    \includegraphics[width=0.22\textwidth,trim=0cm 0cm 0cm -0.3cm]{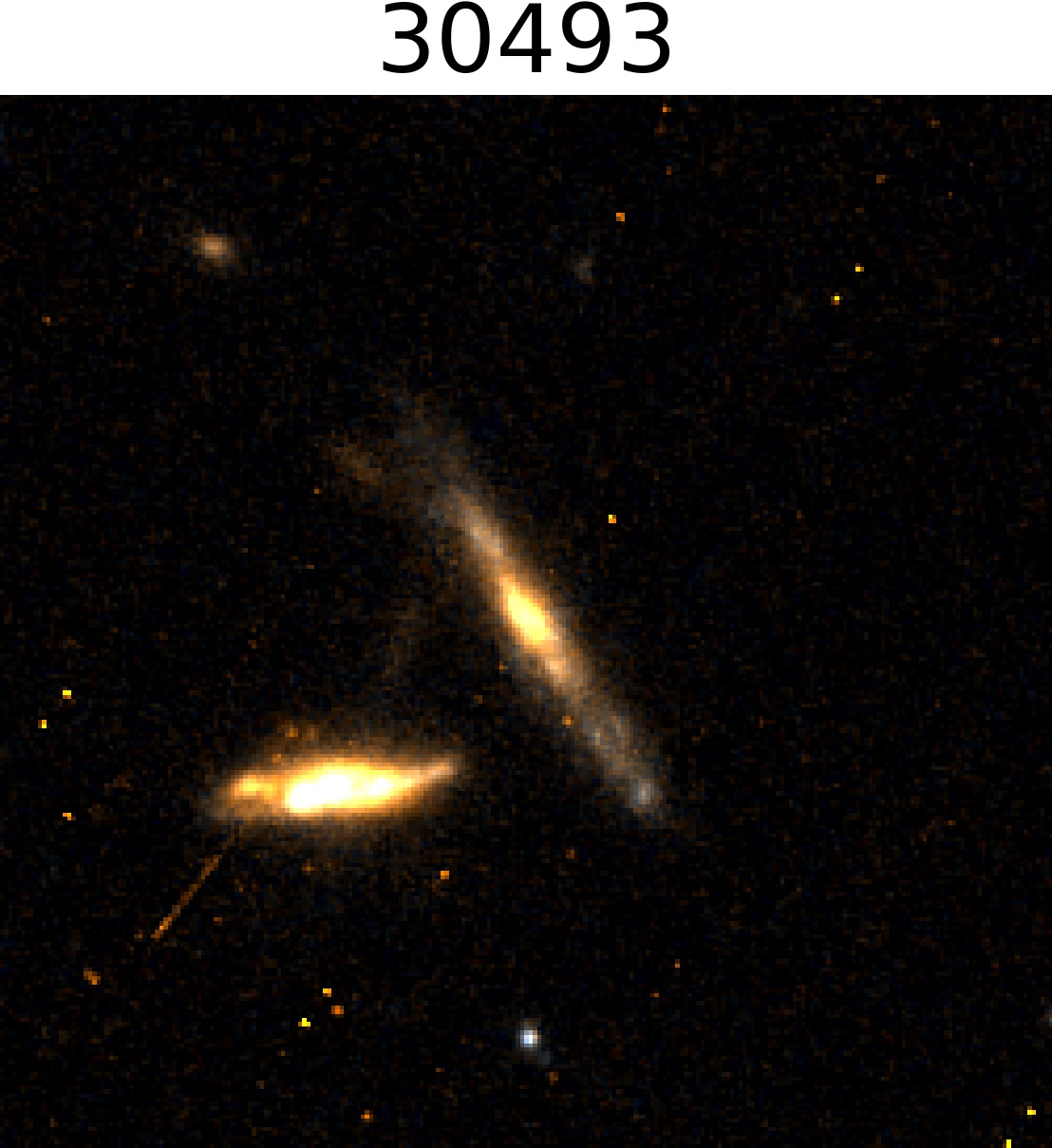}
    \includegraphics[width=0.22\textwidth,trim=0cm 0cm 0cm -0.3cm]{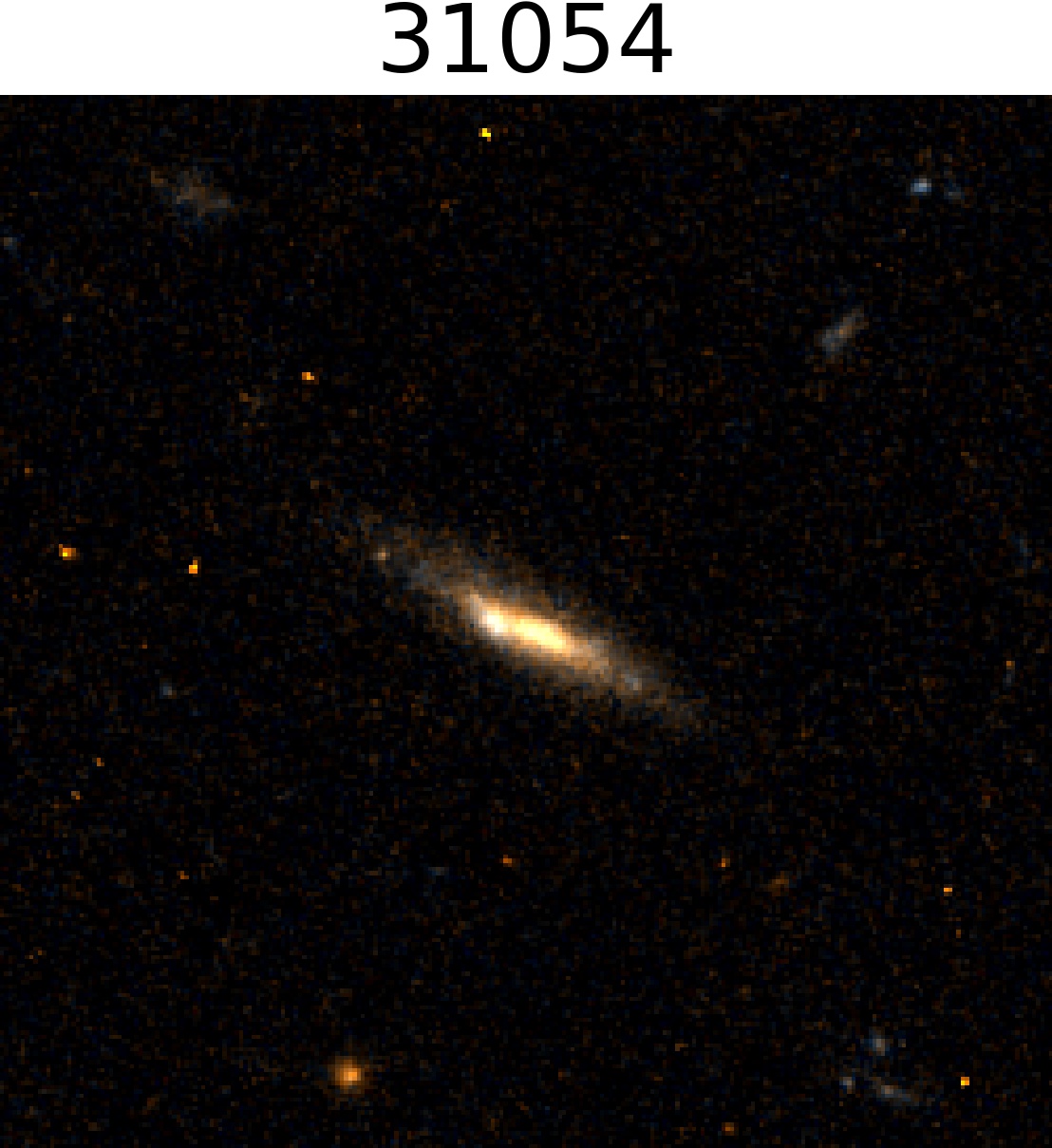}
    \includegraphics[width=0.22\textwidth,trim=0cm 0cm 0cm -0.3cm]{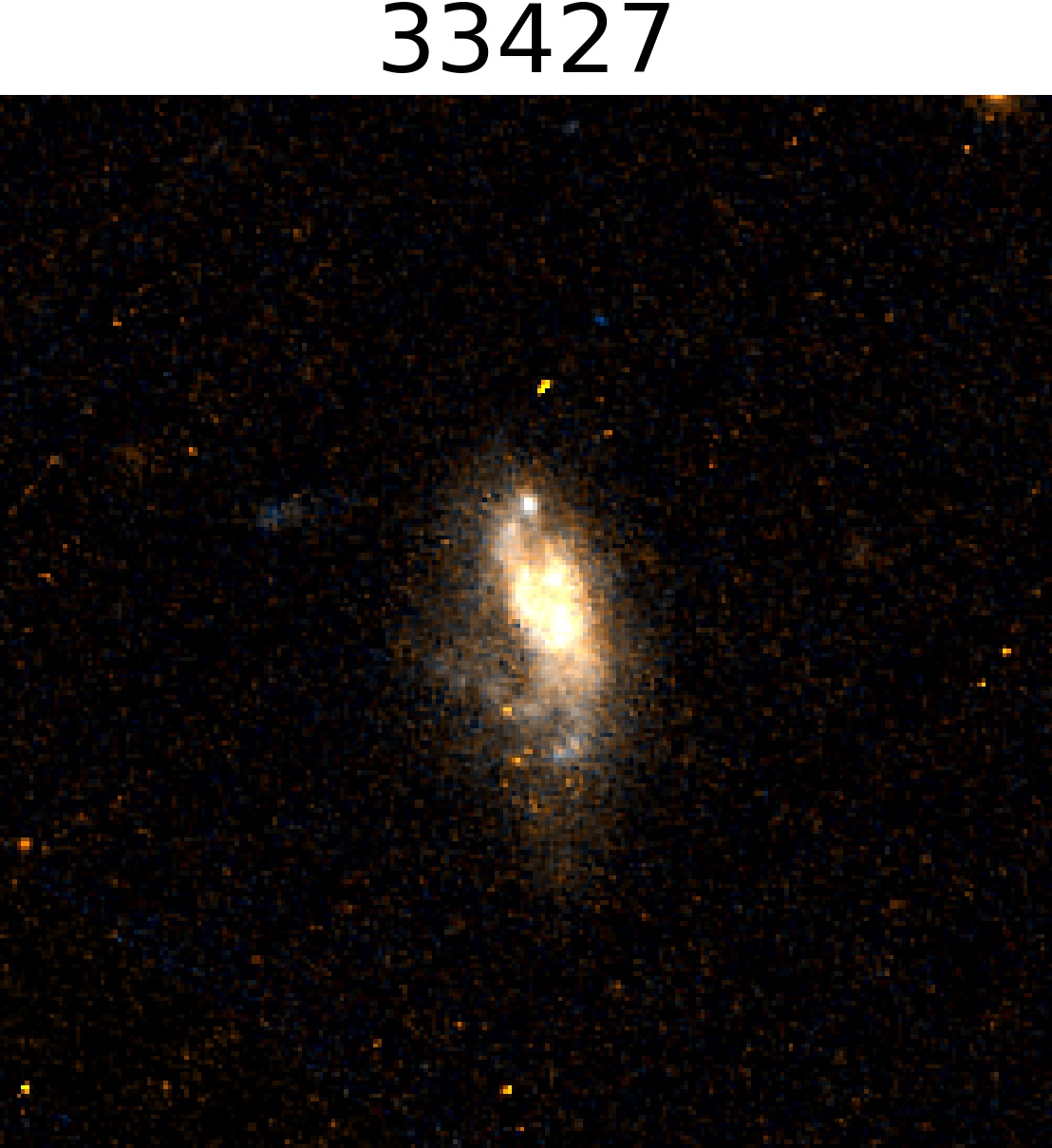}
    \caption{HST color images of galaxies rejected from our sample by the $M_{\mathrm{tot}}-\mu_e$ selection cutoff with a spiral morphology covered by HST archival images.}
    \label{fig:spiralsrejected}
\end{figure*} 
We found 14 galaxies with a spiral morphology in our A1656 sample before applying the $M_{\mathrm{tot}}-\mu_e$ selection cutoff using HST images (see Section \ref{sec:HSTmorph}). We want to stress that the spiral morphology does not necessarily mean that these galaxies are background galaxies contaminating our A1656 galaxy sample. Their colors are consistent with those of cluster members. We show their positions in the $u'-g'$ vs. $g'-r'$  color--color diagram in Figure \ref{fig:colorcolorspirals}. These 14 galaxies are depicted in red, our final galaxy sample is depicted in blue, and our total sample before selecting the cluster members using the bicolor and red sequence selection is depicted in black. Most of these spiral galaxies are significantly separated from the star-forming sequence, and hence, cannot be random interlopers from the star-forming sequence. The HST color images of the eight galaxies with a spiral morphology from our final A1656 sample are shown in Figure \ref{fig:spirals} and the six of those that were rejected by the $M_{\mathrm{tot}}-\mu_e$ selection cutoff are shown in Figure \ref{fig:spiralsrejected}. Blue corresponds to F475W, green to (F475W+F814W)/2, and red to F814W.
\section{Tables}

We present our \verb+SExtractor+ catalogs  for A262, A1656, the reference field analyzed for A262, and the reference field analyzed for A1656 in Tables \ref{tab:SExA262}, \ref{tab:SExA1656}, \ref{tab:SExRefA262}, and \ref{tab:SExRefA1656}. The full tables are available in machine-readable form. Note here that we do not publish uncertainties for our \verb+SExtractor+ catalogs, as they are drastically underestimated due to using smoothed images and \verb+SExtractor+ estimating the error from the background rms. All magnitudes and surface brightnesses in the \verb+SExtractor+ catalogs are given for $\mathrm{ZP_{10}}$.

Our final catalogs for A262, A1656, the reference field analyzed for A262, and the reference field analyzed for A1656 are presented in Tables \ref{tab:finalA262}, \ref{tab:finalA1656}, \ref{tab:finalRefA262}, and \ref{tab:finalRefA1656}. The full tables are available in machine-readable form. $g'$-band magnitudes and surface brightnesses are given for $\mathrm{ZP_{\mathrm{inf}}}$. The colors are given for $\mathrm{ZP_{10}}$.

\begin{deluxetable*}{cccccccccccc}[ht]
    \tabletypesize{\small}
    \tablecaption{SExtractor Catalog A262}
    \label{tab:SExA262}
    \tablehead{
        \colhead{ID} & \colhead{R.A.} & \colhead{Decl.} & \colhead{$m_{g',\,\mathrm{aper}}$} & \colhead{$m_{g',\,\mathrm{auto}}$} & \colhead{$m_{u',\,\mathrm{aper}}$} & \colhead{$m_{u',\,\mathrm{auto}}$} & \colhead{$m_{r',\,\mathrm{aper}}$} & \colhead{$m_{r',\,\mathrm{auto}}$} & \colhead{\ldots} & \colhead{S/G} & \colhead{flags} \\
         \colhead{} & \colhead{(J2000)} & \colhead{(J2000)} & \colhead{(mag)} & \colhead{(mag)} & \colhead{(mag)} & \colhead{(mag)} & \colhead{(mag)} & \colhead{(mag)}  & \colhead{\ldots} & \colhead{} & \colhead{} \\
         (1) & (2) & (3) & (4) & (5) & (6) & (7) & (8) & (9) & \ldots & (18) & (19)
    }
    \startdata
    \hline
    1  & 28.2580497 & 35.7526859 & 99.00 & 24.69 & 25.57 & 21.53 & 24.30 & 19.76 & \ldots &  0.65 & 1\\
    \ldots & \ldots & \ldots & \ldots & \ldots & \ldots & \ldots & \ldots & \ldots & \ldots & \ldots & \ldots \\	
    31556 & 27.9705783 & 36.4898063 & 23.76 &18.97 &25.27 &20.44 &22.86 &17.56 & \ldots & 0.00  & 3
    \enddata
    \tablecomments{Parameters for the sample of 31556 objects detected by SExtractor in the A262 field. The columns are: object number in the SExtractor catalog (1), R.A. (2), decl. (3),  $g'$-band circular aperture magnitude (4), $g'$-band Kron-like automated aperture magnitude (5), $u'$-band circular aperture magnitude (6), $u'$-band Kron-like automated aperture magnitude (7), $r'$-band circular aperture magnitude (8), $r'$-band Kron-like automated aperture magnitude (9), $g'$-band peak surface brightness above background (10), $g'$-band effective model surface brightness above the background (11), $g'$-band mean effective model surface brightness above the background (12), Spheroid Sérsic index from fitting (13), directly integrated half-light radius (14), FWHM assuming a gaussian core (15), elongation a/b (16), position angle (17), S/G classifier output (18), and extraction flags (19). The magnitudes and surface brightnesses are given for $\mathrm{ZP_{10}}$. The full table is available in machine-readable form.}
\end{deluxetable*} 

\begin{deluxetable*}{cccccccccccc}[ht]
    \tabletypesize{\small}
   \tablecaption{SExtractor Catalog A1656}
    \label{tab:SExA1656}
    \tablehead{
        \colhead{ID} & \colhead{R.A.} & \colhead{Decl.} & \colhead{$m_{g',\,\mathrm{aper}}$} & \colhead{$m_{g',\,\mathrm{auto}}$} & \colhead{$m_{u',\,\mathrm{aper}}$} & \colhead{$m_{u',\,\mathrm{auto}}$} & \colhead{$m_{r',\,\mathrm{aper}}$} & \colhead{$m_{r',\,\mathrm{auto}}$} & \colhead{\ldots} & \colhead{S/G} & \colhead{flags} \\
        \colhead{} & \colhead{(J2000)} & \colhead{(J2000)} & \colhead{(mag)} & \colhead{(mag)} & \colhead{(mag)} & \colhead{(mag)} & \colhead{(mag)} & \colhead{(mag)}  & \colhead{\ldots} & \colhead{} & \colhead{} \\
        (1) & (2) & (3) & (4) & (5) & (6) & (7) & (8) & (9) & \ldots & (18) & (19)
    }
    \startdata
    1 &195.0933199 &27.6235220 &19.39 &16.83 &21.29& 18.54& 18.80 &16.17 &\ldots & 0.03 &  0\\
    \ldots & \ldots & \ldots & \ldots & \ldots & \ldots & \ldots & \ldots & \ldots & \ldots & \ldots & \ldots \\	
    45163 &195.0257842 &28.2517214 &22.15 &16.98 &23.51 &18.05 &21.79 &16.70 & \ldots & 0.00 &  2 \\
    \enddata
    \tablecomments{Parameters for the sample of 45163 objects detected by SExtractor in in the A1656 field. The columns are: object number in the SExtractor catalog (1), R.A. (2), decl. (3),  $g'$-band circular aperture magnitude (4), $g'$-band Kron-like automated aperture magnitude (5), $u'$-band circular aperture magnitude (6), $u'$-band Kron-like automated aperture magnitude (7), $r'$-band circular aperture magnitude (8), $r'$-band Kron-like automated aperture magnitude (9), $g'$-band peak surface brightness above background (10), $g'$-band effective model surface brightness above the background (11), $g'$-band mean effective model surface brightness above the background (12), Spheroid Sérsic index from fitting (13), directly integrated half-light radius (14), FWHM assuming a gaussian core (15), elongation a/b (16), position angle (17), S/G classifier output (18), and extraction flags (19). The magnitudes and surface brightnesses are given for $\mathrm{ZP_{10}}$. The full table is available in machine-readable form.}
\end{deluxetable*} 

\begin{deluxetable*}{cccccccccccc}[ht]
    \tabletypesize{\small}
    \tablecaption{SExtractor Catalog Reference A262}
   \label{tab:SExRefA262}
    \tablehead{
        \colhead{ID} & \colhead{R.A.} & \colhead{Decl.} & \colhead{$m_{g',\,\mathrm{aper}}$} & \colhead{$m_{g',\,\mathrm{auto}}$} & \colhead{$m_{u',\,\mathrm{aper}}$} & \colhead{$m_{u',\,\mathrm{auto}}$} & \colhead{$m_{r',\,\mathrm{aper}}$} & \colhead{$m_{r',\,\mathrm{auto}}$} & \colhead{\ldots} & \colhead{S/G} & \colhead{flags} \\
        \colhead{} & \colhead{(J2000)} & \colhead{(J2000)} & \colhead{(mag)} & \colhead{(mag)} & \colhead{(mag)} & \colhead{(mag)} & \colhead{(mag)} & \colhead{(mag)}  & \colhead{\ldots} & \colhead{} & \colhead{} \\
        (1) & (2) & (3) & (4) & (5) & (6) & (7) & (8) & (9) & \ldots & (18) & (19)
    }
    \startdata
    1 &218.4746111 &60.0144578 &23.25 &20.05 &25.31 &20.01 &22.42 &17.12 &\ldots & 0.00 & 3\\
    \ldots & \ldots & \ldots & \ldots & \ldots & \ldots & \ldots & \ldots & \ldots & \ldots & \ldots & \ldots \\	
    9824 &218.2629421 &60.3935867 &24.88 &23.46 &25.31 &23.92& 24.17& 22.88 &\ldots &0.12 & 0\\
    \enddata
    \tablecomments{Parameters for the sample of 9824 objects detected by SExtractor in the reference field when analyzed for A262. The columns are: object number in the SExtractor catalog (1), R.A. (2), decl. (3),  $g'$-band circular aperture magnitude (4), $g'$-band Kron-like automated aperture magnitude (5), $u'$-band circular aperture magnitude (6), $u'$-band Kron-like automated aperture magnitude (7), $r'$-band circular aperture magnitude (8), $r'$-band Kron-like automated aperture magnitude (9), $g'$-band peak surface brightness above background (10), $g'$-band effective model surface brightness above the background (11), $g'$-band mean effective model surface brightness above the background (12), Spheroid Sérsic index from fitting (13), directly integrated half-light radius (14), FWHM assuming a gaussian core (15), elongation a/b (16), position angle (17), S/G classifier output (18), and extraction flags (19). The magnitudes and surface brightnesses are given for $\mathrm{ZP_{10}}$. The full table is available in machine-readable form.}
\end{deluxetable*}    

\begin{deluxetable*}{cccccccccccc}[ht]
    \tabletypesize{\small}
    \tablecaption{SExtractor Catalog Reference A1656}
    \label{tab:SExRefA1656}
    \tablehead{
        \colhead{ID} & \colhead{R.A.} & \colhead{Decl.} & \colhead{$m_{g',\,\mathrm{aper}}$} & \colhead{$m_{g',\,\mathrm{auto}}$} & \colhead{$m_{u',\,\mathrm{aper}}$} & \colhead{$m_{u',\,\mathrm{auto}}$} & \colhead{$m_{r',\,\mathrm{aper}}$} & \colhead{$m_{r',\,\mathrm{auto}}$} & \colhead{\ldots} & \colhead{S/G} & \colhead{flags} \\
        \colhead{} & \colhead{(J2000)} & \colhead{(J2000)} & \colhead{(mag)} & \colhead{(mag)} & \colhead{(mag)} & \colhead{(mag)} & \colhead{(mag)} & \colhead{(mag)}  & \colhead{\ldots} & \colhead{} & \colhead{} \\
        (1) & (2) & (3) & (4) & (5) & (6) & (7) & (8) & (9) & \ldots & (18) & (19)
    }
    \startdata
    1 &218.4746143 &60.0143847 &23.63 &20.50 &25.79 &20.43 &22.88 &17.50 &\ldots & 0.03 &  1\\
    \ldots & \ldots & \ldots & \ldots & \ldots & \ldots & \ldots & \ldots & \ldots & \ldots & \ldots & \ldots \\	
    11488 &218.4746113 &60.0144582 &23.25 &20.10 &25.33 &20.06 &22.42 &17.15 & \ldots & 0.00 & 3
    \enddata
    \tablecomments{Parameters for the sample of 11488 objects detected by SExtractor in the reference field when analyzed for A1656. The columns are: object number in the SExtractor catalog (1), R.A. (2), decl. (3),  $g'$-band circular aperture magnitude (4), $g'$-band Kron-like automated aperture magnitude (5), $u'$-band circular aperture magnitude (6), $u'$-band Kron-like automated aperture magnitude (7), $r'$-band circular aperture magnitude (8), $r'$-band Kron-like automated aperture magnitude (9), $g'$-band peak surface brightness above background (10), $g'$-band effective model surface brightness above the background (11), $g'$-band mean effective model surface brightness above the background (12), Spheroid Sérsic index from fitting (13), directly integrated half-light radius (14), FWHM assuming a gaussian core (15), elongation a/b (16), position angle (17), S/G classifier output (18), and extraction flags (19). The magnitudes and surface brightnesses are given for $\mathrm{ZP_{10}}$. The full table is available in machine-readable form.}
\end{deluxetable*} 

\begin{deluxetable*}{cccccccccccc}[ht]
    \tabletypesize{\small}
    \tablecaption{Final Catalog A262}
    \label{tab:finalA262}
    \tablehead{
        \colhead{ID} & \colhead{R.A.} & \colhead{Decl.} & \colhead{$m_{g'}$} & \colhead{$\delta m_{g'}$} & \colhead{$M_{g'}$} & \colhead{$\delta M_{g'}$} & \colhead{$M_{V}$} & \colhead{$\delta M_{V}$}  &\colhead{...} & \colhead{P.A.} & \colhead{$\delta$ P.A.} \\
        \colhead{} & \colhead{(J2000)} & \colhead{(J2000)} & \colhead{(mag)} & \colhead{(mag)} & \colhead{(mag)} & \colhead{(mag)} & \colhead{(mag)} & \colhead{(mag)} & \colhead{\ldots} & \colhead{} & \colhead{} \\
        (1) & (2) & (3) & (4) & (5) & (6) & (7) & (8) & (9) & \ldots & (34) & (35)
    }
    \startdata
    12& 28.1613427  & 35.7664418 & 20.2122 & 0.0295 & -14.0938 & 0.0295 & -14.35 & 0.03 &  \ldots & 3.7435 & 4.1859\\
    \ldots & \ldots & \ldots & \ldots & \ldots & \ldots & \ldots &  \ldots & \ldots & \ldots & \ldots\\	
    31496  &28.0781689 &36.5186676 &20.9076 &0.0528 &-13.4001 &0.0528 &-13.67 &0.06 &\ldots  &-9.4010  &2.5291\\
    \enddata
    \tablecomments{GALFIT parameters and elliptical aperture colors for the final sample of 185 dwarf galaxies in A262. The columns are: object number in the SExtractor catalog (1), R.A. (2), decl. (3), apparent $g'$-band magnitude (4), error of apparent $g'$-band magnitude (5), absolute $g'$-band magnitude (6), error of absolute $g'$-band magnitude (7), absolute $V$-band magnitude (8), error of absolute $V$-band magnitude (9), absolute $B$-band magnitude (10), error of absolute $B$-band magnitude (11), $u'-g'$ (12), error of $u'-g'$ (13), $g'-r'$ (14), error of $g'-r'$ (15), $u'-r'$ (16), error of $u'-r'$ (17), mean surface brightness within $R_e$ (18), error of mean surface brightness within $R_e$ (19), $g'$-band surface brightness at $R_e$ (20), error of $g'$-band surface brightness at $R_e$ (21), $V$-band surface brightness at $R_e$ (22), error of $V$-band surface brightness at $R_e$ (23), central surface brightness $g'$-band (24), error of central surface brightness $g'$-band (25), central surface brightness $B$-band (26), error of central surface brightness $B$-band (27), Sérsic index (28), error of Sérsic index (29), half-light radius (30), error of half-light radius (31), axis ratio (b/a) (32), error of axisratio (b/a) (33), position angle (34), and error of position angle (35). The $g'$-band magnitudes and surface brightnesses are given for $\mathrm{ZP_{\mathrm{inf}}}$ and the colors are given for $\mathrm{ZP_{10}}$. The full table is available in machine-readable form.}
\end{deluxetable*} 

\begin{deluxetable*}{cccccccccccc}[ht]
    \tabletypesize{\small}
    \tablecaption{Final Catalog A1656}
    \label{tab:finalA1656}
    \tablehead{
        \colhead{ID} & \colhead{R.A.} & \colhead{Decl.} & \colhead{$m_{g'}$} & \colhead{$\delta m_{g'}$} & \colhead{$M_{g'}$} & \colhead{$\delta M_{g'}$} & \colhead{$M_{V}$} & \colhead{$\delta M_{V}$}  &\colhead{...} & \colhead{P.A.} & \colhead{$\delta$ P.A.} \\
        \colhead{} & \colhead{(J2000)} & \colhead{(J2000)} & \colhead{(mag)} & \colhead{(mag)} & \colhead{(mag)} & \colhead{(mag)} & \colhead{(mag)} & \colhead{(mag)} & \colhead{\ldots} & \colhead{} & \colhead{} \\
        (1) & (2) & (3) & (4) & (5) & (6) & (7) & (8) & (9) & \ldots & (34) & (35)
    }
    \startdata
     2 & 194.9256560 & 27.6129024 &20.9263 &0.0667 &-14.2489 &0.0667 &-14.53 &0.07 & \ldots & -2.9467  &2.1811\\
    \ldots & \ldots & \ldots & \ldots & \ldots & \ldots & \ldots &  \ldots & \ldots & \ldots & \ldots\\	
    45157 &194.6432860 &27.9393896 &19.2738 &0.0230 &-15.9154 &0.0230 &-16.26 &0.03 &\ldots &-41.9550  &1.1489\\
    \enddata
    \tablecomments{GALFIT parameters and elliptical aperture colors for the final sample of 900 dwarf galaxies in A1656. The columns are: object number in the SExtractor catalog (1), R.A. (2), decl. (3), apparent $g'$-band magnitude (4), error of apparent $g'$-band magnitude (5), absolute $g'$-band magnitude (6), error of absolute $g'$-band magnitude (7), absolute $V$-band magnitude (8), error of absolute $V$-band magnitude (9), absolute $B$-band magnitude (10), error of absolute $B$-band magnitude (11), $u'-g'$ (12), error of $u'-g'$ (13), $g'-r'$ (14), error of $g'-r'$ (15), $u'-r'$ (16), error of $u'-r'$ (17), mean surface brightness within $R_e$ (18), error of mean surface brightness within $R_e$ (19), $g'$-band surface brightness at $R_e$ (20), error of $g'$-band surface brightness at $R_e$ (21), $V$-band surface brightness at $R_e$ (22), error of $V$-band surface brightness at $R_e$ (23), central surface brightness $g'$-band (24), error of central surface brightness $g'$-band (25), central surface brightness $B$-band (26), error of central surface brightness $B$-band (27), Sérsic index (28), error of Sérsic index (29), half-light radius (30), error of half-light radius (31), axis ratio (b/a) (32), error of axisratio (b/a) (33), position angle (34), and error of position angle (35). The $g'$-band magnitudes and surface brightnesses are given for $\mathrm{ZP_{\mathrm{inf}}}$ and the colors are given for $\mathrm{ZP_{10}}$. The full table is available in machine-readable form.}
\end{deluxetable*} 

\begin{deluxetable*}{cccccccccccc}[ht]
    \tabletypesize{\small}
    \tablecaption{Final Catalog Reference A262}
    \label{tab:finalRefA262}
    \tablehead{
        \colhead{ID} & \colhead{R.A.} & \colhead{Decl.} & \colhead{$m_{g'}$} & \colhead{$\delta m_{g'}$} & \colhead{$M_{g'}$} & \colhead{$\delta M_{g'}$} & \colhead{$M_{V}$} & \colhead{$\delta M_{V}$}  &\colhead{...} & \colhead{P.A.} & \colhead{$\delta$ P.A.} \\
        \colhead{} & \colhead{(J2000)} & \colhead{(J2000)} & \colhead{(mag)} & \colhead{(mag)} & \colhead{(mag)} & \colhead{(mag)} & \colhead{(mag)} & \colhead{(mag)} & \colhead{\ldots} & \colhead{} & \colhead{} \\
        (1) & (2) & (3) & (4) & (5) & (6) & (7) & (8) & (9) & \ldots & (34) & (35)
    }
    \startdata
    2090 &218.3857989 &60.0720155 &21.6090 &0.0531 &-12.7081 &0.0531 &-13.05 &0.06 &\ldots &-74.1852  &0.9548\\
    \ldots & \ldots & \ldots & \ldots & \ldots & \ldots & \ldots &  \ldots & \ldots & \ldots & \ldots\\	
    7516 &217.9862720 &60.3021937 &22.1762 &0.1272 &-12.1346 &0.1272 &-12.42 &0.13 &\ldots &-81.7223  &4.3159\\
    \enddata
    \tablecomments{GALFIT parameters and elliptical aperture colors for the remaining final sample of 11 galaxies in reference field when analyzed for A262. The columns are: object number in the SExtractor catalog (1), R.A. (2), decl. (3), apparent $g'$-band magnitude (4), error of apparent $g'$-band magnitude (5), absolute $g'$-band magnitude (6), error of absolute $g'$-band magnitude (7), absolute $V$-band magnitude (8), error of absolute $V$-band magnitude (9), absolute $B$-band magnitude (10), error of absolute $B$-band magnitude (11), $u'-g'$ (12), error of $u'-g'$ (13), $g'-r'$ (14), error of $g'-r'$ (15), $u'-r'$ (16), error of $u'-r'$ (17), mean surface brightness within $R_e$ (18), error of mean surface brightness within $R_e$ (19), $g'$-band surface brightness at $R_e$ (20), error of $g'$-band surface brightness at $R_e$ (21), $V$-band surface brightness at $R_e$ (22), error of $V$-band surface brightness at $R_e$ (23), central surface brightness $g'$-band (24), error of central surface brightness $g'$-band (25), central surface brightness $B$-band (26), error of central surface brightness $B$-band (27), Sérsic index (28), error of Sérsic index (29), half-light radius (30), error of half-light radius (31), axis ratio (b/a) (32), error of axisratio (b/a) (33), position angle (34), and error of position angle (35). The $g'$-band magnitudes and surface brightnesses are given for $\mathrm{ZP_{\mathrm{inf}}}$ and the colors are given for $\mathrm{ZP_{10}}$. The full table is available in machine-readable form.}
\end{deluxetable*} 

\begin{deluxetable*}{cccccccccccc}[ht]
    \tabletypesize{\small}
    \tablecaption{Final Catalog Reference A1656}
    \label{tab:finalRefA1656}
    \tablehead{
        \colhead{ID} & \colhead{R.A.} & \colhead{Decl.} & \colhead{$m_{g'}$} & \colhead{$\delta m_{g'}$} & \colhead{$M_{g'}$} & \colhead{$\delta M_{g'}$} & \colhead{$M_{V}$} & \colhead{$\delta M_{V}$}  &\colhead{...} & \colhead{P.A.} & \colhead{$\delta$ P.A.} \\
        \colhead{} & \colhead{(J2000)} & \colhead{(J2000)} & \colhead{(mag)} & \colhead{(mag)} & \colhead{(mag)} & \colhead{(mag)} & \colhead{(mag)} & \colhead{(mag)} & \colhead{\ldots} & \colhead{} & \colhead{} \\
        (1) & (2) & (3) & (4) & (5) & (6) & (7) & (8) & (9) & \ldots & (34) & (35)
    }
    \startdata
    648 &217.7855009 &60.0220749 &20.9178 &0.0077 &-14.2700 &0.0077 &-14.61 &0.01 &\ldots &-80.1322  &0.7365\\
    \ldots & \ldots & \ldots & \ldots & \ldots & \ldots & \ldots &  \ldots & \ldots & \ldots & \ldots\\	
    11440 &218.4717012 &60.3977951 &21.1158 &0.0079 &-14.0648 &0.0079 &-14.37 &0.02 &\ldots &-50.0793 &15.8807
    \enddata
    \tablecomments{GALFIT parameters and elliptical aperture colors for the remaining final sample of 33 galaxies in reference field when analyzed for A1656. The columns are: object number in the SExtractor catalog (1), R.A. (2), decl. (3), apparent $g'$-band magnitude (4), error of apparent $g'$-band magnitude (5), absolute $g'$-band magnitude (6), error of absolute $g'$-band magnitude (7), absolute $V$-band magnitude (8), error of absolute $V$-band magnitude (9), absolute $B$-band magnitude (10), error of absolute $B$-band magnitude (11), $u'-g'$ (12), error of $u'-g'$ (13), $g'-r'$ (14), error of $g'-r'$ (15), $u'-r'$ (16), error of $u'-r'$ (17), mean surface brightness within $R_e$ (18), error of mean surface brightness within $R_e$ (19), $g'$-band surface brightness at $R_e$ (20), error of $g'$-band surface brightness at $R_e$ (21), $V$-band surface brightness at $R_e$ (22), error of $V$-band surface brightness at $R_e$ (23), central surface brightness $g'$-band (24), error of central surface brightness $g'$-band (25), central surface brightness $B$-band (26), error of central surface brightness $B$-band (27), Sérsic index (28), error of Sérsic index (29), half-light radius (30), error of half-light radius (31), axis ratio (b/a) (32), error of axisratio (b/a) (33), position angle (34), and error of position angle (35). The $g'$-band magnitudes and surface brightnesses are given for $\mathrm{ZP_{\mathrm{inf}}}$ and the colors are given for $\mathrm{ZP_{10}}$. The full table is available in machine-readable form.}
\end{deluxetable*}


\end{document}